\documentclass[final,towcolumn,10pt,5p,number]{elsarticle}
\usepackage{graphicx} % have images not redneread
\usepackage[inkscapelatex=false]{svg}
\usepackage[colorinlistoftodos,prependcaption,textsize=tiny]{todonotes} %annotation
\usepackage{miller} %miller indices
\usepackage{lineno} %numbers of lines
\usepackage{amsmath,amssymb,latexsym}
\usepackage{textgreek}
\usepackage{textalpha}
\usepackage{chemformula}
\usepackage{orcidlink}% to make AB2 look AB_2
\usepackage[mode=text]{siunitx} %Si-Einheiten
\DeclareSIUnit{\angstrom}{\textup{\AA}}
\DeclareSIUnit{\atpercent}{at.-\%}
\DeclareSIUnit{\wtpercent}{wt.-\%}
\usepackage[greek,english]{babel}
\usepackage{placeins}
\usepackage{acronym}

\usepackage{hyperref}
\newcommand*{\doi}[1]{\href{https://doi.org/#1}{\nolinkurl{#1}}}
\usepackage{comment}

\begin{document}
\newacro{OR}[OR]{orientation relationship}
\newacro{HP}[HP]{habit plane}
\newacro{HT}[HT]{heat-treated}
\newacro{AC}[AC]{as-cast}
\newacro{BSE}[BSE]{backscattered electron}
\newacro{odf}[odf]{orientation distribution function}
\newacro{mdf}[mdf]{misorientation distribution function}
\newacro{mrd}[m.r.d.]{multiples of random distribution}
\newacro{CRSS}[CRSS]{criticial resolved shear stress}
\newacro{UTS}[UTS]{ultimate tensile stress}
\newacro{UE}[UE]{uniform elongation}
\newacro{EDX}[EDS]{energy dispersive x-ray spectroscopy}
\newacro{SEM}[SEM]{scanning electron microscope}
\newacro{SE}[SE]{secondary electron}
\newacro{OPS}[OPS]{oxidal polishing suspension}
\newacro{EBSD}[EBSD]{electron backscatter diffraction}
\newacro{SURF}[SURF]{speeded up robust features}
\newacro{XRD}[XRD]{X-ray diffraction}
\newacro{CT}[CT]{computer tomography}
\newacro{RE}[RE]{rare-earth element}
\newacro{FIB}[FIB]{focused ion beam}
\newacro{TEM}[TEM]{transmission electron microscopy}
\newacro{XRM}[XRM]{X-ray microscopy}
\newacro{FOV}[FOV]{field of view}
\newacro{EDM}[EDM]{electrical discharge machining}
\newacro{SAED}[SAED]{selected area electron diffraction}
\newacro{HR}[HR]{high resolution}
\newacro{CI}{confidence index}
\newacro{ILS}[ILS]{interlayer spacing}
\newacro{IL}[IL]{interlayer }
\newacro{SLO}[SLO]{site lattice occupancy}
\newacro{PCA}[PCA]{principal component analysis}
\newacro{EDS}[EDS]{energy dispersive x-ray spectroscopy}
\newacro{NCC}[NCC]{normalised cross correlation}
\newacro{WDS}[WDS]{wavelength dispersive x-ray}
\newacro{IPF}[IPF]{inverse pole figure}
\newacro{SNR}[SNR]{signal to noise ratio}
\newacro{DFT}[DFT]{density functional theory}
\newacro{HR-TEM}[HR-TEM]{high resolution transmission electron microscopy}
\newacro{HR-STEM}[HR-STEM]{high resolution scanning transmission electron microscopy}
\newacro{HAADF}[HAADF]{high angle annular dark field}
\newacro{HOLZ}[HOLZ]{higher order Laue zones}
\newacro{EBSP}[EBSP]{electron backscatter pattern}
\newacro{PSNR}[PSNR]{peak signal to noise ratio}
\newacro{STEM}[STEM]{scanning transmission electron microscopy}
\newacro{PCX}[PCX]{pattern centre, x}
\newacro{PCY}[PCY]{pattern centre, y}
\newacro{PCZ}[PCZ]{pattern centre, z}
\begin{frontmatter}
% Supplementary Information:
\title{EBSD and Subtle Crystallographic Differences - A
Study of Resolving Interlayer Spacings in Nb-Ni and Nb-Co $\mu$ Phases}
% \author{lukas.berners }
\author[IMM]{Lukas Berners \corref{cor1}\,\orcidlink{0009-0006-3418-0355}}
\cortext[cor1]{Corresponding author}
\ead{berners@imm.rwth-aachen.de}
% \author[GFE]{Joshua Spille}
\author[IMM]{Nisa Ulumuddin \,\orcidlink{0000-0002-5926-6749}}

\author[AICES]{Annika Baum \,\orcidlink{0009-0005-9763-8071}}
\author[GFE]{Silvia Richter}
\author[IMM]{Khalil Rejiba \,\orcidlink{0009-0009-6465-4023}}
\author[GFE]{Joshua Spille \, \orcidlink{0000-0001-8708-7214}}
\author[MPIE,HHB]{Siyuan Zhang \, \orcidlink{0000-0001-7045-0865}}
\author[MPIE]{Christina Scheu \, \orcidlink{0000-0001-7916-1533}}
\author[GFE]{Joachim Mayer \, \orcidlink{0000-0003-3292-5342}}
\author[AICES]{Benjamin Berkels \,\orcidlink{0000-0002-6969-187X}}
\author[UBC]{T. Ben Britton\,\orcidlink{0000-0001-5343-9365}}
\author[IMM]{Sandra Korte-Kerzel\,\orcidlink{0000-0002-4143-5129}}
% \author[HHB]{Helmhol}
\affiliation[IMM]{organization={Institute for Physical Metallurgy and Materials Physics, RWTH Aachen University},%Department and Organization
            addressline={Kopernikusstr. 14}, 
            city={Aachen},
            postcode={52074}, 
            country={Germany}}
\affiliation[AICES]{organization = {Aachen Institute for Advanced Study in Computational Engineering - AICES, RWTH Aachen University}, city={Aachen},postcode={52074},country={Germany}}
 
\affiliation[GFE]{organization={Central Facility for Electron Microscopy, RWTH Aachen University},%Department and Organization
addressline={Ahornstr. 55}, 
city={Aachen},
postcode={52074}, 
country={Germany}}
\affiliation[MPIE]{organization = {Max Planck Institute for Sustainable Materials},adressline={Max-Planck-Str. 1},city={Düsseldorf},postcode={40237},country={Germany}}
 \affiliation[HHB]{organization = {Helmholtz-Zentrum Berlin für Materialien und Energie GmbH}, city={Berlin},postcode={12489},country={Germany}}  
 % \affiliation[FGEO]{organization = {Faculty of Georesources and Materials Engineering, RWTH Aachen University},city={Aachen},country={Germany}}
\affiliation[UBC]{ organization = {University of British Columbia, Department of Materials Engineering},adressline={309-6350 Stores Road}, city={Vancouver},postcode={V6T 1Z4},country={Canada}}       

\date{}
% \begin{comment}

\begin{abstract}

In ordered intermetallics, slight variations in lattice site occupancy and specific interlayer spacings have been identified as the sources of significant changes in critical resolved shear stress and therefore how a given phase may affect alloy properties. So far, atom positions and lattice site occupancies have traditionally been characterised by high-resolution transmission electron microscopy (HR-TEM) and X-ray diffraction (XRD), which are methods which offer either local detail or high statistical significance but not both. Electron backscatter diffraction (EBSD), by contrast, provides high spatial resolution across large sample areas and therefore, has the potential to enable the local investigation of interlayer spacing and site lattice occupancy with improved statistical reliability. The objectives of the study are to benchmark EBSD’s capability for resolving these subtle features and to correlate them with compositional and mechanical properties. In this case study, we therefore show that EBSD can resolve key crystallographic features of $\mu$‑phase intermetallics, specifically interlayer spacings. We combine pattern matching with large‑scale dynamical simulations of template libraries guided by XRD based information on lattice parameters. For this, we generate structures that vary in the spacing between triple-layer and Kagome layer and in the site lattice occupancy of the 3a site. 
This approach successfully predicts the change of interlayer spacing between Kagome and triple layers in Nb–Co and Nb–Ni $\mu$‑phases, in good agreement with XRD and HR-TEM.

\end{abstract}

\begin{highlights}

\item EBSD is utilised to investigate crystallographic features which only lead to subtle changes of EBSD pattern intensity.

\item Trends in variation of the interlayer spacing inside the triple layer spacing of Nb-Ni and Nb-Co $\mu$-Phases are observed by this method.

\item These trends are in agreement with literature observations obtained by STEM and XRD.

\end{highlights}
% \end{comment}

\end{frontmatter}
% \linenumbers
\section{Introduction}

A central challenge in materials characterisation is to detect subtle crystallographic variations that can strongly influence material properties, while at the same time retaining sufficient spatial resolution and area coverage to study them within heterogeneous microstructures or in direct correlation with local property measurements. In practice, this is difficult because techniques with very high structural sensitivity are often limited to small local volumes, such as \ac{TEM}, whereas methods that provide more statistically representative information, e.g. \ac{XRD}, commonly average over larger regions and therefore lose local structural detail. This raises the question of whether \ac{EBSD}, as a scanning and diffraction-based method, can detect such subtle structural variations in a spatially resolved manner. This question is particularly relevant for the intermetallic $\mu$-phases, in which small changes in \ac{ILS} and \ac{SLO} are known to influence mechanical properties \cite{Luo.2023} (for the \ac{SLO}, this is in particular the occupancy of the 3a site). Therefore, we use the $\mu$-phase as an example material in this study.

$\mu$-phases are typically hard and brittle phases, observed as a detrimental precipitate phase in refractory-metal super alloys \cite{Chen.2025}. They are topologically close-packed phases, characterised by close packing in atomic arrangements with high coordination number \cite{Frank.1959}, also known as Frank-Kasper phases. The $\mu$-phases have a rhombohedral crystal structure ($R\bar{3}m$, int. tables space group 166 with \ch{Fe7W6} prototype), but are typically investigated with respect to their trigonal unit cell \cite{Sluiter.2003}. The general composition is described as \ch{A6B7}, with A being the larger and B being the smaller atom. However, they are not stoichiometric phases and have varying composition ranges depending on the A and B elements \cite{Joubert.2004}. The unit cell has 5 different lattice sites: 3a, 6c1, 6c2, 6c3, and 18h. For the 6c sites, only the z coordinate can vary, and for the 18h site, the x and z coordinates can vary, while all other coordinate values are dependent on symmetry elements. 

Typically, the larger A atoms in these ordered intermetallics occupy the sites with larger coordination polyhedra (namely 6c1 (CN15), 6c2 (CN16), and 6c3 (CN14)), while B atoms are expected on the 3a and 18h sites (CN12) \cite{Forsyth.1962}. In the co-existence range of most $\mu$ -phases between \qty{40}{\atpercent} and \qty{60}{\atpercent} of A-atom content, the change in composition is expected to be accomplished by occupation of the 3a site with the A element, followed by occupation of the 18h site with the A element \cite{Joubert.2004}. Recent \ac{DFT} calculations by Wu et al. \cite{Wu.2022} indicate that for the system Nb-Zn, this occupancy change can also occur concurrently, meaning that the A atom occupies 18h and 3a sites equally with increasing concentration. As a result of these occupation changes, the interplanar spacings also expand and contract accordingly.

In intermetallic $\mu$-phases, two structural features are therefore of particular interest: \ac{ILS} (with $d_t$ the \ac{ILS} in the triple layer) and \ac{SLO} (i.e. the way the different lattice sites are occupied in an ordered or random manner as the composition deviates from \ch{A6B7}). Both are known to influence mechanical properties   \cite{Luo.2023}. So far, these features have been investigated either locally by \ac{HR-STEM}, for example, in Nb-Co \cite{Luo.2023}, Co-Cr-W-Ta-Al-Ti \cite{Chen.2025}, and Ni-Cr-Mo \cite{Dong.2023}  systems, by means  of Rietveld-refinement of \ac{XRD} data at a more macroscopic scale \cite{Joubert.2002}, or computed using \ac{DFT} for individual unit cells \cite{Wu.2022}. In this work, we will ask the question whether such analyses may also be performed by means of \ac{EBSD} in order to combine its ability to scan across large areas with its potential to reveal crystallographic detail at the level of the spacing of individual layers in the unit cell and occupancy of individual sites with changes in chemical composition.

\Ac{EBSD} is a diffraction-based technique in the \ac{SEM}, that is commonly used to map crystal orientations, classify different phases, and interpret microstructure morphology \cite{Wilkinson.2012} and texture \cite{Randle.2000}. In this method, the diffracted signal is formed by quasi-elastic scattering of the primary electron beam \cite{Reimer.1998} and can be detected using different types of detectors, such as phosphor screens \cite{Schwartz.2009} or direct electron detectors \cite{S.Vespucci.2015,Zhang.2024}. Direct electron detectors offer the advantage of high pattern acuity with no additional optical distortions of the \ac{EBSP} as no additional optics are needed, and also high efficiency of pattern capture. For most \ac{EBSD} experiments, the diffraction pattern typically covers a large angular range (> \qty{90}{°}), which is useful as it provides information about the whole unit cell.

\begin{figure*}
    \centering
    \includegraphics[width=0.8\linewidth,trim = 0cm 7.7cm 0cm 0cm, clip]{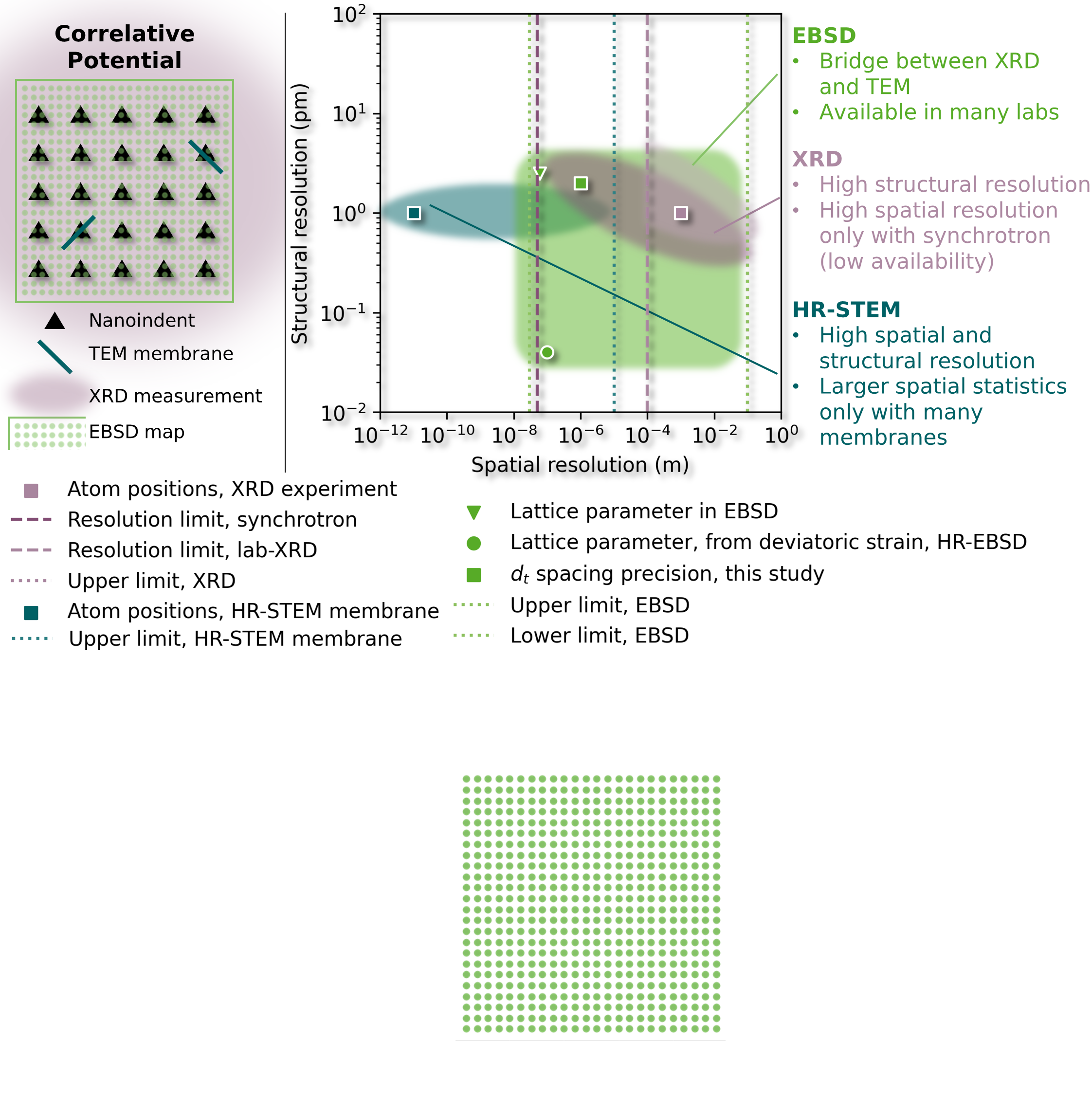}
    \caption{Comparison of spatial and structural resolution of different methods. See discussion in the main text for our considerations on the respective ranges of values. The range for (HR)-STEM is highlighted in teal and that for \ac{EBSD} in green. In the case of \ac{XRD}, the range of synchrotron radiation sources is highlighted in a darker violet colour than lab-based sources. For the structural resolution, the coloured area gives an indication of the range of potential limits, which is relatively large, as such methods are just recently emerging for \ac{EBSD}. For the spatial resolution, the coloured patches give both the resolution limit and also an approximate upper bound, which is practical for investigation. In the "correlative Potential", we showcase the potential application of the methods to nano-indentation arrays. Lab-scale XRD is represented by a diffuse violet circle in the background, while TEM membranes are indicated as dark green streaks and an EBSD map is shown as a grid of bright green dots.}
    \label{fig:ResolutionComparison}
\end{figure*}

 The effective resolution of EBSD depends on the analysis method and how the data is interpreted (e.g. the effective resolution for Hough-based analysis depends on the classification of one pattern with respect to its neighbour). Tong et al. \cite{Tong.2015} demonstrated an effective resolution of $\approx$ \qty{61}{nm} for high angular resolution EBSD in zirconium using cross-correlation methods, for \qty{20}{kV} EBSD patterns captured on an indirect detector. We note that the resolution is, of course, further dependent on the voltage and geometry of the experiment, such that the resolution might be further improved by alternative setups \cite{Marshall.2021, Britton.2026AngRes}. In addition, the chosen approach of using PCA for averaging of EBSD patterns reduced the resolution artificially but does not indicate a limitation of the method itself. Thus, EBSD has a resolution in the range of some other correlative mapping methods, such as nanoindentation. The effective resolution of EBSD for the present study is set at the smallest physical area of any of the PCA components in \autoref{fig:ResolutionComparison}, which is in the micro-meter regime.
% For a sample tilted to \qty{70}{°} for the EBSD analysis, the resolution in $y$ is approximately three times larger in $y$ (due to the elongated electron interaction volume). The penetration depth of the information that contributes to the EBSD patterns depends on the inelastic mean free path of the electrons, for an accelerating voltage of 30 kV Wisniewski et al. \cite{Wisniewski.2016} demonstrated that the parts of the EBSD pattern originate from electron interactions of up to \qty{142}{nm} in the sample. Variations in camera technology (e.g. use of a direct electron detector) enables lower kV EBSD to be routinely performed \cite{Zhang.2025PractCons}, tilt free EBSD geometries further alter this interaction volume and the scattering effects that contribute to the EBSD signal significantly \cite{Marshall.2021, Britton.2026AngRes}. Finally, EBSD-like approaches can be used with an electron transparent thin film to unlock ‘transmission Kikuchi diffraction’ (TKD) which realizes an even smaller interaction volume that can, for a sufficiently thin foil, extends to the electron beam width and consists of only the bottom $\approx$ \qty{30}{nm} of the lamella (with the exact depth related to the Z-number of the material studied) \cite{Sneddon.2024}. 
As we did not find any studies so far investigating measurements of \ac{ILS} from EBSD, we can only compare with resolutions reported for similar quantities, such as the lattice parameter.
For the resolution of lattice parameters by means of EBSD, Nolze et al. \cite{Nolze.2017} reported accuracies of  \qty{0.025}{\angstrom} to \qty{0.1}{\angstrom}. Similarly, mapping of changes in the lattice parameter with $\Delta d / d = 0.02$ are now possible \cite{NOLZE.2021_tetragonal}. 
In contrast, the precision from deviatoric strain tensors using HR-EBSD may approach 1e-4 \cite{Britton.2012}, which would result in 0.04 pm resolution of lattice parameters for simple fcc or bcc metals, or 0.3 pm for larger unit cells such as the $µ$-phase. Both measurements need to be distinguished from measuring \ac{ILS}, as for lattice parameter measurements the major band widths will change, while for local shifts of atoms with a given lattice parameter, they will not. The lattice parameter from the Nolze method is indicated with a green triangle in \autoref{fig:ResolutionComparison}, and the lattice parameter based on deviatoric strains with a green square.
The question of spatial and structural resolution in XRD is similarly complicated. Modern, synchrotron-based radiation sources can achieve spatial resolutions as low as \qty{50}{nm} - \qty{100}{nm} \cite{Schulli.2018} and lab-based XRDs can achieve lattice parameter and atom position or bonding length resolution in the \unit{pm} range \cite{Hill.1992, SanjuanSzklarz.2020}, while the minimum spot size of lab-based XRDs can approach \qty{50}{µm} e.g. \cite{Cao.2026MicroXray}.
%Typically, the structure resolution limit decreases, when the spatial resolution increases, due to the impact on beam convergence angle \cite{Schulli.2018}. 
While strain-mapping of e.g. nano-indentations in the sub-100 nm range is possible using modern beam lines \cite{Lotze.2024}, we did not find any study providing mappings of lattice parameters or atomic spacings in the unit cell at that resolution.% thus we can just presume, based on e.g. the work from Schülli et al. \cite{Schulli.2018}, that the structural resolution in this range is reduced.

Further,  \ac{HR-STEM} offers local investigation of the \ac{SLO} and \ac{ILS} with high statistical significance possible within the given volume of a TEM membrane \cite{Azulay.2025}, but with very low statistical significance with respect to much larger volumes or surface areas investigated in terms of properties, e.g. by indentation. This is due to the limited number and size of possible samples. XRD offers a high significance in terms of the volume or surface to be correlated with properties, but in typical laboratory settings does not offer enough spatial resolution of the information.  In contrast, \ac{EBSD} is known to have a high spatial resolution, e.g. \cite{Steinmetz.2010, Zaefferer.2011}, but can be applied to large sample areas without the need for time-consuming sample preparation specific to each investigated site (as is the case for TEM). \ac{TEM}-based diffraction for a single sample can also be difficult, as these diffraction patterns subtend a narrow capture angle, and therefore the crystal must be rotated to multiple angles to sufficiently access information regarding the full unit cell. For a complete overview of advantages and disadvantages, as well as indications of the potential spatial resolutions, see \autoref{fig:ResolutionComparison}. In this figure, we further indicated the potential of the different methods to be correlated with local probing methods such as nano-indentation. While hypothetically, a TEM membrane could be milled at every indent (dark green lines in \autoref{fig:ResolutionComparison}), this would require more than a complete work day for milling and analysis per membrane. In contrast to this, lab-scale XRD (violet blurred patch) would only offer one point of information at the position of the indent array, while EBSD (grid of green dots) could easily sample multiple points per indent.

In recent years, pattern matching and similar routines in \ac{EBSD} have been adopted to several very subtle problems that use the diffraction signal of an \ac{EBSD} pattern to obtain information about the crystal lattice. This includes: phase identification of crystallographically similar phases on examples of ferrite and cementite \cite{Ram.2018}, $\delta$-and $\eta$-phases in Ni-based super alloys \cite{Sharma.2021}, $\gamma$-and $\gamma'$-phases \cite{T.P.McAuliffe.2020_darkfield}, carbides in Ni-Co-based super alloys \cite{McAuliffe.2021}, identification of crystallographic point group \cite{Winkelmann.2015}, chirality \cite{Winkelmann.2023_chirality,Cios.2025}, polarity \cite{Wilkinson.2019,NareshKumar.2017}, enantiomorphism \cite{Burkhardt.2020}, or pseudosymmetry \cite{Martin.2022}. Further, \ac{EBSD} has been employed extensively to determine local lattice strains in the crystal lattice, as induced by e.g. martensitic tetragonality \cite{Cios.2023} or to analyse the strain caused by dislocations \cite{Abdellaoui.2021} or stacking faults \cite{ABDELLAOUI2019135} in the crystal lattice \cite{Britton.2011}. 
The method has also been used to compare the backscatter coefficient, estimated from \ac{EBSD} intensities, to the atomic number of elements \cite{Nolze.2017}. Sometimes, these distinctions are even possible without the use of advanced pattern matching techniques, e.g. for distinguishing Feldspar members \cite{Wallace.2019}. Further, the atomic structure can be retrieved using advanced reconstruction methods, such as electron holography \cite{Luhr.2016}.

In light of these recent advances, we investigate whether \ac{EBSD} is capable of measuring small changes in the \ac{ILS} that result from changing site-lattice occupancy in intermetallic Nb-Ni and Nb-Co $\mu$-phases (specifically the variation of the triple-layer spacing as a result of 3a site occupancy). To address this question, we use dynamical template matching based on simulations with EMSoft \cite{Callahan.2013} and assess the influence of simulation parameters, noise, and projection-centre misalignment on the detectability of small changes in site-lattice occupancy and specific \ac{ILS}. We use an EDAX Clarity\textsuperscript{\texttrademark} direct electron \ac{EBSD} detector, which offers the advantage of optically undistorted detection of \acp{EBSP}. Analysis of these high-quality patterns is further augmented with large-area projection centre calibration, noise and dimensionality reduction by means of \ac{PCA}, and \ac{XRD}- informed template matching, to retrieve information about atomic spacings inside the triple layer, as well as site lattice occupancies of various lattice sites.  For this template-based \ac{EBSD} pattern matching, we build structures that either vary in the spacing of the triple layer (by modification of the 6c2 position) or in the occupancies of the 3a crystal sites. The best \ac{ILS} or occupancies are then obtained by fitting a 4$^{th}$-degree polynomial to the various fits between experimental pattern and simulated structure.

\section{Materials and Methods}

\subsection{Experimental procedure}
\subsubsection{Sample synthesis}
The Nb-Ni and Nb-Co samples, utilised in this study to investigate \ac{SLO} and \ac{ILS}, were synthesised using arc-melting of the raw materials Ni, Nb and Co (\qty{99.95}{\%}, \qty{99.99}{\%} and \qty{99.98}{\%} purity, each) using a MAM-1 arcmelter (Edmund Buehler, Bodelshausen, D). The materials were re-melted at least three times for each sample. The composition of the samples was chosen to be at the edges of the maximum solubility of the $\mu$-phase coexistence range in the Nb-Co \cite{Stein.2008} and Nb-Ni \cite{Okamoto.2008} phase diagrams, respectively (compositions: 51Nb-49Ni, 58Nb-42Ni, 48Nb-52Co and 56Nb-44Co, all in \unit{\atpercent}). No further heat treatment was performed.

\subsubsection{Metallographic preparation}
The samples investigated in this study were cut using electrical discharge machining. Subsequently, the samples were ground using SiC paper (Struers) with Grit 1200 (US \#500), Grit 2000 (US \#800) and grit 4000 (US \#1200), followed by polishing on MD Dac (Struers) polishing cloths with \qty{6}{µm},  \qty{3}{µm} and  \qty{1}{µm} polycrystalline diamond suspension, supplying isopropyl alcohol + PEG400 as a lubricant for multiple hours on each cloth. Final polishing was achieved using OPS for  \qty{60}{s} followed by cleaning using a detergent-water solution for  \qty{120}{s}. Electropolishing using A2 electrolyte from Struers was tested on one sample, but it led to strong surface relief compared to the OPS-based preparation route, which was considered insufficient for the following EBSD measurements.

\subsubsection{Electron microscopy}
For electron microscopy, a Clara \textsuperscript{\texttrademark} SEM (Tescan, Brno, CZ) equipped with a Clarity\textsuperscript{\texttrademark} direct electron detector-based EBSD camera (Edax Ametek, Tilburg, NL) was used at \qty{20}{kV} acceleration voltage and beam current of \qty{10}{nA}. The Clarity\textsuperscript{\texttrademark} was operated at a detector threshold value of 407, which is slightly above the detector noise level. For the Nb-Co samples, the initial micro-structure characterisation was performed on a FEI-Helios 600 i (FEI, Eindhoven, NL), using a Hikari\textsuperscript{\texttrademark} camera, operated at \qty{20}{kV} and \qty{11}{nA}. For initial microstructure and texture determination, EBSD maps with \qty{500}{µm} $\times$ \qty{500}{µm} size and \qty{2}{µm} step size were recorded.%  For the Nb-Co samples, this was done on a cross-section of the sample, parallel to the bottom. For further investigations (apart from initial microstructure characterisation), these samples were re-cut, as later also indicated in the microstructure overview, \autoref{fig:Microstructure}). 
For all the following \ac{EBSD} maps, the scan mode analysis was used. Alignment of the samples was done based on a straight edge of the sample holder with a pre-tilt of \qty{70}{°}. For initial pattern centre calibration, a large map of \qty{1490}{µm} $\times$ \qty{1490}{µm} at a step size of \qty{10}{µm} and frame averaging of \num{3} was used. 
For the smaller EBSD maps used in the analysis of \ac{ILS} and site lattice occupancies, maps of \qty{15}{µm} $\times$ \qty{15}{µm} in size with a step size of \qty{0.2}{µm} were used. Due to large grain sizes in the 56Nb-44Co sample, some of the maps were taken at a larger size of \qty{25}{µm} $\times$ \qty{25} {µm}.
For all maps, we performed static background division, dynamic background subtraction and intensity histogram normalisation in the EDAX APEX software.

For \ac{EDS} investigation of the samples, a Helios 600i (FEI, Eindhoven, NL) equipped with an Octane Elite EDS Detector (EDAX Ametek, Tilburg, NL) operated at \qty{20}{kV} and \qty{2.7}{nA} was used. For quantification, at least 3 EDS spot measurements of 30 s at multiple positions of the sample were used. The data was processed in PeBaZAF mode \cite{Eggert.2018,Eggert.2019}.
For \ac{WDS} analysis, elemental line scans were acquired on a Schottky field-emission gun electron microprobe, JEOL JXA-8530F (JEOL Ltd., Tokyo, Japan), using an accelerating voltage of 15 kV and a beam current of \qty{100}{nA}. X-ray lines Nb L $\alpha$ and Ni K $\alpha$  were selected and measured with the crystals PETH and LiF. Line scans were performed with a stage movement of \qty{2}{µm} and about \num{100} measuring points. At each step, the peak and background X-ray intensities were acquired with a measuring time of \qty{10}{s}. Quantification of the line scans was performed using pure standards for Ni and Nb. NbC was used as a second standard for a deeper validation of the Nb concentration values. Calibrated X-ray intensities were transferred to elemental concentrations by using a matrix correction procedure based on the Phi(rhoz)-model developed by Pouchou and Pichoir \cite{Pouchou1991}.

Membranes for \acf{HR-STEM} have been prepared using an FEI Strata 400 dual beam \ac{FIB}. \ac{HR-STEM} was conducted on a Spectra microscope (Thermo Fisher, Waltham, US) operated at \qty{300}{kV}. Aberration correction of the STEM probe with \qty{21}{mrad} semi-convergence angle enables imaging at \qty{0.1}{nm} resolution. \Ac{HAADF}-\ac{STEM} images were collected using electrons scattered to the range of 78-200 mrad.

\subsubsection{XRD analysis}
\ac{XRD} on Nb-Co samples was performed using a D8 Advance X-ray texture diffractometer (Bruker, Karlsruhe, D) with an FeK\textsubscript{\textalpha} target operated at \qty{30}{kV} and \qty{25}{mA} using a Vantec D800 detector. XRD on Nb-Ni samples was performed using an XRD 3000 PTS (Seifert, Mannheim, D) with a chromium X-ray source and using a 2$\theta$  range from \qty{20}{°} to \qty{168}{°} with a resolution of \qty{0.1}{°}, operating at \qty{40}{kV}. The lattice parameters were refined using MAUD (\cite{LucaLutterotti.1999}).

\subsection{DFT calculations}

The lattice parameters of Nb$_7$Ni$_6$ and Nb$_6$Ni$_7$ were calculated via iterative full cell relaxation in the Vienna Ab Initio Simulation Package (VASP) \cite{kresse1996efficient,kresse1993ab}. Planewave basis sets via the Projector Augmented Wave (PAW) potential \cite{kresse1999ultrasoft} represented the one-electron orbital basis sets. The Generalised Gradient Approximation (GGA) Perdew-Burke-Ernzerhof (PBE) \cite{perdew1996generalized} was employed as the exchange-correlation functional. The kinetic cut-off for the plane-wave basis set was \qty{550}{eV}, where Nb 4f$^0$ 4d$^4$ 4p$^6$ 5s$^1$ and 4s$^2$ and Ni 4f$^0$ 4p$^0$ 4s$^1$ and 3d$^9$ are set as the valence orbitals. The Fermi level was smeared using the 1st-order Methfessel-Paxton level with a $\sigma$ value of \qty{0.005}{eV} \cite{methfessel1989high}. Geometric optimisation was considered converged when the energy difference in SCF cycles reaches 10$^{-7}$ eV and forces in ionic movement reach 0.001 eV/Å. The Brillouin zone was sampled using the Monkhorst-Pack $k$-point scheme \cite{monkhorst1976special}, with a K-point per Reciprocal Atom (KPPRA) value of \num{20000} to ensure that the internal stresses reach below \qty{0.05} {kbar}.

\subsection{Analysis methodology}

\subsubsection{Image processing}
Each EBSD pattern consists of a signal from four timepix chips, providing an image size of 514 by 514 pixels, including 2 pixels in the centre carrying no image signal. The EBSD patterns as received from EDAX APEX were further normalised by subtracting the mean of all intensity values and dividing by the standard deviation. A mask of \qty{2}{px} width covering the pixels where each quadrant is stitched was applied. Further, a Gaussian envelope was applied to the image to reduce the influence of pixels at the edge of the image.
\subsubsection{Noise / dimension reduction (\ac{PCA})}
To reduce both noise and the amount of data to be analysed, we perform the \ac{PCA} approach as first described by \cite{Wilkinson.2019} with the code made available by McAuliffe et al. in the AstroEBSD repository \cite{McAuliffe.2020}. We set the variance retained to $v_t=0.1$ for all datasets, while the crop factor was typically chosen to be one. In some cases, we needed to choose a crop factor of 2 due to larger map sizes. The PCA approach used here will return one pattern per multiple map points, which typically represents a single orientation and phase.

\subsection{Optimisation approach}
\label{sec:Crystal_setup}
In the present study, for orientations, the conventions from \cite{Britton.2016} are used in optimisation routines internally, and the conventions from MTEX \cite{Krakow.2017} for plotting of EBSD data and handling the proprietary formats of EDAX. Throughout this study, the conventions for the projection centre ($PC$) from AstroEBSD \cite{Britton.2016} are used. We perform an orientation refinement using global optimisation based on the interior point algorithm, which was implemented in AstroEBSD \cite{Qaiser.2026}, with a maximum allowed number of iterations of 300 and step and optimality tolerances of $10^{-12}$, each. The optimisation is based on the \acf{NCC} metric of two images $A$ and $B$ \cite{Gonzalez.2002}:
\begin{equation}
NCC = \frac{ \sum_{i,j} {(A_{i,j}-\bar{A})(B_{i,j} - \bar{B} )} }{ {(||A -\bar{A}||\cdot ||B-\bar{B}||)}}
\label{eq:NCC}
\end{equation},
with $||A -\bar{A}|| = \sqrt{ \sum_{i,j} {(A_{i,j}-\bar{A})^2}}$ and $\bar{A}$ is the mean of $A$ (and for $B$ vice versa). Further, for evaluation of small \ac{EBSD} maps, a large \ac{EBSD} map of \num{1490}x\qty{1490}{µm} and \qty{10}{µm} step size is used to calibrate the projection centres using an affine transformation, as specified in \cite{Winkelmann.2019}. The projective transformation from there was also implemented in this study for comparison, but led to worse fitting results.

\subsubsection{Simulation of dynamical templates}
Dynamical templates of \ac{EBSD} Patterns in this study have been simulated using EMSoft \cite{Callahan.2013} with an acceleration voltage of \qty{19.45}{kV}, minimum lattice spacing of \qty{0.2}{\angstrom} and Bethe parameters of $c1$=\num{5.0}, $c2$=\num{10.0} and $c3$=\num{50}, which corresponds to the EMSoft recommendation  \cite{Jackson.2019} of increasing these values slightly from default parameters for heavier elements, such as Nb. For the initial screening of optimal voltage, we used the same parameters but varied the acceleration voltage in the range between \qty{18}{kV} and \qty{21}{kV} for testing on sample 51Nb-49Ni (see also \autoref{fig:VoltageTesting} in the supplementary material), the thus derived voltage of \qty{19.45}{kV} was used for all subsequent simulations. Further, for final testing of the influence of Bethe parameters, another simulation dataset using $c1$=\num{40.0}, $c2$=\num{50.0} and $c3$=\num{50.0} was used, as \cite{Wang.2016b} showed a strong influence of the Bethe parameters on the intensities of the EBSD patterns. To visualise the influence of minimum $d$ spacing, dynamical templates with minimum $d$ spacings of \num{1}, \num{0.5}, \num{0.2} and \qty{0.1}{\angstrom} have been simulated.

\subsubsection{Generation of crystal structures}

One of the core parts of this study to investigate the potential of EBSD to resolve \acp{ILS} and \acp{SLO} is the creation of crystal structure models, which are used to generate dynamical templates, against which we can compare the experimental EBSD patterns:
For initial screening of optimal voltage (comparing against the 51Nb-49Ni sample), the crystal structures from \cite{Joubert.2002}  for \qty{51.5}{\atpercent} Nb have been used. For further "matrix searches", the initial structures from Joubert et al. \cite{Joubert.2002}  have been modified by applying the experimentally determined lattice parameters for the 51Nb-49Ni and 58Nb-42Ni samples.  The atom positions for Nb-Ni were taken to be the ones from the \qty{53.5}{\atpercent} structure from Joubert et al. \cite{Joubert.2002} (see also \autoref{tab:CSSetup} for details). For Nb-Co, the initial atom positions were taken from the DFT-relaxed cells from \cite{Luo.2023}, after applying the correct crystal symmetry and removing duplicate sites with a deviation of \qty{10e-4}{\angstrom}. The lattice spacing was chosen from the experimental data from the XRD measurements of the 48Nb-52Co and 56Nb-44Co samples. 

\begin{figure}[t]
    \centering
    \includegraphics[width=\linewidth, trim = 0cm 13.7cm 6.2cm 0cm,clip]{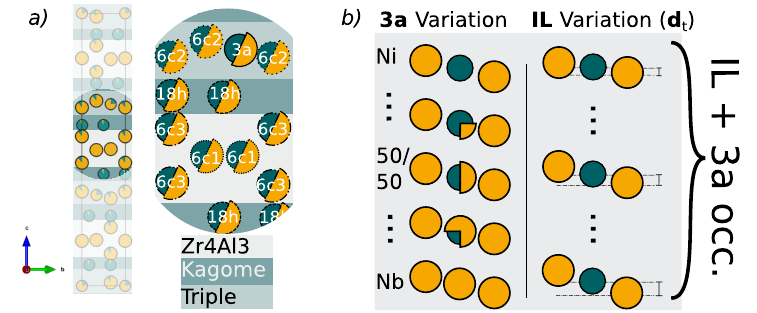}
    \caption{Overview of structural properties of candidate structures, which are varied throughout this study: \textit{(a)}: typical $\mu$-phase with labelling of the 5 distinct lattice sites in the $\mu$-phase. Green atoms are Ni, and yellow atoms are Nb. \textit{(b)}: Systematic change of the 3a site occupancy and the \ac{ILS} of the triple layer (performed by changing the 6c2 positions). %\textit{(c)}: Variation of atom site lattice occupancies for best matching interlayer spacing.}
    }\label{fig:CrystalStructSetup}
\end{figure}

To be able to build a structure pool which can be compared against experimental patterns, \ac{ILS} and \ac{SLO} are varied systematically. To modify the 
\ac{ILS} inside the triple layer ($d_t$), the z value of the 6c2 position was modified (see also \autoref{fig:CrystalStructSetup}). The site lattice occupancy of the 3a site was changed between 0 and 1 in steps of 0.25. Meanwhile, the \ac{ILS} of the triple layer ($d_t$) was varied between \qty{0.25}{\angstrom} and \qty{0.45}{\angstrom} (according to the expected range of IL parameters from \cite{Joubert.2002}), with a step size of \qty{0.025}{\angstrom}.

\begin{table}
\caption{Initial crystal structures used for modification of $d_t$ and 3a occupancy, given as a fraction of the unit cell (only positions not determined by crystal symmetry (Wyckoff position) are listed.} \label{tab:CSSetup}

\scriptsize
    \begin{tabular}{l|c|c|c|c|c}
    system&  6c1, z & 6c2, z & 6c3, z & 18h,x & 18h, z\\
    \hline \hline
    Nb-Ni &  0.165 & var & 0.4516 & 0.8355 &  0.2562 \\
    \hline
    Nb-Co &  0.1656 & var & 0.4498 & 0.3305 &  0.2563 \\
    % \hline
    % 56Nb44Co &  0.1662 & var & 0.4505 & 0.3249 &  0.2552 \\

\end{tabular}
\end{table}
%\FloatBarrier

\begin{figure*}
    \centering
    \includegraphics[width=\linewidth,trim= 0cm 13.7cm 0cm 0cm, clip]{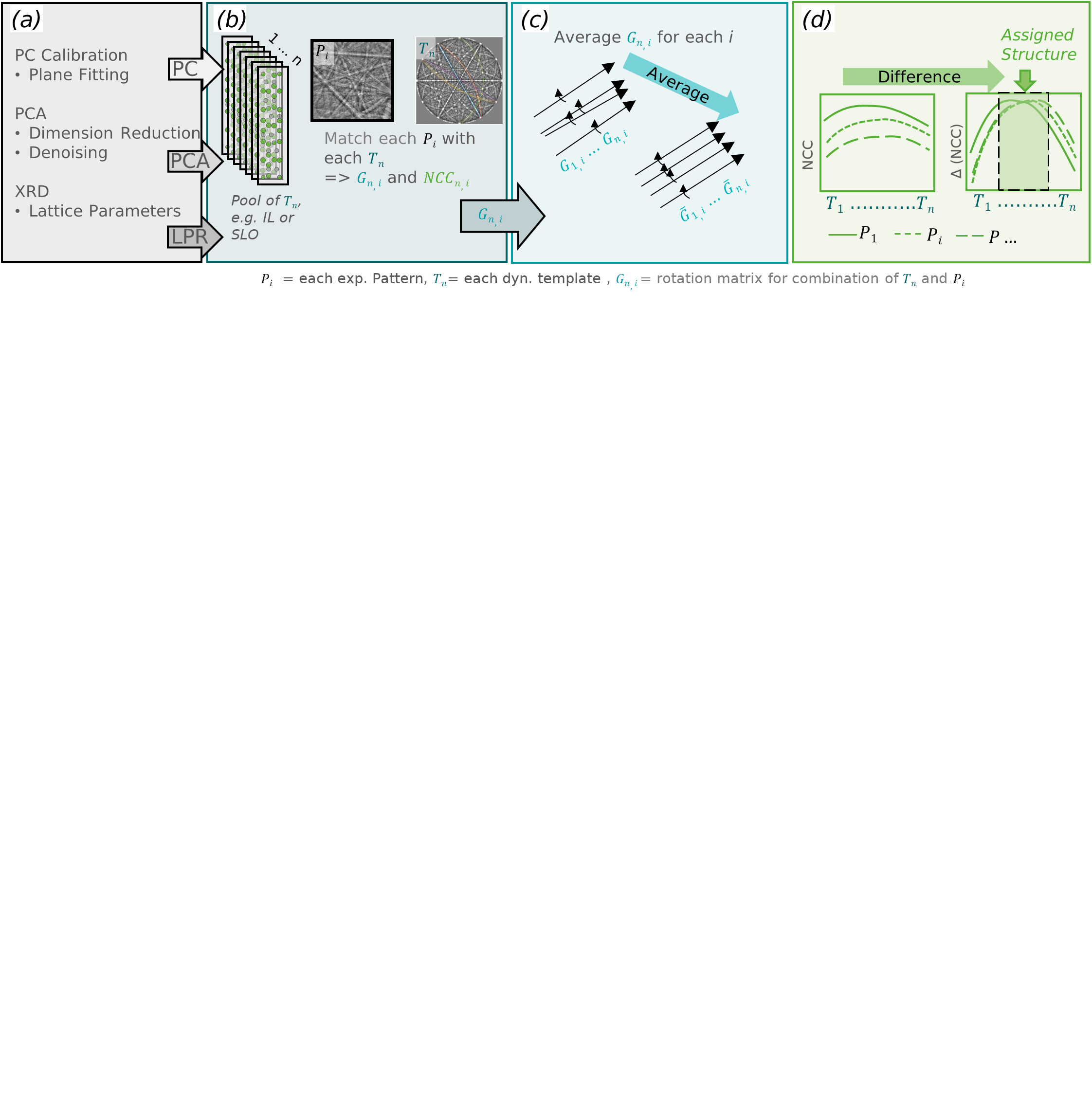}
    \caption{Schematic principle of the methodology employed in this study: \textit{(a)}: Inputs for the pattern matching routine. \textit{(b)}: Pattern matching of all structures of a given set with all experimental patterns obtained from \ac{PCA}. \textit{(c)}: Averaging of refined orientations for each experiment from the mean of the crystal structure set to reduce scatter of $NCC$ results. \textit{(d)}: Determination of best structure by fitting a 4$^{th}$ degree polynomial to specific sets for each experimental pattern.}

    \label{fig:AnalysisConcept}
\end{figure*}
\subsection{Structure resolution strategy}
To investigate the feasibility of resolving subtle crystallographic details such as site-lattice occupancy and specific \acp{ILS}, we combine the above-described processes in the following way to obtain maximum match metrics, as given by the normalised cross-correlation values (\autoref{eq:NCC}): Initially, the projection centre of each sample is refined on a large map. Then, pattern matching based on the interior-point algorithm is performed between the patterns obtained from the \ac{PCA} and the dynamical templates, which are generated based on the structure model description and \ac{XRD} lattice parameters as given before. This procedure is initially performed on a set of crystal structures, where only the 3a site and the \acp{ILS} are varied. 
The detailed procedure is as follows: To distinguish between the experimental pattern (as obtained from \ac{PCA}) and the dynamical template (reprojected to the same gnomonic coordinates as the experimental pattern), the labels $P_i$ and $T_n$ will be introduced, where $i$ and $n$ represent the number of experimental patterns and crystal structures used in the comparison, respectively. After pattern matching of each pair of $P_i$ and $T_n$, the orientation matrices $G_{n,i}$ still scatter between the different $T_n$. Therefore, all orientations above a threshold of $NCC$ = 0.65  for $G_{1..N,i}$ are averaged. The value of 0.65 was found to be a good value to filter for, e.g. too blurred patterns, patterns with overlapping twins obtained by the PCA, or patterns from secondary phases in this context. The thus obtained averaged orientations are used again to build comparison templates ($T_n$) to compare against the experimental patterns ($P_i$). To obtain the best matching crystal structure, for each $P_i$, the $NCC$ value can now be plotted (see \autoref{fig:AnalysisConcept} \textit{(d)}).
 To better visualise the dependence of the $NCC$ metric (\autoref{eq:NCC}) on each set of $P_i$, for each set of $P_i$, the $NCC$ values are normalised against the maximum $NCC$ value from each set (by calculating the difference). Finally, to obtain an averaged structure value for each sample, all sets of normalised $NCC$ values are fitted using a 4$^{th}$ degree polynomial and the position of the maximum of each fit is used to assign the $d_t$ spacing for each structure or the \ac{SLO}, respectively. This approach allows us to find values of $d_t$ and \ac{SLO} which do not lie on a discretised grid of matched patterns, but rather find an optimum structure. Then, a weighted mean for each sample can be calculated based on the number of pixels that contribute to each $P_i$ (as the \ac{PCA} will give back one pattern for multiple points of the \ac{EBSD} map).

\subsection{Motif extraction from HR-STEM images}
\label{sec:Motif_extraction}
To obtain accurate positions of the atomic columns in a 2D HR-STEM image and their deviation from the crystal structure, we use an atom and motif detection technique \cite{AlZhBe23} followed by a local optimisation step. 
We briefly describe the two methods below.

The motif detection finds the underlying periodic structure in the image in the form of the unit cell and extracts a motif image.
The atom detection then uses a Gaussian bump model to recreate the atomic columns and therefore identify their positions.

We first choose a representative region in the image that is (nearly) periodic and crop and rotate the image manually, such that this region is in the top left corner. On this region, we obtain the underlying periodicity by estimating two lattice vectors which span the unit cell in real space using \cite{MeBe15}. With these vectors, we define a unit-cell coordinate system, i.e. the crystal space. 
This enables the extraction of a representative motif: a single ``average'' unit-cell image.
This motif image is then projected to each cell of the Bravais lattice, resulting in a tiled and fully periodic reconstruction of the image. 
Further, in this reconstruction, we model the atomic columns with 2D Gaussian bumps. 
Now, the motif image can be remodelled as a sum of Gaussian bumps with individual parameters.
The output of the motif and atom detection is therefore (i) the unit-cell vectors, (ii) 
a clean motif image, and (iii) Gaussian parameters for each atomic column in the motif (see \cite{AlZhBe23} for a detailed description).
\par
To capture deviations from ideal periodicity, we then perform a local refinement.
We mask the area of each unit cell indexed by the tuple ($i, j$) and optimise for each cell individually.
On this masked area $M_{i,j}$, we compute the error between the original image $f$ and an affine transformed version of the projected motif $u \circ P_{EC}$, i.e.,
\[
        \int_{\tilde{\Omega}}  \bigl(\chi_{M_{i,j}}(\mathbf{x}) \cdot \bigl[N_{M_{i,j}}[u\circ P_{EC}\circ \phi_{A,b}](x)-N_{M_{i,j}}[f](x)\bigr]\bigr)^2 \, \mathrm{d}\mathbf{x}
    \]
with the affine transformation $\phi_{A,b}(x):= A\mathbf{x} + b.$ \par
The parameters $A_{i,j}$ and $b_{i,j}$ are obtained by a non-linear Fletcher-Reeves conjugate gradient descent with Armijo step size control. 
We start the optimisation in the corner of the image that was used to perform the motif detection and proceed from there, cell by cell, to the point furthest from that corner.
The optimisation is stabilised by initialising the solution on each cell from the best already obtained neighbouring solution.
To make this comparison robust, we normalise the mean to zero and the standard deviation to one in each cell, denoted by $N_{M_{i,j}}$. 
That way the optimisation focuses on the pattern of the atomic columns rather than on slow brightness changes or background variations.
For efficiency, the comparison is only computed on a bounding box $\tilde{\Omega}$ that is just big enough to contain the cell. %
\par
Applying the obtained per-cell transformations to the positional parameters of the Gaussian model gives us a list of transformed atom positions across the entire image. Because each unit cell is allowed its own ($A_{i,j}, b_{i,j}$), the resulting atom coordinates are not constrained to a single global periodic lattice. They can differ from the ideal periodic structure in a physically meaningful way.
\FloatBarrier
\section{Results}
\subsection{Microstructure of investigated samples}
\begin{figure*}[t]
    \centering
    \includegraphics[width=\linewidth,trim = 0cm 0cm 0cm 0cm, clip]{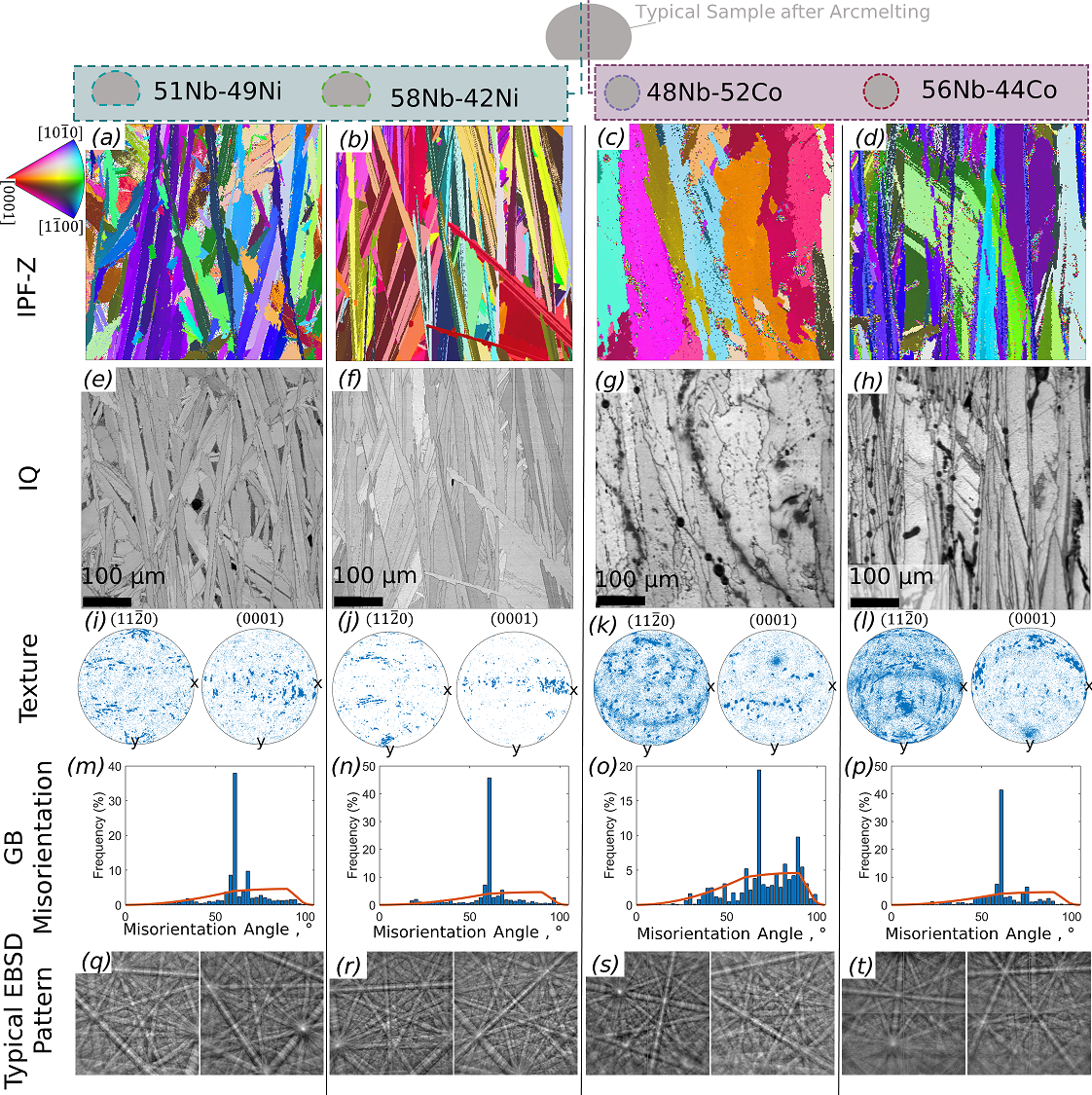}
    \caption{Microstructure of investigated samples, \textit{(a)-(d)}: \ac{IPF} map with respect to sample normal (same scalebar as image-quality maps), \textit{(e)-(h)}: Image quality map of investigated samples corresponding to \ac{IPF} maps in \textit{(a)-(d)}, \textit{(i)-(l)}: Texture of the sample represented by \hkl(11-20) and \hkl(0001) Pole figures, \textit{(m)-(p)}: Misorientation angle distribution calculated from the \acp{IPF} maps, the blue bins correspond to the misorientation angles and the orange curve to the uniform distribution function.\textit{(q)-(t)}: Typical \ac{EBSD} Patterns for the investigated samples (taken from \ac{EBSD} maps shown in the supplementary material, \autoref{fig:PCAexp})
    \label{fig:Microstructure}}
\end{figure*}

An overview of the morphology of the as-cast samples used in the present study is given in \autoref{fig:Microstructure}.
The samples are viewed with the presumed solidification direction lying in the plane of observation (indicated by dark green and purple lines overlaid on a sketch of a typical sample (top centre of \autoref{fig:Microstructure} )). 
The 48Nb-52Co and 56Nb-44Co samples are viewed along the solidification direction (indicated by the purple line in the sample sketch).

The Nb-Ni samples and the 56Nb-44Co show elongated grains, with a mixture of straight boundaries and serrated boundaries, as is evident by both the IPF-Z (\autoref{fig:Microstructure} \textit{(a)} to \textit{(d)}) and image quality maps (\textit{(e)} to \textit{(h)}). The images in \autoref{fig:Microstructure} \textit{(a)} and \textit{(b)}, \textit{(e)} and \textit{(f)} (that is, of the Nb-Ni samples) reveal grains with the long direction approximately along the expected thermal gradient during solidification. The IPF-Z maps further indicate that the grains must be interwoven in 3 dimensions, as there are many identical orientations which disappear and reappear along their longest axis (visible for Nb-Ni as well as the 56Nb-44Co sample). 
The long axis can reach several hundred of \unit{µm} in some cases, while the short axis is typically on the order of \qty{50}{µm} or less. Thin twins extend through the entire length of many grains and can approach the step size of the overview EBSD scans (\qty{1}{µm}) for the Nb-Ni samples ((\textit{(a)} \& \textit{(b)}) and 56Nb-44Co (\textit{(d)}). In contrast, the 48Nb-52Co sample (\textit{(c)}) shows more uniform and also larger grains (with the diameter being about \qty{200}{µm}). 

All investigated samples show the presence of small volume fractions of secondary phases (see \autoref{fig:Microstructure}, \textit{e)}-\textit{h)}) which have not been investigated in this study, as the $\mu$-phase is of primary interest. The investigated Nb-Ni and 56Nb-44Co samples show tendencies of \hkl(11-20) fibre textures (with (11-20) oriented towards the presumed solidification direction (cf. also \autoref{fig:Microstructure} \textit{i)}-\textit{j)}.
The investigated samples show a strong presence of misorientation angles of \qty{60}{°}  (51Nb-49Ni,58Nb-42Ni and 56Nb-44Co) and \qty{68}{°} (48Nb-52Co), cf. \autoref{fig:Microstructure} \textit{m)} - \textit{h)}. These correspond to basal and pyramidal twins previously observed in $\mu$-phases \cite{Zhao.2023,Ma.2018, gasper2025planar}.
% Note that for the investigations following hereafter, the Nb-Co samples were re-cut and re-polished with the view direction now being the same as for the Nb-Ni samples.

\FloatBarrier

\subsubsection{Chemical analysis}
\begin{figure*}[t]
    \centering
    \includegraphics[width=\linewidth,trim = 0cm 10.0cm 0cm 0cm, clip]{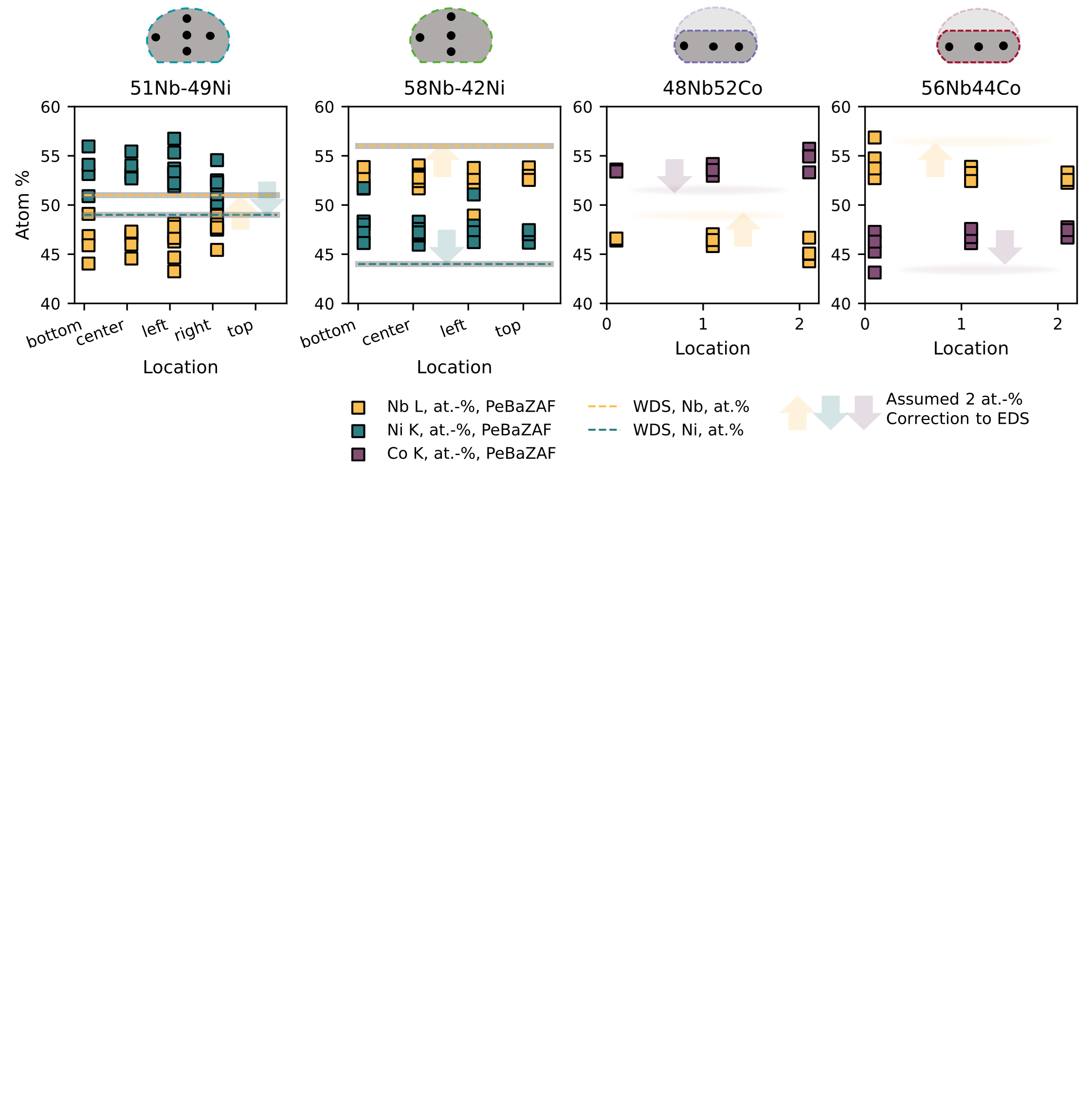}
    \caption{EDS and WDS results for Nb-Ni and Nb-Co samples. The sample locations are indicated on sketches of the investigated samples above the figure (Left corresponds to the left-most, top to the top-most black dot and so on. For Nb-Co locations, 0 to 2 correspond to the positions from left to right.) Squares indicate \ac{EDS} measurements quantified using the PeBaZAF method. The horizontal lines for the two Nb-Ni samples indicate the composition as measured using multiple line scans with \ac{WDS}. Further, the corrections based on the assumed error of \ac{EDS} measurements compared to \ac{WDS} measurements are indicated by arrows.}
    \label{fig:EDS}
\end{figure*}

\begin{figure*}[h]
    \centering
    \includegraphics[width=\linewidth,trim = 0cm 10.2cm 0cm 0cm, clip]{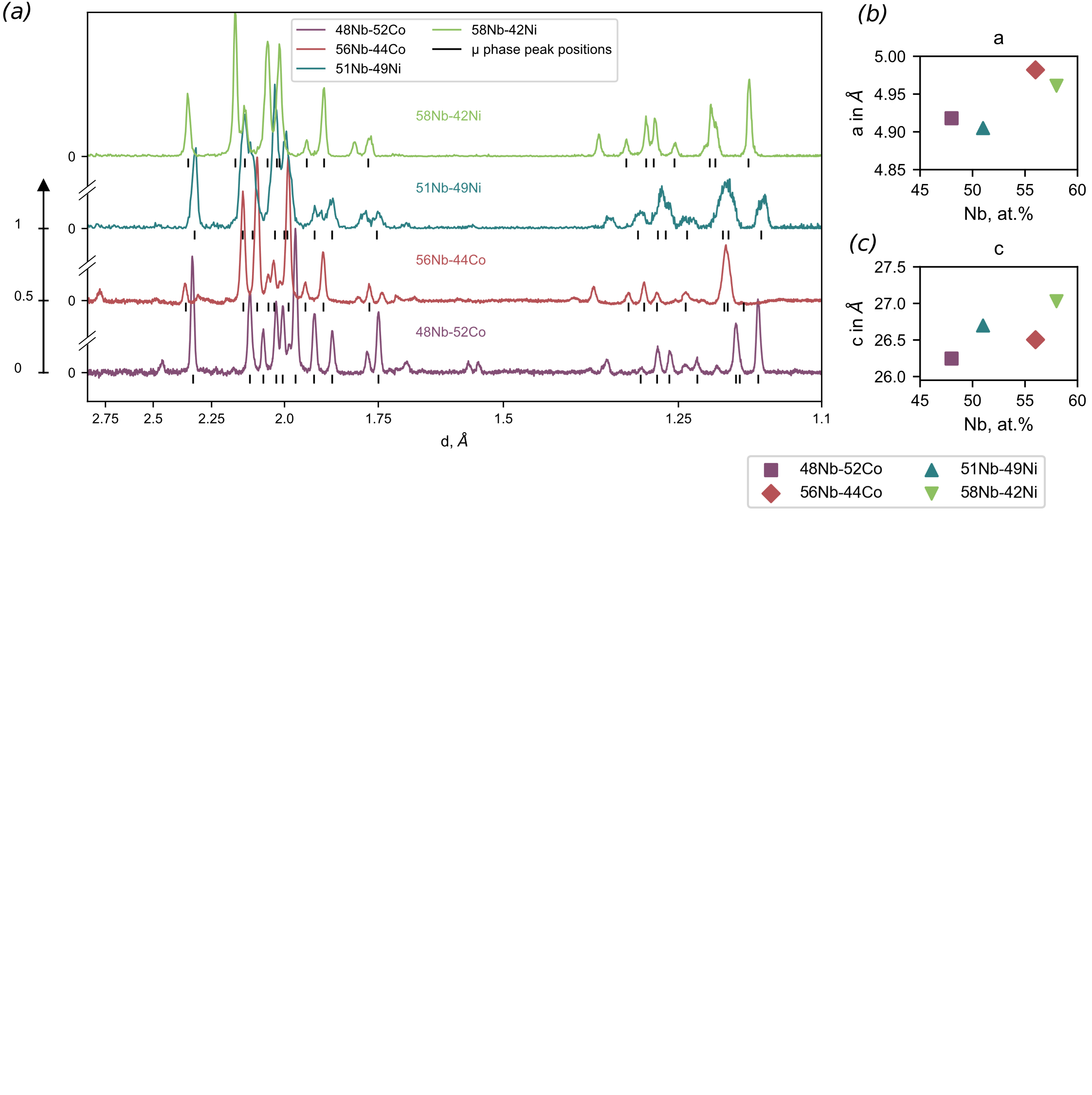}
    \caption{\ac{XRD} analysis: \textit{(a)}: \ac{XRD} intensity for Nb-Ni and Nb-Co samples in dependency of the \ac{ILS}. The black lines mark the peak positions of the $\mu$-phase after fitting (shown up to an intensity of \qty{10}{\%} of the maximum intensity). The data is normalised by the maximum intensity and shifted by \num{0.5} along the relative intensity axis to enhance clarity. \textit{(b)}: Dependency of lattice parameter $a$ on the Nb content in all samples. \textit{(c)}: Dependency of lattice parameter $c$ on the Nb content in all samples. }
    \label{fig:XRD}

\end{figure*}

Quantification results of Nb-Ni and Nb-Co samples using both \ac{WDS} (for Nb-Ni only) and \ac{EDS} are shown in \autoref{fig:EDS}. Here, each data point corresponds to an \ac{EDS} spot measurement on the sample. The measured composition of the Nb-Ni samples using EDS scatter by about \num{2} - \qty{3}{\atpercent}. For the Nb-Co samples, the scatter is a bit less (about \num{1} - \qty{2}{\atpercent}). There is no strong dependence of the chemical composition of samples on the position of measurement. The measured quantities show an excess of the light element compared to the nominal sample composition. As \ac{WDS} can be more accurate and does not suffer from overlap of peaks \cite{Herrington.1985,Camus.2015} for chemical quantification, we can assume the WDS measurements in the Nb-Ni samples as the ground truth.  From this, it can be seen that there is a trend to underestimate the content of the heavier element Nb. E.g. in the 58Nb-42Ni sample, the Nb contents are \num{2} to  \qty{3}{\atpercent} below the ones from \ac{WDS}, when measured using the PeBaZAF correction method for quantification of the \ac{EDS} data. 

As Co and Ni are very close in terms of their characteristic X-ray energies, we can assume that also for Nb-Co samples we are underestimating the content of Nb by \num{2} to \qty{3}{\atpercent}. With a correction of \qty{2}{\atpercent}, the Nb-Co samples also approach values close to their nominal composition at the centre of the samples (\qty{51.5}{\atpercent} Co for the 48Nb-52Co sample and \qty{44.5}{\atpercent} Co for the 56Nb-44Co sample.

Further, in the \ac{WDS} line scans, an increasing Ni content is observed towards the two-phase regions. The horizontal lines here indicate the composition away from the two-phase regimes. (See also \autoref{fig:WDS_gradients} in the supplementary material for further details).
From this, the composition of the central parts of the 51Nb-49Ni samples' $\mu$-phase is $48.5\pm0.2$\unit{\atpercent} Ni and $50.5\pm0.2$\unit{\atpercent} Nb, while for the 58Nb-42Ni sample, the $\mu$-phase composition is $44.4\pm0.5$\unit{\atpercent} Ni and $55.6\pm0.5$\unit{\atpercent} Nb.

% \FloatBarrier
\subsubsection{XRD analysis}
XRD analysis is performed, to measure the lattice parameters, based on which we build the structures to simulate the dynamical templates for each sample. \autoref{fig:XRD} depicts diffraction data for the Nb-Ni and Nb-Co samples. The intensity is normalised by division by the maximum values for each diffractogram. The data is plotted against the lattice spacing in \unit{\angstrom}, as the wavelength for the Nb-Co and Nb-Ni samples was different and thus the 2$\theta$- values would not be comparable.  The theoretical position of each peak of the $\mu$-phase up to 10\% of the maximum intensity is shown for each structure by a black line. All observed peak positions agree well with the peak positions expected for the $\mu$-phase after refining the lattice parameters. There is no strong peak from any of the secondary phases present, which is likely due to the low phase fraction of these phases. The lattice parameters strongly depend on the composition, as visible from \autoref{fig:XRD} \textit{(b)} and \textit{c)}. For the Nb-Ni samples, $a$ increases by \qty{0.06}{\angstrom} from \qty{4.90}{\angstrom} at \qty{51}{\atpercent}{Nb}, while $c$ increases from \qty{26.65}{\angstrom} to \qty{27.1}{\angstrom} (from \qty{51}{\atpercent} to \qty{58}{\atpercent}{Nb}). The increase for Nb-Co is similar, but the lattice parameter $c$ is smaller with an increase from  \qty{26.25}{\angstrom} at \qty{48}{\atpercent} Nb to \qty{26.5}{\angstrom} at \qty{56}{\atpercent} Nb (and \qty{4.918}{\angstrom} to \qty{4.982}{\angstrom} for $a$).

\subsection{Influence of simulation parameters}

\begin{figure*}[t]
    \centering
    \includegraphics[width=0.7\linewidth,trim = 0cm 0.9cm 0cm 0cm, clip]{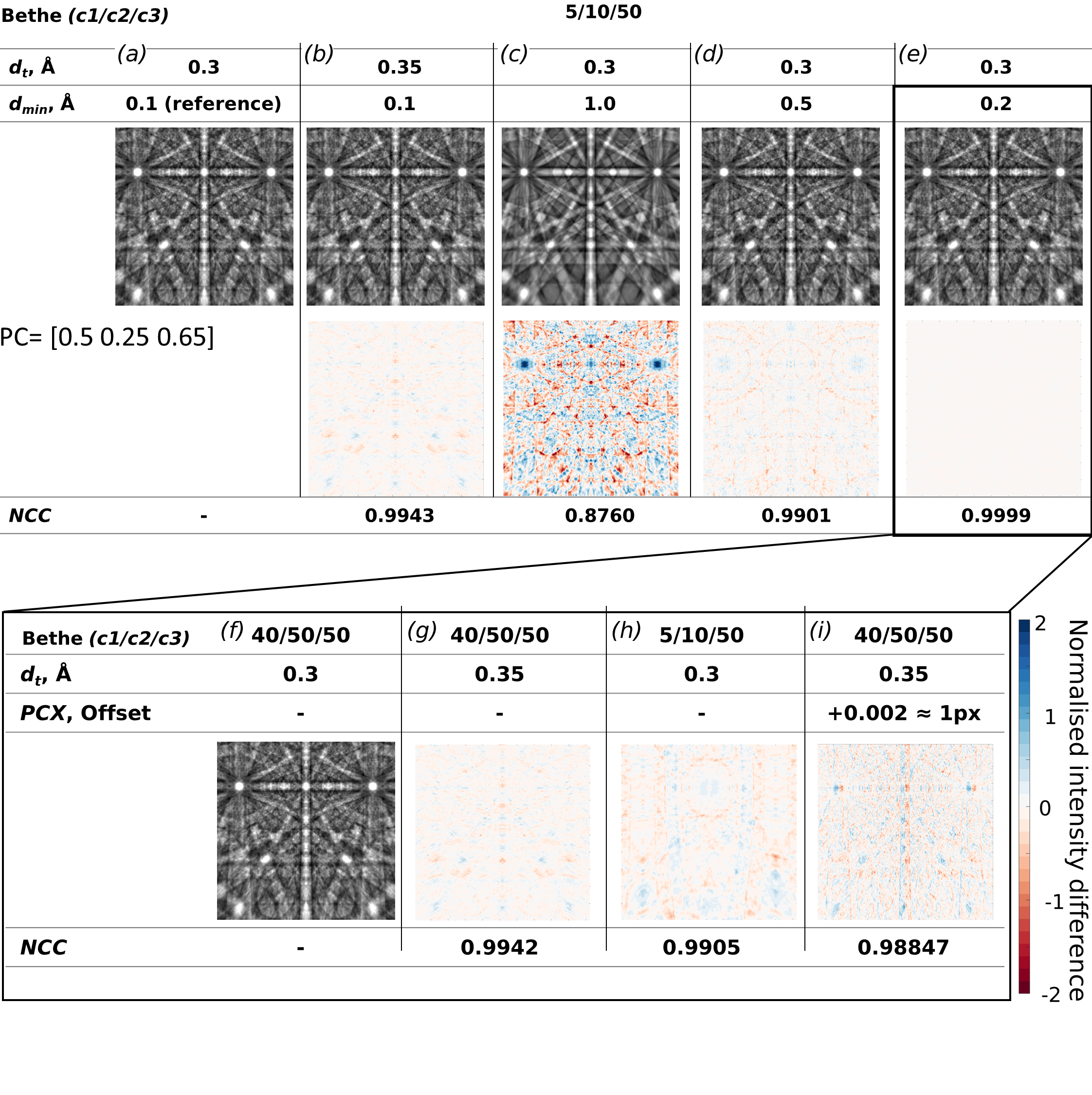}
    \caption{Visualisation of various influences of simulation parameters on the patterns. The colour bar gives the intensity values of the difference plots (according to normalising by mean and standard deviation, see experimental section). Top row (\textit{(a)} to \textit{(e)}): comparison of dynamical templates with different minimum $d$ spacings ranging between \qty{0.1}{\angstrom} and \qty{1.0}{\angstrom}. (Simulated pattern on top of the difference pattern between the respective simulated one and the reference pattern) Further, the influence of a change of the triple-layer interlayer spacing by \qty{0.05}{\angstrom} (in \textit{a)} from \qty{0.3}{\angstrom} to \qty{0.35}{\angstrom}) is shown in the left-most column (\textit{(b)}). As the simulation with $d_{min}$ should in principle contain the physically most complete set of information, we choose this for further investigation of the influence of Bethe Parameters, which have been shown to influence intensities of dynamical template simulations. Bottom row (\textit{(f)} to \textit{(i)}): Influence assessment of Bethe parameters. For these patterns, the minimum considered lattice spacing is always \qty{0.2}{\angstrom}. There is a significant difference between the low and high Bethe parameter settings. \textit{(h)}: To further highlight the influence that even a slight offset of the projection centre can have, this is added here in the last column (\textit{(i}). The $PC$ coordinates are given according to the AstroEBSD convention \cite{Britton.2016}, with $PCX$ being the first value of the $PC$ row vector. $NCC$ is the normalised cross correlation, as defined in \autoref{eq:NCC}. 
    \label{fig:parameterInfluence}}
\end{figure*}
To assess the influence of simulation parameters in EMSoft, we visually present the influence of minimum $ d$-spacing and Bethe parameters on the simulated EBSD patterns. The top row of \autoref{fig:parameterInfluence} highlights the influence of the minimum lattice spacing used in the simulations. For this, the simulated patterns are depicted in greyscale in the topmost row. The respective differences to the reference pattern \textit{(a)} are shown in a blue-red colour map directly below. Here we compare lattice spacings of \qty{1}{\angstrom}, \qty{0.5}{\angstrom}  and \qty{0.2}{\angstrom} against a reference of \qty{0.1}{\angstrom}. For all these simulations, Bethe parameters of $(c1/c2/c3) = (5/10/50)$ according to the recommendations given in \cite{Jackson.2019} are used. The absence of many high-order reflections for $d_{min}$=\qty{1}{\angstrom} (cf \autoref{fig:parameterInfluence}, \textit{(c)}) is evident even from the simulated EBSD pattern. The difference to the reference pattern shows strong intensity variations,  most prominently at the two \hkl(0-1-10) zone axes.  For the $d_{min}$=\qty{0.5}{\angstrom} (\autoref{fig:parameterInfluence}, \textit{(e)}) pattern, the differences are already very faint, which is also represented by a high similarity metric of $NCC$ = \num{0.9901}. There is hardly any difference between the pattern simulated using $d_{min}$=\qty{0.2}{\angstrom} (\autoref{fig:parameterInfluence}, \textit{(f)}) and \qty{0.1}{\angstrom}. We can further see that the subtle difference between the reference pattern and a pattern simulated using an \ac{ILS} of \qty{0.35}{\angstrom} instead of \qty{0.3}{\angstrom} at constant minimum lattice spacing (\autoref{fig:parameterInfluence}, \textit{(a)} vs. \textit{(b)}) is smaller than the difference between the structures of minimum lattice spacing \qty{0.5}{\angstrom} and \qty{0.1}{\angstrom} at constant interlayer spacing (\autoref{fig:parameterInfluence}, \textit{(d)}).

As the Bethe parameters are also known to influence the visual appearance of EBSD patterns \cite{Wang.2016b}, we further tested their influence(\autoref{fig:parameterInfluence} \textit{(f)}-\textit{(i)}). \autoref{fig:parameterInfluence} \textit{(h)} shows the comparison of a structure simulated with the Bethe parameter combination of $(c1/c2/c3) = (5/10/50)$ compared to $(40/50/50)$ (\autoref{fig:parameterInfluence} \textit{(f)}). The difference in NCC and visual impression of the difference plot (\textit{(h}) is bigger compared to two different $d_t$  spacings for both the low and high Bethe parameter settings (\textit{(g)} vs. \textit{(b})). Interestingly, the $NCC$ values as well as the difference image for comparing the $d_t=$\qty{0.35}{\angstrom} and \qty{0.3}{\angstrom} structures appear to be close to identical regardless of the choice of Bethe parameters (compare \autoref{fig:parameterInfluence}\textit{(b)} and \textit{(g)}). 

Finally, \autoref{fig:parameterInfluence},\textit{(i)} shows the influence of misalignment of the projection centre. The offset is chosen to be 0.002 in the x direction, which corresponds to \qty{1}{px}. The difference in $NCC$ is bigger than expected, e.g. for differences in Bethe parameters or \ac{ILS}. This highlights the relevance of very carefully calibrating the projection centre.

For the subsequent simulations, which are used to extract the $d_t$ and 3a - \ac{SLO}, a minimum lattice spacing ($d_{min}$) of 0.2 Å and Bethe parameters of c1=5.0, c2=10.0 and c3=50.0 were chosen, while the voltage was fixed to \qty{19.45}{kV}. The only parameters varied in this final analysis are $d_t$ and 3a \ac{SLO}, as described in \autoref{sec:Crystal_setup}.

\subsection{Assessment of distinctiveness of $d_t$ and SLO on simulated patterns}
\begin{figure*}[t]
    \centering
    \includegraphics[width=\linewidth, trim=0cm 5.9cm 0cm 0cm,clip ]{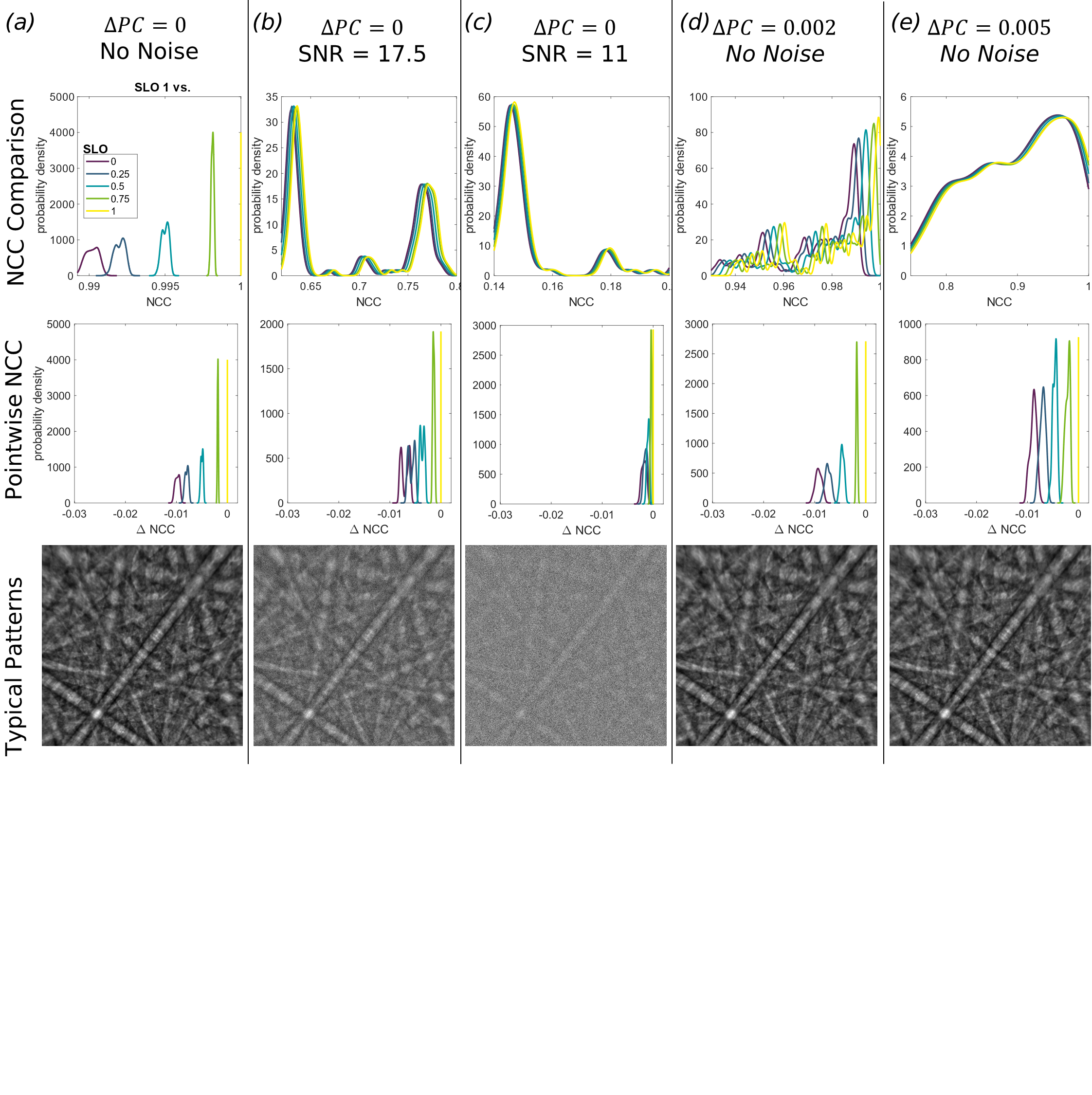}
    \caption{Assessment of the influence of noise and $PCX$ offsets on 100 random orientation differences in $NCC$ on the distinctiveness of dynamical templates with varying site-lattice occupancy: \textit{(a)}: No noise and offset, \textit{(b)}: Poisson noise with a \acf{SNR} of \qty{17.5}{dB}. \textit{(c)}: Poisson noise with \ac{SNR} \qty{11}{dB}. \textit{(d)}: No noise, but $PCX$ offset by \num{0.002} $\approx$ \qty{1}{px}
    \textit{(e)}: No noise and $PCX$ offset by \num{0.005} $\approx$ \qty{2.5}{px}}
    \label{fig:accAssesment}
    
\end{figure*}

In the following, we investigate whether it is in principle possible to distinguish the very similar crystallographic structures that vary only by small changes, such as fractions of an \unit{\angstrom} of an atom position (\ac{ILS}) or by changes in the site lattice occupancy, but are otherwise identical and how this depends on noise and misalignments. For this, we generate a dataset of simulated EBSD patterns for 100 random orientations using AstroEBSD \cite{Britton.2018} and MTEX \cite{Bachmann.2010}. We do this to test specifically the effect of variations in the site lattice occupancy of the 3a site (\autoref{fig:accAssesment}) and for variations in the $d_t$ spacing (given in the supplementary material, \autoref{fig:accAssesmentIL}). For each orientation, the $NCC$ metric between the reference structure (\ac{SLO} of 3a =1 in (\autoref{fig:accAssesment}) and $d_t$ = 0.35 in (\autoref{fig:accAssesmentIL}) is calculated, and both the total distribution of $NCC$ values (top rows of each figure) as well as the "point-wise" $NCC$ difference are given (bottom rows of each figure). \footnote{The "point-wise difference" is the NCC change caused by SLO or $d_t$ variations for one given orientation. This means, if a simulated pattern for $d_t$=0.35 and \ac{SNR}=17.5 gives $NCC$=0.75, and the one for $d_t$=0.3 and the same noise ratio gives $NCC$=0.749, the resulting $\Delta NCC$ is 0.001.} The bottom row of \autoref{fig:accAssesment} further gives typical patterns to show the influence of noise and pattern centre offset. The noise is quantified by the \acf{PSNR}.% \cite{}.    

\autoref{fig:accAssesment}, \textit{(a)} shows that if no additional noise or misalignments are applied, the distribution of $NCC$ values is easy to distinguish, and the same holds true for this kind of analysis for the $d_t$ spacing (\textit{(a)} in the supplementary material, \autoref{fig:accAssesmentIL}). \autoref{fig:accAssesment} \textit{(b)} shows that if we start to add Poisson noise with a \ac{SNR} of \qty{17.5}{dB}, the direct comparison of $NCC$ values already suffers from strong overlap of the distributions. However, when comparing the point-wise comparison in the second row, it becomes clear that the distributions of the simulated patterns for site-lattice occupancies of the 3a site of \num{0.75} are clearly distinguishable. On the other hand, if we increase the noise to SNR values of \qty{11}{dB} (higher SNR values mean less noise), the peaks start to overlap strongly (see \autoref{fig:accAssesment}, \textit{(c)}). The same trend is observed for the variation of $d_t$ spacings. 

To further test what the influence of misaligned patterns will be, we show the same distributions for misalignments of the projection centre ($\Delta PCX$) coordinate (in the absence of noise) by \num{0.002} and \num{0.005} (corresponding to about \qty{1}{px} and \qty{2.5}{px} for the given detector size of \qty{514}{px} of the Clarity\textsuperscript{\texttrademark} detector used in the present study). Although the distributions of $NCC$ values strongly overlap, the point-wise comparison still shows clearly separate distributions even for an offset of 0.005. However, in the case of the interlayer variation, the pointwise distributions for $d_t$ = \qty{0.375}{\angstrom} and \qty{0.325}{\angstrom} start to overlap with the distributions of the \qty{0.35}{\angstrom}, this overlap is, however, not as severe as for image noise of \qty{11}{dB}. Further, the combination of PC offset with noise reveals that, in general, the influence of noise is stronger than the influence of PC offsets (cf. \autoref{fig:accAssesmentOCCADD} in the supplementary material).

\subsection{Principal component analysis}
The \ac{PCA} approach in this study serves the main purpose of reducing the number of patterns to be matched while decreasing the level of noise (and therefore acts as a non-local accumulation method), as the goal is to compare the experimental \ac{EBSD} patterns against a large library of dynamical templates. The effect of the PCA on two selected patterns is shown in \autoref{fig:PCA_pat_improv} in the supplementary material. The \ac{PCA} gives rise to several components for the same grain, and thus several unique EBSD patterns, based on the \ac{IPF} maps shown in the supplementary material, \autoref{fig:PCAexp}. These components align with variations in the image quality map that correspond to different orientations, different phases, the topography of the sample or potential scratches.
\FloatBarrier

\subsection{Projection centre refinement}
\begin{figure*}[t]
    \centering
    \includegraphics[width=\linewidth,trim = 0cm 10cm 0cm 0cm, clip]{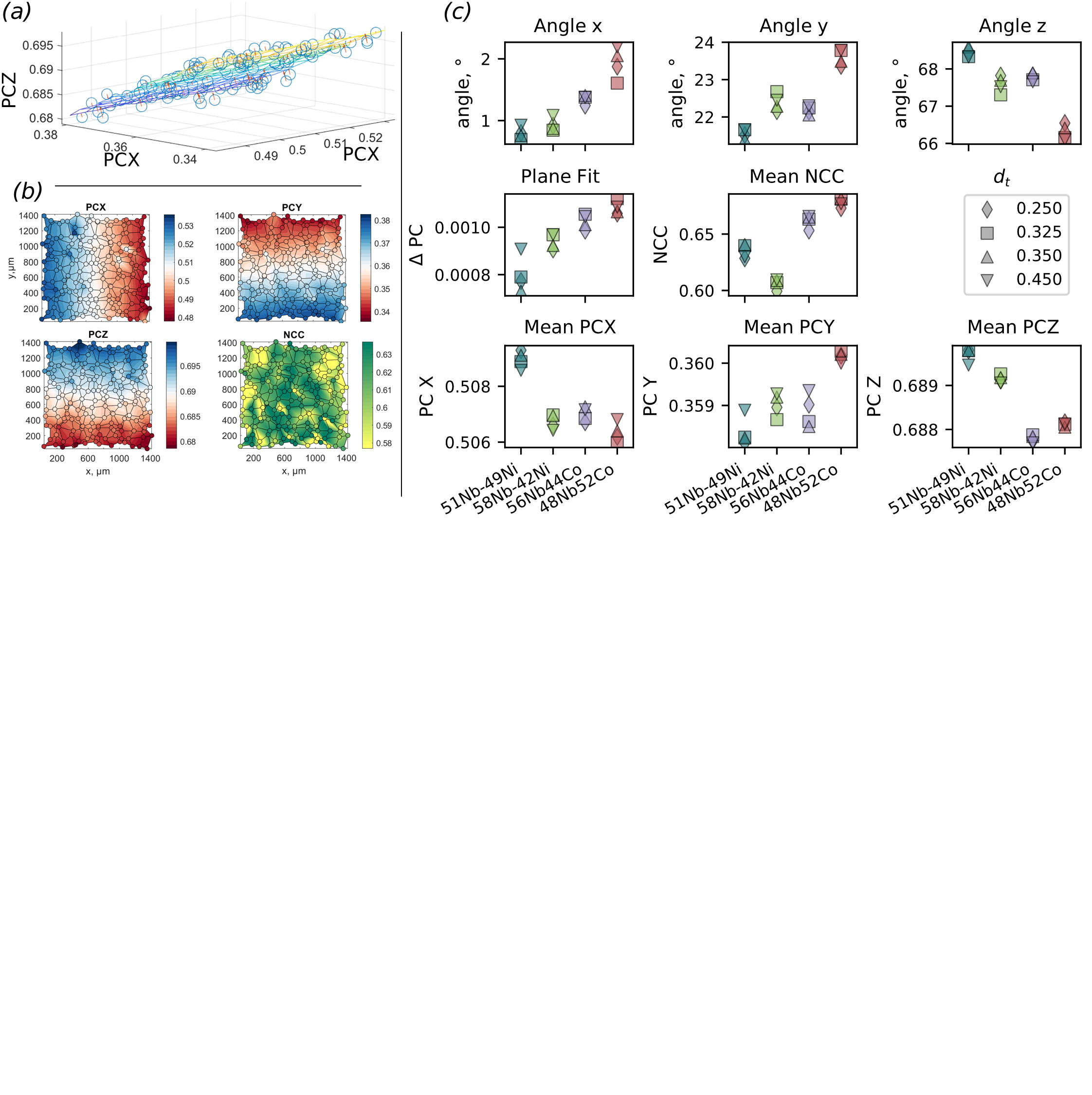}
    \caption{Projection centre ($PC$) refinement \textit{(a)}: Plane fitted to pattern centre points from global optimisation routine, as parametrised by $PCX$, $PCY$ and $PCZ$. Typical result, example taken from 58Nb-42Ni for $d_t=0.325$ \textit{(b)}: Typical variation of projection centre values and $NCC$  over the sample, example taken from 58Nb-42Ni for $dt=0.325$ \textit{(c)}: Results from fitting of a plane to projection centre values: Top row: x, y and z angles between sample normal and EBSD camera normal. Middle row: Deviation of refined $PC$ values from fitted plane and mean $NCC$ values for each sample - crystal structure pair. Bottom row: mean $PCX$, $PCY$ and $PCZ$ values for each fit.}
    \label{fig:projCRef}
\end{figure*}
Careful calibration of the projection centre is necessary, as misalignments will lead to indistinguishability of either $d_t$ spacing or \ac{SLO} as shown before.
The process of fitting a plane to multiple refined pattern centres is visualised using an example from the 58Nb-42Ni sample (\autoref{fig:projCRef}). The fitted plane parametrised by $PCX$, $PCY$ and $PCZ$ is given in \autoref{fig:projCRef} \textit{(a)}. Further, the distribution of $PCX$, $PCY$, $PCZ$ and the $NCC$ values for each point are given in dependence of beam position on the sample (\autoref{fig:projCRef}, \textit{(b)}). The grid is not regular, as a subset of points showing the highest image quality values within a 5$\times$5 pixel neighbourhood around loosely spaced points on a rectangular grid was chosen.
The refinement of the projection centre leads to mostly consistent results for each sample, leading to mean deviations from the fitted plane on the order of \num{1e-3} (in fractions of the pattern width) or smaller, corresponding to less than half a pixel of scatter (cf. \autoref{fig:projCRef}, \textit{(c)}. We observe that the alignment of the samples with respect to the detector is not perfect. (based on \qty{1}{°} camera tilt, the optimal alignment would be at x = \qty{0}{°}, y= \qty{21}{°} z= \qty{22}{°}. These angles are off by \qty{1}{°} - \qty{2}{°} depending on the sample (in general, for the Nb-Co samples, a higher deviation is observed). There is also a slight variation of the mean projection centre for each map, which varies around $\pm$ \num{0.003} for $PCX$  and around $\pm$ \num{0.002} for $PCY$and $PCZ$ between different maps.

\FloatBarrier
\subsection{Structure identification for the variation of interlayer spacings ($d_t$) and 3a sites}
\begin{figure*}
    \centering
    \includegraphics[width=\linewidth,trim = 0cm 2.1cm 0cm 0cm,clip]{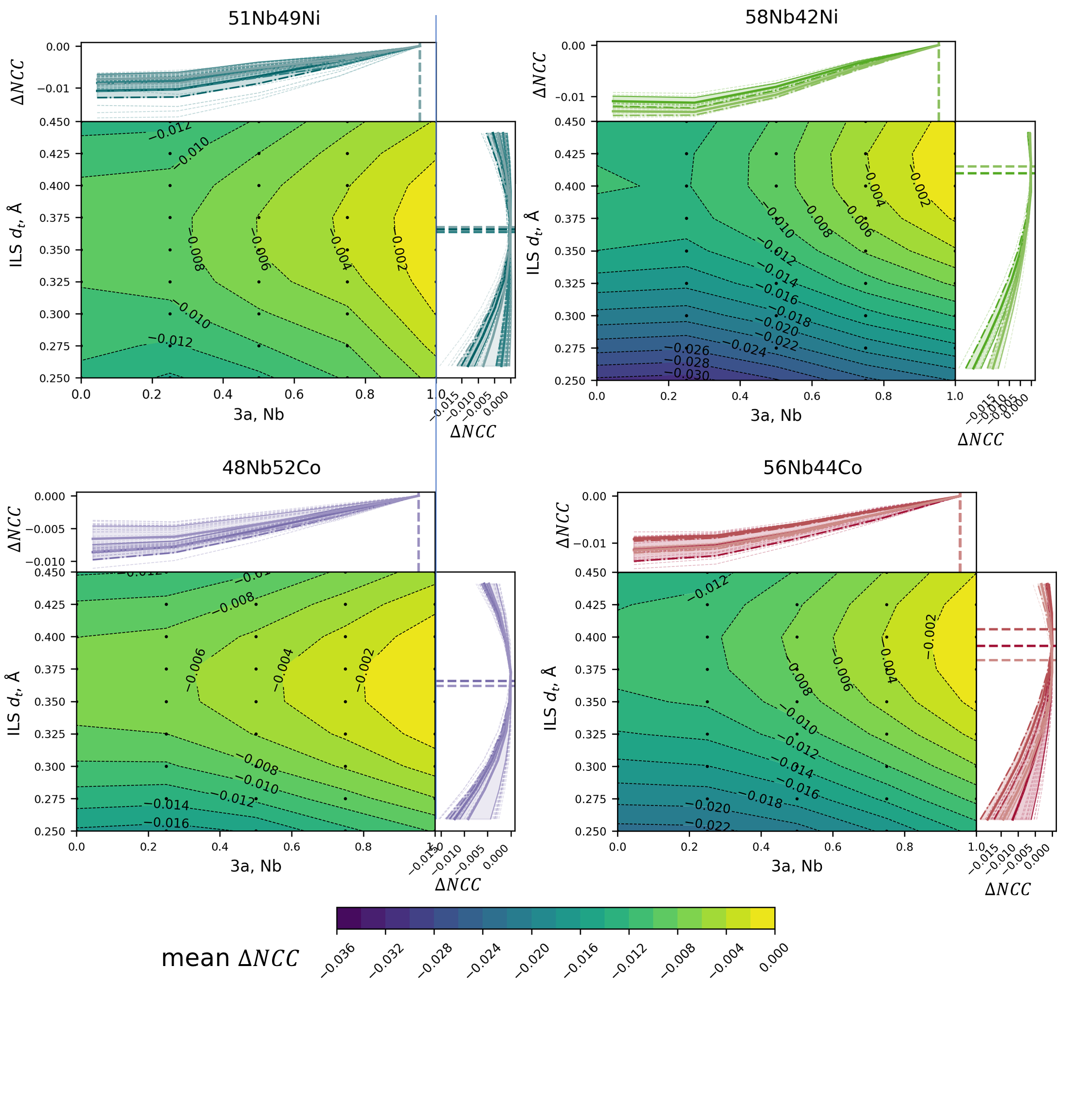}
    \caption{Results after comparing experimental and simulated patterns for multiple components identified by the \ac{PCA}. Contour plots of mean $\Delta NCC$ for 5 \ac{SLO} × 9 $d_t$ combinations. Each black dot represents a candidate structure (e.g., Nb on 3a site = 1.0, $d_t$ = \qty{0.35}{\angstrom}). On the sides of the plots, the curves for the respective maximum section of the landscape are given to visualise the scatter of individual data points. The dashed lines in the side-plots give the respective maximum of the $\Delta NCC$ values per sample.}
    \label{fig:ILSpacingRefinement}
\end{figure*}

\begin{figure}
    \centering
    \includegraphics[width=0.7\linewidth,trim = 0cm 0cm 0cm 0cm,clip]{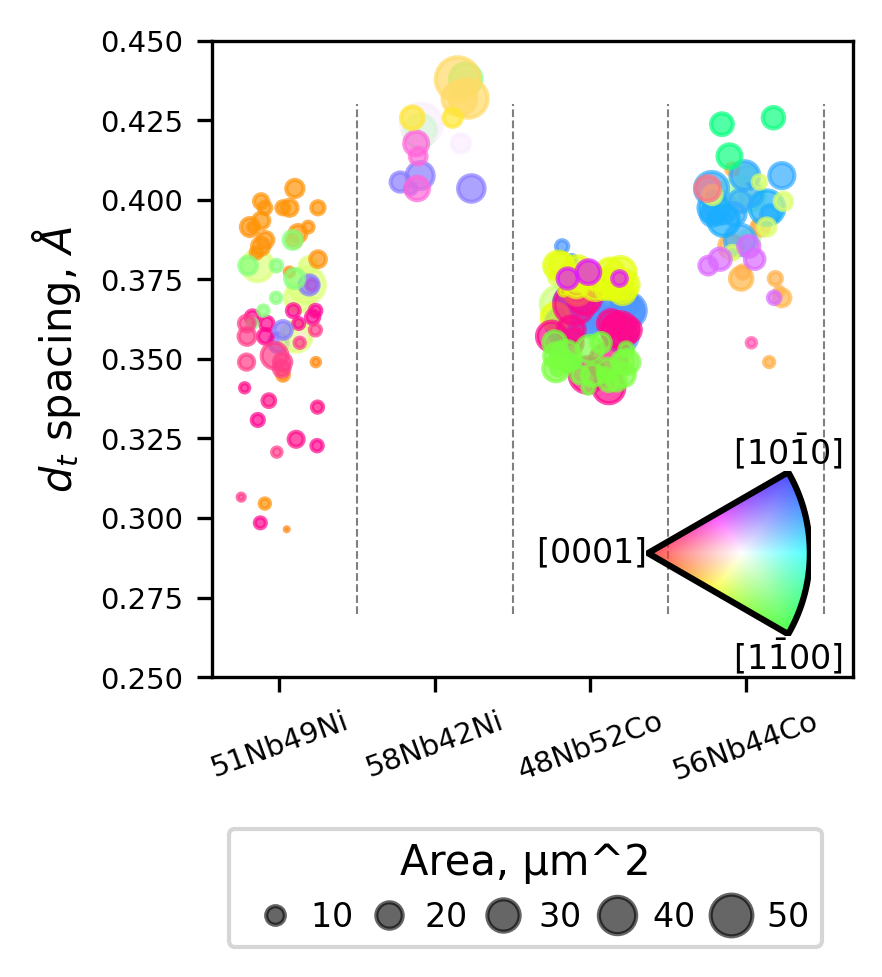}
    \caption{Variation of  $d_t$ with orientation. The scatter in the horizontal direction is artificially added to reduce the overlap of data points. The size of the markers corresponds to the size of PCA component, where the EBSD pattern was from}
    \label{fig:IL_ori}
\end{figure}

The main goal of this study is to "measure" \ac{ILS} - specifically $d_t$ - and site lattice occupancies of the Nb-Ni and Nb-Co $\mu$-Phases (in this case the occupation of the 3a site). The results from the pattern-matching between structures with variation in $d_t$ spacing and occupancy of the 3a site are shown in 
\autoref{fig:ILSpacingRefinement}. The figure shows that the metric based approach in this study leads to "measuring" a full occupancy of the 3a site by Nb for all samples and further $d_t$ spacings of \num{0.363} and \qty{0.396}{\angstrom} for 48Nb-52Co and 56Nb-44Co and $d_t$ spacing of \num{0.368} and \qty{0.424}{\angstrom} for 51Nb-49Ni and 58Nb-42Ni, respectively. This can be read from the diagram as follows:  for each pattern resulting from the \ac{PCA}, the crystal orientation is fitted using the interior-point algorithm, as described before, for all pairs of candidate phases and experimental patterns. In this case, there are 45 (5(\ac{SLO}) $\times$ 9 ($d_t$) different candidate "phases" and the experimental patterns are given by the components of the \ac{PCA} analysis. %The projection centre in this case is propagated from the previous analysis by the mean x and y values of each component identified by the \ac{PCA}. 
The blue-green contour plots in \autoref{fig:ILSpacingRefinement} show the mean normalised $NCC$ values, where the mean is taken from all sets of $NCC$ between one dynamic template out of the 45 crystal structures and all experimental patterns. The normalisation of the $NCC$ values is performed against the maximum $NCC$ value for each experimental pattern, as described before (the best matching phase thus having a difference of zero). Each black dot in \autoref{fig:ILSpacingRefinement} represents one of these candidate "phases" corresponding to a total of 45 "phases" for each sample. From the contour maps, it can be seen that the 58Nb-42Ni and 56Nb-44Co samples show higher values of the \ac{ILS} than the sample containing less Nb (51Nb-49Ni and 48Nb-52Co). Further, for all the samples, the best match is obtained for a full occupation of the 3a site by Nb. At the edges of the contour plot, the distribution of $\Delta NCC$ values for the respective maximum value of the other plot property is given (i.e. for the $d_t$ spacings, the values for $3a, Nb = 1.0$ are shown). For each sample, multiple smaller EBSD maps are taken, which are shown by different colouring of the curves. Further, each curve belonging to one experimental pattern is fitted using a 4$^{th}$ degree polynomial to obtain the maximum value. The perpendicular dashed lines indicate the position of the mean maximum for each map. From this, it can be seen that the deviation between different maps is smaller than the distance between sampling points (in terms of $d_t$ spacing for the Nb-Ni samples and 48Nb-52Co). In contrast, the 56Nb-44Co sample shows a stronger scatter of the \ac{ILS}, spanning about \qty{0.02}{\angstrom} between individual maps. The averaged values for each sample are shown in \autoref{tab:IL_results}. The standard deviations in this table are calculated from all $d_t$ values of the fitted 4$^{th}$ grade polynomial to the NCC values of each PCA pattern, weighted with the area of the PCA component, the pattern originated from. \autoref{fig:IL_ori} further shows how the $d_t$ spacing depends on the orientation of each pattern. There is a slight trend visible that, for more basal orientations (meaning that the \hkl(0001) is oriented along the sample normal), lower \acp{ILS} are observed. The larger scatter for the 56Nb-44Co sample can be seen as well from this figure. Further, prismatic orientations, meaning either \hkl[10-10], \hkl[1-100] or \hkl[11-20], tend to lead to observations of higher \ac{ILS}, especially for the Nb-Co samples.
To show that possible averaging of patterns based on the PCA approach does not influence the results significantly, we further performed pattern matching for a hundred individual experimental patterns of the 58Nb-42Ni and 51Nb-49Ni samples with structures of a 3a \ac{SLO} by Nb = 1 and $d_t$ spacings in the range of 0.25-0.45. As indicated by \autoref{fig:PatMatchIndividual} in the supplementary material, the trend of increased $d_t$ spacing towards a higher Nb content is preserved.

\begin{table}[]
    \centering
        \caption{Results (mean with standard deviation) from fitting 4$^{th}$ grade polynomial to $NCC$ values and averaging. \vspace{0.2cm}}
    \begin{tabular}{l|r|r}
% \toprule
Sample & \multicolumn{2} {c}{d\textsubscript{t} spacing (\unit{\angstrom})}   \\  &  mean   & std  \\
\hline
\hline

% \midrule
48Nb-52Co & 0.363 & 0.01 \\
\hline
51Nb-49Ni & 0.368 & 0.02 \\
\hline
56Nb-44Co & 0.396 & 0.012 \\
\hline
58Nb-42Ni & 0.424 & 0.012 \\
% \bottomrule
\end{tabular}

    \label{tab:IL_results}
\end{table}

\subsection{Results from HR-STEM Motif extraction}
\begin{figure*}
    \centering
    \includegraphics[width=\linewidth,trim=0cm 7.1cm 0cm 0cm, clip]{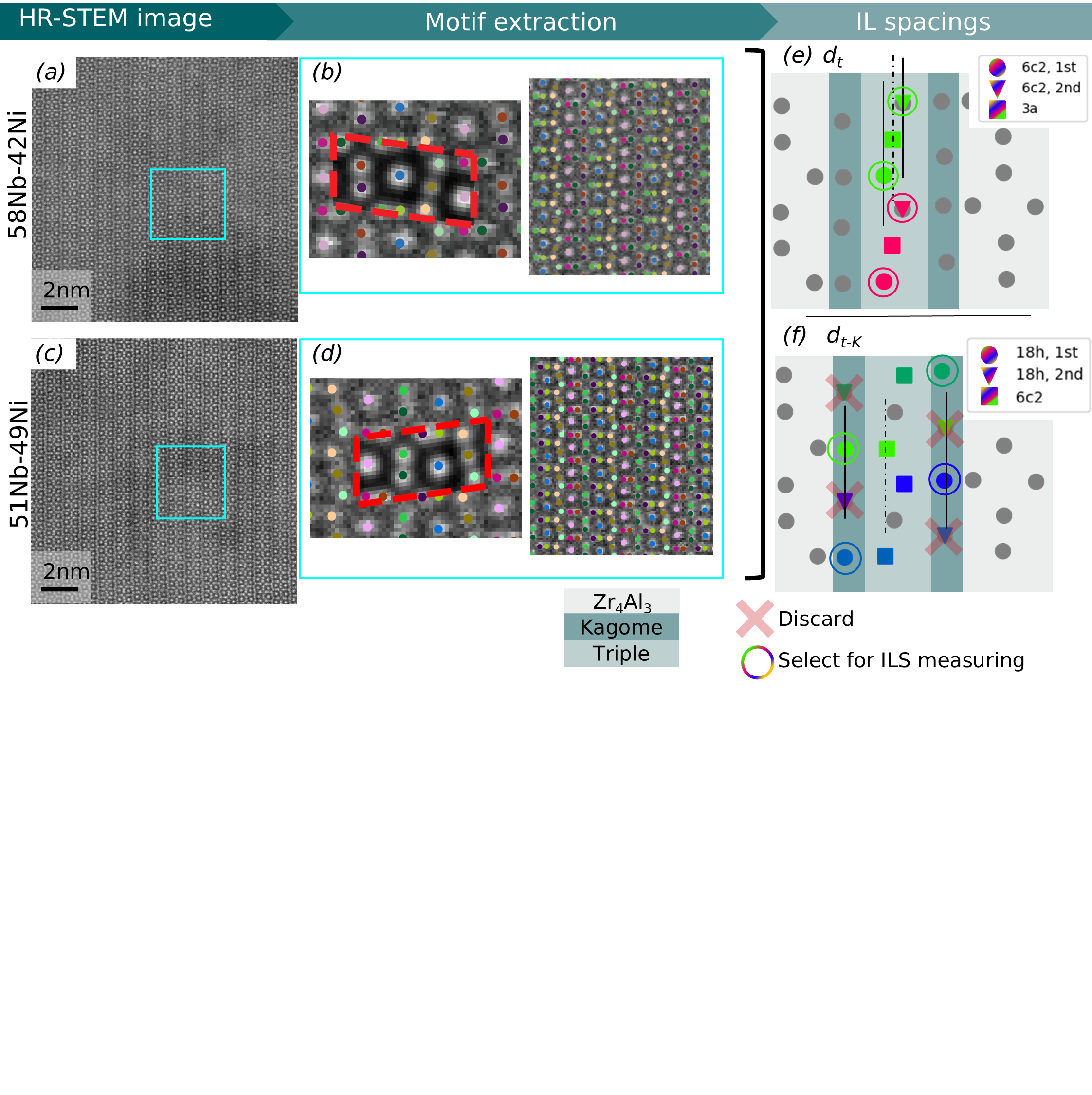}
    \caption{Results from \ac{ILS} analysis of the \ac{HR-STEM} data: \textit{(a)} and \textit{(c)}: HR-STEM image of thin foils from 58Nb-42Ni and 51Nb-49Ni sample, respectively. \textit{(b)} and \textit{(d)}: Motif extraction from the \ac{HR-STEM} images. The dashed red line visualises the outline of the extracted motif, as explained in \autoref{sec:Motif_extraction}, each coloured dot represents one of the 11 possible different sites found in the motif. The right panel in \textit{(b)} shows the application of the motif to extract the position of all atoms in the \ac{HR-STEM} image. \textit{(e)} and \textit{(f)}: Principle of extracting the interlayer spacings for $d_t$ (\textit{e}) and $d_{t-K}$ (\textit{f}). The colours correspond to the sets of 6c2 atoms, with the closest distance to the 3a site or the sets of 18h atoms to the 6c2 site. The open circles and the red cross visualise the sites respectively selected and discarded for further analysis. (Discarding the red-crossed sites is necessary due to the asymmetry of 6c2 with respect to the two next 18h sites). The dash-dotted line marks the position of the 3a or 6c2 site, and the solid lines mark the position of the candidate sites to measure the \ac{ILS}. Atoms not considered in this analysis are shown in grey, and the \ch{Zr4Al3}, triple- and Kagome-layer building blocks are indicated below the atoms.}
    \label{fig:HR_TEM_maintext}
\end{figure*}

\autoref{fig:HR_TEM_maintext} visualises the results from \ac{HR-STEM}  investigation of thin foils from 51Nb-49Ni and 58Nb-42Ni. It can be seen that the motif extraction is capable of extracting the atom positions from the \ac{HR-STEM} images  and that this motif is replicated over the complete image. After manual assignment of the sites in the motif to the five different Wyckoff positions in the $\mu$ -phase, the two closest 6c2 neighbours are identified for each 3a site (\autoref{fig:HR_TEM_maintext},\textit{(e)}, triangle and circle). The colour of each set of 3a + 6c2 indicates a set, which is used to calculate the \ac{ILS}. The same analysis is performed for the $d_{t-K}$ spacing. As this Interlayer spacing is not mirror symmetric for the distance between the 6c2 and the two closest 18h sites, only the closest 18h site is selected in this analysis (\autoref{fig:HR_TEM_maintext}, \textit{(f)}). Here, discarded sites are marked with a red cross, and selected sites per each set are marked with a circle in the colour of the set The results from this investigation are presented in \autoref{tab:HR-Stem_motif}. This analysis delivers two sets of  $d_t$ and one set of $d_{t-K}$ per each 3a site in a given \ac{HR-STEM} image. The standard deviations given in \autoref{tab:HR-Stem_motif} are directly calculated from these populations of $d_t$ and $d_{t-K}$ spacings. For each sample, one \ac{HR-STEM} image is analysed. 
%The standard deviation of the data result from the deviations of the difference calculated here. 

\begin{table}[]
    \centering
        \caption{Results (mean with standard deviation) from measuring \ac{ILS} from extracted \ac{HR-STEM} motifs. \vspace{0.2cm}}
    \begin{tabular}{l|r|r|r|r}
% \toprule
Sample & \multicolumn{2} {c}{$d_t$ spacing (\unit{\angstrom})}  & \multicolumn{2} {c}{$d_{t-K}$ spacing (\unit{\angstrom})}   \\ &  mean   & std    &  mean   & std  \\
\hline
\hline

% \midrule

51Nb-49Ni & 0.386 & 0.035  &  1.623 & 0.067\\
\hline
58Nb-42Ni & 0.436 & 0.035 & 1.62 & 0.017 \\
% \bottomrule
\end{tabular}

    \label{tab:HR-Stem_motif}
\end{table}

\FloatBarrier
\FloatBarrier
\section{Discussion}
As the investigation of site-lattice occupancies and \ac{ILS} in this study is approaching limits in the detectability of subtle intensity changes in EBSD patterns, we will discuss the following points in this context:
\begin{itemize}
    \item How well do we need to know the alignment of our EBSD system, and how well do we know it?
        \item What is the accuracy and precision of our measurement in comparison to available literature data?

    \item Can we identify crystallographic features that lead to the difference in $NCC$ between EBSD patterns?
        \item What is the likely influence of approximations made by the dynamical template simulation?
    \item What do our observations imply for further correlative investigations of subtle crystallographic effects and, e.g., mechanical properties?
    \item What are the new capabilities of identifying interlayer spacings using \ac{EBSD}?
\end{itemize}

\subsection{Alignment of the EBSD system - projection centre calibration} 
While there are multiple available techniques for projection centre calibration, such as moving-screen \cite{Li.2024}, shadow-casting \cite{Mingard.2011} and iterative approaches \cite{Pang.2020,Winkelmann.2020}, we focused on the latter, utilising a pattern matching routine, as iterative approaches allow the calibration of the projection centre at multiple points of the map easily. For the moving screen technique, this would only be possible if multiple maps were collected with varying positions of the projection centre. Further, the shadow-casting method is also not sufficient to detect potential misalignments of the sample. 

The key question is how accurately the alignment of the EBSD system must be known for the present analysis, and whether our calibration reaches this level. Our results from projection centre calibration indicate that there is no strong dependence of the mean projection centre coordinates of each affine transformation on the choice of structures with varying \ac{ILS}. The observed variation is well below 0.001, corresponding to at most 0.5 px of the pattern in the most extreme case and more typically around 0.1 px.  Further, the deviation of individual points from the fitted plane is also of the order of 0.001 or lower, which is similar to the affine model reported by Winkelmann et al. \cite{Winkelmann.2020}, but higher than their results obtained using a projective model, which achieved mean deviations between fitted and experimental pattern of \num{0.0003}. Although we also implemented the projective model, we achieved worse fits compared to the implemented affine model. The deviations of the projection centre values compared to the fitted plane (Plane Fit plot in \autoref{fig:projCRef}, \textit{(c)}) are also lower than other typical standard deviations reported by, e.g. Britton et al.\cite{Britton.2010}, who measured standard deviations of \num{0.005} for $PCX$, \num{0.007} for $PCY$ and \num{0.004} for $PCZ$, on multiple grains of alpha Ti.

In addition, we assessed the influence of misalignments for crystal structures that only had a difference of \qty{25}{\%} occupation of the 3a site or \qty{0.025}{\angstrom} of the triple layer spacing. We observed that, for $PC$ shifts of \qty{1}{px}, the distributions of point-wise $NCC$ metric differences of the aforementioned crystal structures are clearly distinct from each other. This provides a practical criterion for the required alignment accuracy:  misalignments of this magnitude do not prevent discrimination between the subtle structural differences relevant here. In this sense, the calibration appears to be known within the required range. However, we can of course not rule out systematic offsets, which would still lead to good precision of the plane fitting, but uncertain accuracy of the absolute alignment. 

Further, it has been shown by Alkorta et al. \cite{Alkorta.2013} that combinations of strain and change in projection centre can lead to virtually the same \ac{EBSP}. Therefore, it is relevant to consider our PC calibration approach in light of the potential distribution of lattice parameters from our XRD data. Some broadening of the X-ray peaks is expected due to the instrumental setup. The remaining broadening can then be attributed to residual strains (or variations in the lattice parameter). As we do not have access to the instrumental broadening of our diffractometers, we just give a crude approximation of the broadening of the X-ray peaks due to variation of lattice parameters in the supplementary material. From \autoref{fig:XRDLitComparison_broadening}, we can estimate that when neglecting any instrumental broadening, the lattice parameters might vary by about 0.2-0.5 \%. Alkorta et al. \cite{Alkorta.2013} demonstrate that pattern centre offsets would lead to phantom strains proportional to the ratio between the $PC$ offset and detector distance ($pcz$ in the present study). For example,  a 0.5\% strain of the crystal $c$ axis requires an offset of 0.0034 (or 1.7 px) (if the crystal is oriented with the c axis pointing to the projection centre). From \autoref{fig:projCRef}, however, we observe that the mean scatter of the pattern centre is about 0.001, which would correspond to a change of the lattice parameters by only 0.0014, although locally larger values are of course possible. Using XRD to estimate mean lattice parameters, our large-area fitting approach should yield a PC plane within the distribution of false PC coordinates (in positive and negative directions). Further, an extreme strain of 0.5 \% changing the $c$ value of the 51Nb-49Ni sample from \num{26.81} to \qty{26.944}{\angstrom}, leads to a change in $d_t$ spacing of \qty{0.015}{\angstrom} based on the atomic coordinates given in \autoref{tab:CSSetup}. This behaviour might explain the scatter of \autoref{fig:IL_ori} in extreme cases. The crudely approximated value of \qty{0.015}{\angstrom} is further larger than all std. deviations of the \ac{ILS} that have been observed (\autoref{tab:IL_results}).

Although we used a large area of \qty{1.5}{mm} x \qty{1.5}{mm} to achieve a large coverage of grain orientations, we cannot rule out any influence of the presence of texture, because orientation can influence the results of projection centre determination. Further, we observe that, although great care was taken to grind co-parallel sample surfaces, all angles of the surface normal deviate by at least \qty{1}{°} from the desired x-direction (even \qty{2}{°} in the case of the 48Nb-52Co sample) and even more with respect to the angle between the \ac{EBSD} camera and the sample of \qty{21}{°}, which is off by \num{1} to \qty{3}{°}. The angle of deviation agrees with other reports by e.g. \cite{Winkelmann.2020} (\qty{2.16}{°} from parallelness to the camera and \qty{3.24}{°} deviation from the 70° tilt).

In principle, the fit might be further improved by considering dynamical effects imposed by the geometry of the diffraction experiment, such as excess deficiency \cite{Winkelmann.2021}. However, making such simulation routines available and combining them with the global optimisation deployed in this study is beyond the scope of the present work. 

Overall, our results suggest that the $PC$ calibration is sufficiently precise for the screening carried out here and therefore for resolving the subtle crystallographic differences of interest. However, as we currently do not have access to the accuracy of projection centre refinement, we cannot fully exclude residual systematic errors in the absolute alignment.

\subsection{Accuracy and precision of the measured interlayer spacing and site lattice occupancy}

The assessment of precision in light of this study can be performed with respect to the scatter present in the data. Further, we can investigate the influence of orientation on the pattern-matching-based measurement of \ac{ILS} and site lattice occupancies. Although the sample is textured, we tried to capture a variety of orientations,  see e.g. \autoref{fig:IL_ori}. In this figure, it is apparent that there is a slight dependency of the measurement of the $d_t$  spacing on the orientation, where, in general, basal orientations lead to lower values of $d_t$. This is in agreement with expectations from comparing dynamical templates of varying $d_t$ spacings, where a big difference is observed around the rim of the stereographic projection, while the centre portion (being close to \hkl[0001]) shows only subtle differences (cf. supplementary material, \autoref{fig:TripleLayerInfluenceOnMasterpattern}). However, overall, the scatter between orientations is less than the difference between the chemistry of the samples. Further, the area-weighted standard deviations for each system do not overlap (cf. \autoref{tab:IL_results}).

With an \ac{NCC} based approach, it is generally necessary to select a cut-off value to exclude patterns that are too blurred or are otherwise of low quality. Through manually investigating the indexing results, we determined this threshold to be \num{0.65}. To highlight the influence of the \ac{NCC} on the resulting $d_t$ spacing, we have plotted the correlation between the two properties in \autoref{fig:NCC_corr_dt} in the supplementary material. From this, it is visible that for most samples, most of the NCC values are above 0.65. Further, we indicate the influence of \ac{NCC} on the \ac{ILS} in \autoref{fig:NCC_corr_dt},\textit{(c)}, sampling the mean $d_t$ spacing in bins of 0.025 \ac{NCC}. There is a dependence of the $d_t$ spacing on NCC in the 51Nb-49Ni and 56Nb-44Co samples. However, the sampled area, where these low NCC values are from, is small compared to the major component. Further, the scatter of values is larger at values below 0.65 for these two samples. Overall, this figure indicates that the compositional changes tend to be conserved within a range of NCC thresholds that correspond to the majority of patterns. %The NCC values of samples 48Nb52Co and 58Nb42Ni are in the narrow range between 0.75 and 0.8. In contrast to this, for the 51Nb49Ni and 56Nb44Co samples, the distribution of NCC values is more spread out. Essentially, s
Some high values for the 56Nb44Co sample (between 0.75 and 0.8 NCC) align with the $d_t$ range of the 48Nb52Co sample. A detailed investigation of these patterns which have an overlapping distribution of NCC values with the 48Nb52Co sample ( \autoref{fig:NCC_corr_dt}, \autoref{fig:PCA_NbCO_spatial} and \autoref{fig:56Nb44Co_details}), reveals that the patterns leading to high NCC and somewhat lower $d_t$ values in the 56Nb44Co sample compared to the sample average stem from a region on the right side of a larger crack running through the sample. Further, it becomes obvious that the patterns from Nb-Co are generally more blurred than in the other samples, leading to the overall lower NCC values. To assess how suitable the current method is for coupling $d_t$ spacings with \ac{EDS} information in higher detail, more studies on gradient samples like diffusion couples would be needed. 

\begin{figure}
    \centering
    \includegraphics[width=\linewidth,trim = 0cm 9.6cm 9.1cm 0cm, clip]{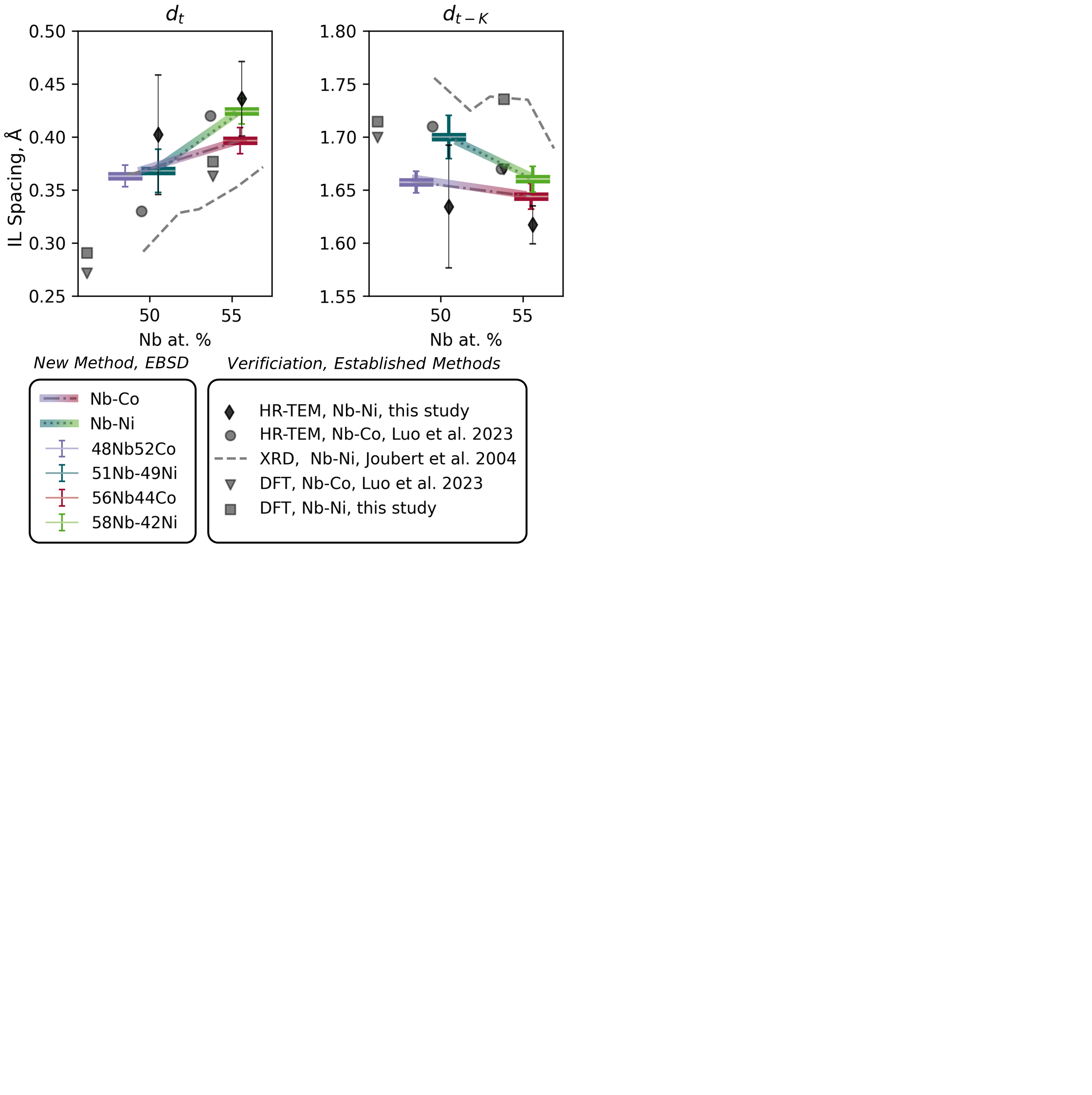}
    \caption{Comparison of \ac{ILS} determined in this study with literature values and additional DFT calculations for Nb-Ni. The data is given for both the $d_t$ and the $d_{t-K}$ spacings, as modification of the 6c2 site naturally leads to a modification of both.}
    \label{fig:ILsLitComparison}
\end{figure}

To assess the accuracy of our measurement, we need to compare our measurements with other methods of quantification or with literature data. In this case, multiple sources, such as calculations based on \ac{DFT} and measurements from \ac{XRD} and \ac{HR-STEM}, are available. In this study, we only varied the position of the 6c2 lattice site for any given lattice parameter,  thus only the \ac{ILS} of the triple layer ($d_t$) and the spacing between triple and Kagomé layer ($d_{t-K}$) will vary. In general, literature values from XRD (\cite{Joubert.2002}) for the Nb-Ni $\mu$-phase and HR-STEM \cite{Luo.2023} for the Nb-Co $\mu$-phase indicate that the triple layer spacing is expected to increase by about \num{0.075} to \qty{0.1}{\angstrom} for an increase in Nb concentration by \qty{7.5}{\atpercent} and a similar trend is expected from DFT values. HR-STEM investigations in the present study reveal an increase of the $d_t$ spacing by about \qty{0.05}{\angstrom}.  The results from the pattern matching roughly follow the trends from literature (an increase by \qty{0.05}{\angstrom} for a chemistry difference of about \qty{5}{\atpercent} in Nb-Ni and the same increase for a difference of \qty{7}{\atpercent} in Nb-Co, see \autoref{fig:ILsLitComparison}). 

The change of $d_t$ in the Nb-Co system, as measured by EBSD is slightly smaller than measured by \ac{HR-STEM} \cite{Luo.2023}, with the absolute \ac{ILS} falling directly onto the gradient between the two measurements reported by \ac{HR-STEM} and the value from EBSD determined here at the higher Nb content, yielding a slightly lower absolute value for the \ac{ILS}. In comparison with the spacings determined by XRD and Rietveld refinement on Nb-Ni performed by Joubert et al. \cite{Joubert.2002}, the determined \ac{ILS} values are about \qty{0.05}{\angstrom} higher than expected, while the lattice parameters determined in the present study agree well with those from Joubert et al. (see \autoref{fig:XRDLitComparison} in the supplementary material). Predictions of the \ac{ILS} for both systems by DFT yield intermediate values and again a consistent gradient of $d_t$ with Nb content.

Further, the $d_{t-K}$ spacing is not separately controlled when building the structures for pattern matching. Nevertheless, we can compare these values with the expectations from literature and DFT. 
In the case of the Nb-Ni system, the trend indicated by pattern matching again agrees with observations from Joubert et al.(\cite{Joubert.2002}) in terms of the direction and gradient of change in \ac{ILS}. The \ac{DFT} data for Nb-Co, as well as the Nb-Co \ac{HR-STEM} data from Luo et al. \cite{Luo.2023}, are also consistent with this and our EBSD data. While the DFT results for Nb–Ni indicate a slight increase in $d_{t-K}$ with increasing Nb content, the magnitude of this change is minimal. Therefore, the \ac{ILS} can be considered effectively constant, consistent with the observations reported by Luo et al \cite{Luo.2023}. The site lattice occupancy of the 3a site, as investigated alongside the $d_t$ spacing does not agree with observations by \ac{HR-STEM} in the Nb-Co system \cite{Luo.2023} for the low Nb sample. This might be due to the low atomic fraction of the 3a site in the unit cell of the $µ$ -phase and thus small contribution to the total scattering. On the other hand, for the Nb-Ni system \cite{Joubert.2004} in principle, the expected change on site-lattice occupancy is smaller than observed in Nb-Co. Further, the change of chemistry from low Nb to high Nb content, might also be accommodated by other site-lattices, which we will investigate in a future study.

\FloatBarrier

\subsection{Potential crystallographic features that lead to differences in NCC values}
\begin{figure*}
    \centering
    \includegraphics[width=0.9\linewidth,trim= 0cm 8cm 0cm 0cm,clip]{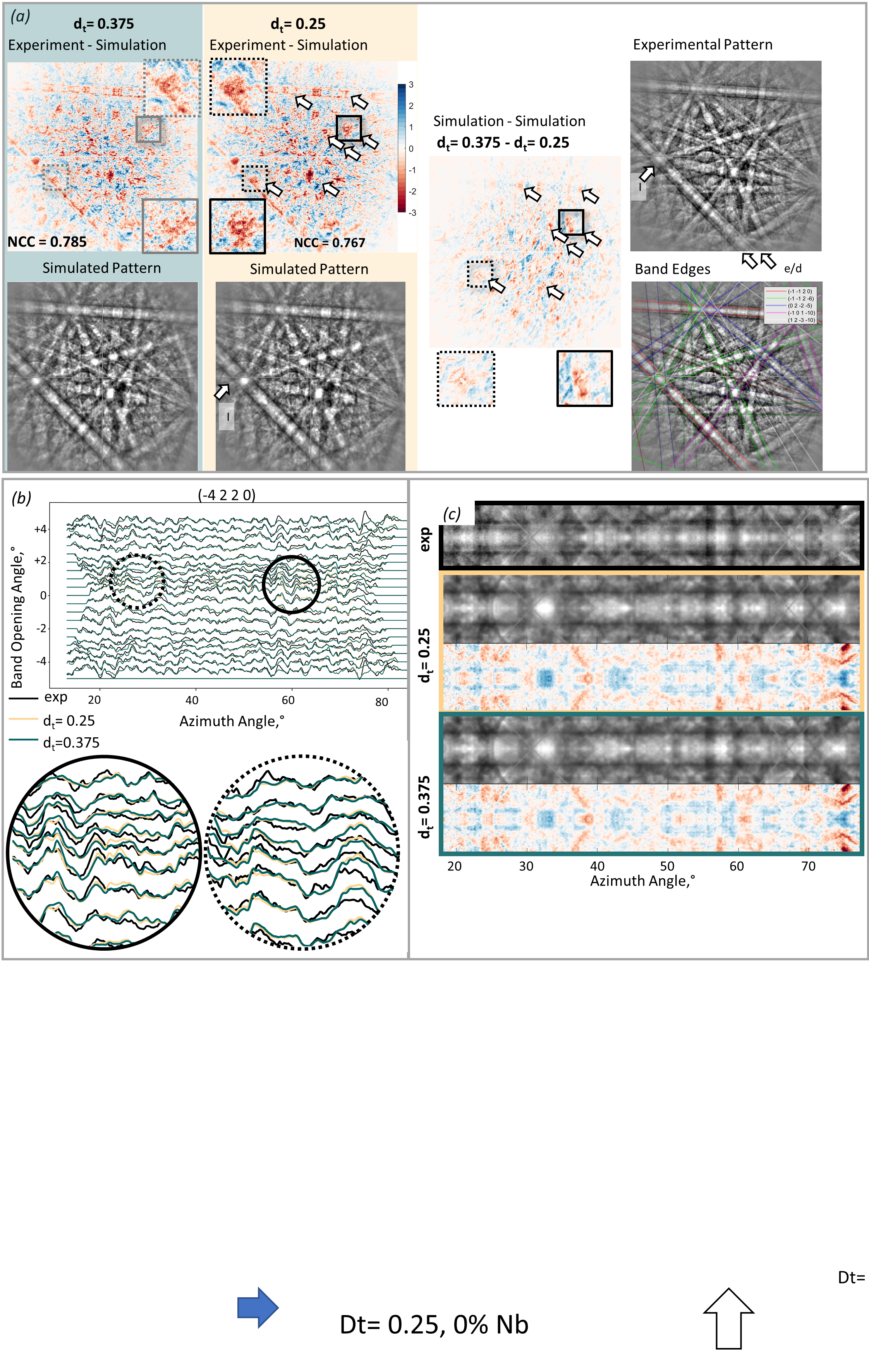}
    \caption{Attempt to highlight key differences between a $d_t=0.375$ and $d_t=0.25$ EBSD pattern, to indicate what might be causing the higher $NCC$ values for these $d_t$ spacings. \textit{(a)}: Direct comparison of experimental pattern and two simulations of structure $d_t=0.375$ and $d_=0.25$ (highlighted in yellow and green, respectively. The white arrows on the difference between experimental and simulated patterns for $d_t=0.25$ indicate regions where a larger difference is visible in the $d_t=0.25$ pattern compared to the $d_t=0.375$ pattern. Two of these regions are further enlarged. Further, the difference between a simulated pattern at $d_t=0.375$ and $d_t=0.25$ is given. The band edges are overlaid over the experimental pattern for judgment of the fit of the refinement in the bottom right corner of \textit{(a)}. Excess deficiency effects are apparent by brighter intensities on top of bands and darker intensities at the bottom of bands (e.g. the lower \hkl(-1-120) band, labelled "e/d"). \textit{(b)}: Plots of the intensity for different band opening angles across the \hkl{-4220} band (intensity values binned with \qty{0.5}{°} resolution). \textit{(c)}: Dotted and solid circles show enlarged sections of the band profile plot in \textit{b}. \textit{(d)}: "Rectified" projection of the \hkl{-4220} band used to extract the binned band profiles in \textit{(b)} for the experimental and simulated patterns, including difference plots as in \textit{a}. }
    \label{fig:FeaturesExpIL}
\end{figure*}

The main objective of this work is to enable the identification of subtle differences within a given crystal structure by means of \ac{EBSD}. We therefore need to consider not only at which level we can identify changing features within \acp{EBSP} but also what gives rise to these features in order to connect to their physical origin for interpretation. 

So far, studies reporting subtle differences in crystallographic features have, to the best of our knowledge, included, apart from the purely pattern-matching based approach, further measures to indicate the reliability of the outcomes of the pattern-matching based routine. Martin et al. \cite{Martin.2022} have used the Funk transform to identify additional features in the experimental pattern directly corresponding to pseudo-symmetry. The difference between individual patterns in the respective study was $NCC$ = 0.03, which is slightly more than the 0.02 observed in this study, e.g. between the \qty{0.35}{\angstrom} and \qty{0.3}{\angstrom} structures. Typically, differences between multiple structural candidates can be indicated in difference plots \cite{Cios.2025}, or by plotting the band profiles \cite{Wilkinson.2019}. While the band profiles can be strongly dependent on the experimental geometry \cite{Zhang.2025}, the difference plots are often hard to interpret, as they can be governed by noise or non-similarities between dynamical simulation and experiment. Nevertheless, we try to give some indications of which distinctions can be made that lead to a change in the $NCC$ values. 

In \autoref{fig:FeaturesExpIL} \textit{(a)}, the white and black arrows indicate manually identified deviations in the difference plots between experiment and simulations of the $d_t=0.375$ and the $d_t=0.25$ structures, which all indicate a darker intensity in the simulated EBSD pattern compared to the experimental one as evident by the red colour indicating a negative difference between experimental and simulated intensity. In these features, the difference between the experiment and simulation for $d_t=0.375$ is smaller than the same difference for $d_t=0.25$. Further, the same regions are highlighted on a difference between $d_t=0.375$ and $d_t=0.25$, indicating that some of these positions indeed coincide with positions where there is a somewhat large difference between $d_t=0.375$ and $d_t=0.25$. Note that several features of the difference plots do not agree very well between experimental and simulated patterns. These mostly include excess deficiency effects and variations of intensity at the zone axes. One example is the bottom \hkl(11-20) band indicated by the arrows labelled "e/d" in the experimental pattern in \autoref{fig:FeaturesExpIL}\textit{(a)}. Similarly, intensities of zone axes vary, e.g., at the crossing of \hkl(11-20) band with \hkl(-1 -1 2 -6) band. This is indicated by the arrow labelled "I" in \autoref{fig:FeaturesExpIL}\textit{(a)} in the experimental pattern and that of the simulated pattern with $d_t = 0.25$). 

The general impression of influence on the simulated patterns due to variations in the \ac{ILS} seems to be movement of bands in the background (cf. \autoref{fig:TripleLayerInfluenceOnMasterpattern} in the supplementary material). This is mostly evident when sequentially viewing the simulated dynamical templates with varying $d_t$ spacing; the features that change between subsequent patterns are, in the context of this work, best described as "sub-bands". From this, it can further be seen that the \hkl(22-40) and \hkl(10-10) bands are most likely to show differences in terms of these moving background bands crossing them.
Thus, we further consider the tangential band of \hkl(22-40) in more detail (\autoref{fig:FeaturesExpIL}, \textit{(b)} and \textit{c)}). In general, the variation between the profiles from the two simulated structures with deviating $d_t$ are more similar than the simulation compared with the experimental pattern. In \autoref{fig:FeaturesExpIL} \textit{(b)} and \textit{(c)}  this is evident from the green and yellow curves representing the two simulated structures running generally closer to each other than to the black curve representing the experimental pattern. Overall, the data shows good agreement in terms of the positions of peaks and valleys, so no features can be identified that can be directly correlated with the better match between simulation and experiment in terms of $NCC$ for the $0.375d_t$ structure.
% ference plots are however, generally difficult to interpret, as they show also variations between experimental and dynamically simulated patterns that are due to noise or assumptions of the simulation, that are not present in the experimental pattern.

We cannot at present identify those or any crystallographic features which are decisive in the pattern matching. However,  we observe a clear and consistent trend for the $NCC$ metric to discriminate changes to the $d_t$ spacings in good agreement with literature values. Therefore, we hope that the detailed understanding needed to physically link subtle changes in the crystal structure to both experimental and simulated patterns will be built as research on this matter progresses.

%, we want to stress that the retrieved results agree very well with the expectations from literature (see again \autoref{fig:ILsLitComparison}). 

\FloatBarrier

\subsection{Influence of approximations in the dynamical template simulation}
We also consider the potential impact of the approximations in the dynamical template simulation on the interpretation of the experimental data.
\autoref{fig:parameterInfluence} visualises the influence of selected simulation parameters as well as that of changes in the crystallographic structure. For the variation in minimum $d$ spacing considered in the simulation, there is only a $NCC$ difference of \num{0.0001} between a $d_{min}$ of \qty{0.2}{\angstrom} and \qty{0.1}{\angstrom}. The Bethe parameters, on the other hand, lead to a difference of \num{0.0095} in $NCC$ for the projection given in \autoref{fig:parameterInfluence} \textit{(h)}. This is the same magnitude that would be expected for the difference between two structures with $d_t$ spacings of \qty{0.35}{\angstrom} and \qty{0.3}{\angstrom} (without added noise, cf. \autoref{fig:accAssesmentIL} \textit{(b)} in the supplementary material). For the added noise with a SNR of \qty{17.5}{dB}, the $NCC$ difference between the mean difference between a \qty{0.35}{\angstrom} and a \qty{0.3}{\angstrom} structure is about \num{0.0038} and for a difference between \qty{0.325}{\angstrom} and \qty{0.35}{\angstrom} about \num{0.0010} (cf. \autoref{fig:accAssesmentIL}). Incidentally, this is also about the range we observe when comparing simulated with experimental patterns using those with $d_t$ spacings of \qty{0.35}{\angstrom} and \qty{0.3}{\angstrom} structures(cf. \autoref{tab:NCC_DIFF_EXP}). These differences between experimental and simulated patterns naturally contain a relatively large scatter of the data points.
\begin{table}
\caption{Difference in $NCC$ metric from comparing the experimental pattern to two simulated patterns belonging to the \qty{0.35}{\angstrom} and \qty{0.3}{\angstrom} structures ($\Delta NCC = NCC(exp,d_t=0.3)-NCC(exp,d_t=0.35)$), for full occupation of the 3a site with Nb. Mean values and standard deviation of the difference are given.}
\label{tab:NCC_DIFF_EXP}
\vspace{0.2cm}
\begin{tabular}{llrr}
% \toprule
  Sample & \multicolumn{2}{r}{$d_t$ \qty{0.35}{\angstrom} and $d_t$ \qty{0.3}{\angstrom}} \\
   & mean & std \\ \hline \hline
% \midrule
 48Nb52Co & 0.0029 & 0.0014 \\
\hline
 51Nb-49Ni & 0.0035 & 0.003 \\
\hline
 56Nb44Co & 0.004& 0.002 \\
\hline
 58Nb-42Ni & 0.0074 & 0.002 \\
% \bottomrule
\end{tabular}
\end{table}
 
However, tests with nine additional structures (\qty{100}{\%} Nb occupancy on the 3a site, $d_t = 0.25$--\qty{0.45}{\angstrom}, c1=40, c2=50, c3=50) showed no significant change in the measured $d_t$ spacings (\autoref{fig:BetheInfluence_IL} in the supplementary material), indicating that this approximation is also unlikely to affect the measured trends in $d_t$ spacing discussed in the present study.

Another approximation made is the choice of only a single voltage for the dynamical templates. This choice is based on the observation that the energy contributing significantly to an EBSD pattern is constrained to approximately \num{0.5} - \qty{1.5}{kV} from the primary beam \cite{Winkelmann.2019}; and thus, the assumption of a continuous slowing down approximation, as done by the EMSoft software \cite{Callahan.2013} if multiple energy bins are considered is not correct. Further, we utilised the energy filtering capabilities of the Clarity\textsuperscript{\texttrademark} \ac{EBSD} detector, and observed that the application of energy filtering with a threshold of \qty{19.5}{kV} leads mostly to the presence of diffuse features in the \ac{EBSP} (see \autoref{fig:EnergyFiltering} in the supplementary material, top row), as is revealed by the difference plot. In contrast, the difference between an unfiltered \ac{EBSP} acquired at \qty{20}{kV} and \qty{19}{kV} acceleration voltages, respectively, shows a lot of high-frequency changes in the difference pattern. (see \autoref{fig:EnergyFiltering}, bottom row). The improvements of energy filtering in this case are not as stark as observed by Vespucci et al. \cite{S.Vespucci.2015} for Si, indicating further that the energy distribution in the Nb-Ni sample is close to the primary beam. In general, the observations for the energy distribution of experimental \acp{EBSP} are conflicting in the literature: Template matching with weighting of the energy bins \cite{Shi.2022b} leads to an energy range of \qty{19.9}{kV} to \qty{18.4} {kV} for a primary beam energy of \qty{20}{kV}.  While direct measurement of the energy distribution on a Si wafer for \qty{12}{keV} primary beam \cite{DellaVentura.2026} leads to large spreads of the energy range of 3 to 4 keV and a significant contribution of electrons with energy losses larger than 4 keV, we note that all of the low-energy patterns displayed in the referenced work seem to have similar diffraction features and band-widths. For lattice parameter evaluation (and sub-lattice spacings), it is important to try to match the signal that contributes to the Kikuchi bands (i.e. quasi-elastic losses). The total signal striking the detector may include higher energy loss electrons (as shown in the energy spectra within \cite{DellaVentura.2026}). These are unlikely to be significant for the analysis method used in the present manuscript. 

The energy being closer to the primary beam in the present study might be due to the higher atomic numbers of Nb and Ni, and thus stronger scattering compared to Si. To reduce the potential influence of broadening of the energy distribution, which is expected to be stronger at the outsides of the patterns, we applied a weighting in the form of a Gaussian distribution around the pattern for the template matching in this study (as evident in e.g. \autoref{fig:FeaturesExpIL}, \textit{(a)}  by the reduced contrast of the outside of the experimental and simulated pattern, as well as the difference plots in red-blue colour-map).
In summary, we conclude that the energy constraint to a single voltage of the dynamical templates is sufficient for the present study.

Lastly, as e.g. Zhang et al. \cite{Zhang.2025} report, the EBSD geometry can generally lead to blurring of band profiles. A further question is therefore whether such blurring could mask the structural differences of interest. We have given the band profiles of all \ac{PCA} patterns of the first map and \ac{PCA} tile used in the analysis of this study in the supplementary material, \autoref{fig:BandProfiles}. From this, it can be seen that, e.g., for the \hkl(-1 -1 2 0) and the \hkl(02-2-5) bands, the width of the bands is generally comparable to the simulated structures. In total, the positions of minima and maxima agree between the simulation and the experiment. The variation between a structure with \qty{0.25}{\angstrom} and \qty{0.35}{\angstrom} is much smaller than the variations between individual band profiles. This suggests that geometric broadening contributes little to the spread of individual profiles and does not obscure the overall correspondence between experiment and simulation.

\subsection{ Implications for future investigations of subtle changes in complex crystals} 

The main objective of this study was to assess whether EBSD can enable correlative analysis of subtle local crystal-structure changes across larger areas and gradients within a sample. For this reason, the Nb-based $\mu$-phase was chosen as a model system, since earlier work had already shown that a small compositional change, leading to full occupation of the 3a site by Nb, causes both measurable changes in interplanar spacing and a dramatic change in the critical stress for basal slip \cite{Luo.2023}. These structural changes also provided a basis for proposing the underlying active dislocation mechanisms, which were indirectly supported by atomistic simulations and TEM \cite{Luo.2023}. More generally, if similarly subtle structural variations also govern properties in other complex intermetallics, EBSD may offer a valuable path to studying them over spatial ranges that are difficult to access with highly local methods such as HR-TEM or non-local methods like laboratory-based XRD.

Naturally, the EBSD analysis also comes with high demands on resources, although in this case, mainly of a computational nature. The effort required to determine the specific $d_t$ spacings and site lattice occupancies is rather high: A total of 45 dynamical templates was needed for the \textit{(IL+3a)\textsubscript{var}} dataset for each sample, and a further 31 structures for the \textit{\ac{SLO}\textsubscript{var}} set. This only sampled a small parameter space out of all possibilities that we have in the $\mu$-phase, which has, considering lattice parameters, site lattice occupancy, and atom positions, a total of 2 + 5 + 5 = 12 parameters that could be controlled. Sampling all of these parameters with 5 different values per parameter would lead to a search space of about \num{244000000} dynamical templates, which would consume 8 times as many core hours ($\approx$\num{2} billion hours, based on $\approx$ \qty{20}{min} per calculation running on \qty{24}{cores}) and \qty{46400}{TB} of memory (based on about \qty{190}{MB} per template) solely for the dynamical calculations without any additionally applied template matching.  

This illustrates that the present approach is only practical for well-constrained problems. Thus, the method to estimate interplanar spacings and site lattice occupancies should ideally be applied to well-constrained problems, where only 2 or 3 parameters need to be tuned (as in the present study, where we focused initially on the two most important parameters (as revealed by Luo et al. \cite{Luo.2023}), namely $d_t$ and occupancy of 3a sites). In the future, reduction to a fraction factorial design or other screening methods may be considered and validated, but in addition, advances in pattern simulation, including potentially the prediction of master patterns by machine learning methods or the acceleration of pattern matching for a large number of alternative patterns forming the potential solution space, may greatly reduce the computational demand of this method or enable studies with a less well-constrained parameter space.

Where the EBSD analysis is combined with local testing of microstructures, by methods such as nanoindentation, and large gradients are present, the inclusion of further methods may be helpful. For example, the lattice parameters could initially be estimated by refinement of the \ac{HOLZ} rings \cite{Nolze.2017}, analysis of the band profiles \cite{Nolze.2023,Saowadee.2017} or distortion of dynamical templates \cite{Cios.2023}. For simpler crystal structures than the Nb-Ni $\mu$-phase, the parameter space is also further decreased, such that maybe only 2 or 3 atom positions would need to be modified. Further, although we could not completely explain why we achieve a good agreement of the $d_t$  and $d_{t-K}$ spacings with literature values, further investigation with dynamical templates corrected for e.g. excess/deficiency effects \cite{Winkelmann.2021} and more detailed corrections of the energy distribution \cite{Shi.2022b} might lead to resolution of these details. More generally, such developments could make the present approach more efficient and more transferable to other material systems. This could make even investigation methods based on straining the individual atomic site dynamical templates possible, which would reduce the amount of data that needs to be simulated and pattern-matched.

\subsection{New capabilities of \ac{EBSD} for identifying interlayer spacings and other subtle crystallographic variations}

As a method being established, we have opted to verify all results against known literature data or other investigations, which are in themselves often not ideal or comprehensive in many respects. However, given the increasing availability of high quality data with respect to measurements or predictions of lattice spacings and site occupancies from experiments and simulations, we expect that the progress of using \ac{EBSD} will now be able to keep pace with those developments for the statistically more meaningful or correlative analysis across many points of interest investigated by other methods, e.g. in terms of mechanical properties (see also the explanation on this in the introduction of the manuscript and \autoref{fig:ResolutionComparison}). In this context, the improvements to the dynamical template matching, as discussed above (especially excess/deficiency effects and assessment of accuracy of the pattern centre rather than just precision), will also likely contribute to further method development, achieving greater accuracy and significant predictions even for difficult lattice sites or spacings in the future. Further, even without these corrections, the pattern-matching routine presented in this study is apparently capable of predicting the "correct" $d_t$ spacings 
in accordance with the literature, as discussed above.

\section{Conclusion}

We investigated the potential of \ac{EBSD} to determine the  triple-layer spacings ($d_t$)  in Nb--Ni and Nb--Co  $\mu$-phases with different compositions. Concurrent with this investigation, we also refined the 3a site-lattice occupancies, as they are considered responsible for the observed change in  $d_t$ spacings of the investigated systems.

\begin{itemize}

    \item We found that, after careful calibration of the EBSD geometry, the projection centre can be determined with a precision of approximately \qty{0.1}{\%} of the pattern width. This level of precision is essential for the subsequent analysis of crystallographic features in EBSD patterns by pattern matching.

    \item By combining this precise projection-centre calibration with additional prior information, this case study demonstrates that the interlayer spacing (especially the $d_t$ spacings) of certain atomic planes can be derived from EBSD in close agreement with evaluations based on STEM or XRD in the literature.
    
    \item The occupancy changes of the 3a lattice sites, which are considered necessary for the change in $d_t$ spacing, could not be resolved by the presented methodology, which could be attributed to the small fraction of total lattice sites in the $\mu$-Phase.
    % \item With regard to site occupancy, the partial occupancies of the 6c sites determined by EBSD are consistent with expectations from the literature. In contrast, the occupancy of the 18h site is predicted to vary, in disagreement with literature expectations and leads to inconsistent estimates of sample composition by approximately \qty{10}{\atpercent}

    \item Differences between experimental and simulated patterns remain. These discrepancies may arise from unconsidered excess--deficiency variations and from ambiguities in the energy distribution of the EBSD patterns. They may also explain why the exact crystallographic features responsible for the pattern differences associated with variations in the $d_t$ spacing could not yet be identified.
    
\end{itemize}

Building on these promising first results in expanding the crystallographic detail accessible by EBSD, future studies will benefit from further reductions in computational cost beyond the parameter restrictions applied here, as well as from continued advances in pattern matching and simulation. 
The approach presented here may then be extended to include energy distributions and additional dynamical effects, such as excess/deficiency. We envision that extracting this new depth of information by EBSD as a surface scanning technique will be particularly valuable for correlative approaches, such as indentation mapping, where the combination of higher spatial resolution and statistical significance from many measurement points can fully leverage the potential of EBSD compared with XRD and STEM.

\subsection*{Acknowledgement}

This work was supported by the German Research Foundation (DFG) within the Collaborative Research Centre SFB 1394 "Structural and Chemical Atomic Complexity—From Defect Phase Diagrams to Materials Properties" (Project ID 409476157), including the project groups C02, A03, A04, A07 and B01. This project has received funding from the European Research Council (ERC) under the European Union’s Horizon 2020 research and innovation programme (grant agreement No. 101168203 TailorPlast). Further support was granted by the MITACS Globallink Research Award. Computation of dynamical templates, as well as pattern matching and projection centre calibration routines, were performed with computing resources granted by RWTH Aachen University under project rwth-1308. The data used in this publication were managed using the research data management platform Coscine with storage space granted by the Research Data Storage (RDS) of the DFG and Ministry of Culture and Science of the State of North Rhine-Westphalia (DFG: INST222/1261-1 and MKW: 214-4.06.05.08 - 139057). The authors express thanks to F. Busch and K.B. Dinh for help with metallographic preparation, F. Busch for sample synthesis of Nb-Co samples and H. Springer for synthesis of Nb-Ni samples. We thank N. Ayeb for providing \ac{XRD} measurements of the Nb-Ni samples. We further want to thank G. Cios for the helpful discussion on the potential influence of lattice parameters on pattern centre calibration and T. Zhang for discussions on measuring pattern quality in the frequency domain. T.B.B. acknowledges funding from the Canada Research Chair program and Natural Sciences and Engineering Research Council of Canada (NSERC) [Discovery grant: RGPIN-202204762, ‘Advances in Data Driven Quantitative Materials Characterisation'].

\subsection*{CRediT authorship contribution statement}
\textbf{Lukas Berners:} Writing – original draft, Visualisation, Software, Methodology, Investigation, Validation, Formal analysis, Data curation, Conceptualisation.
\textbf{Nisa Ulumuddin:} Writing  – review \& editing, Formal analysis, Validation, Data Curation.
\textbf{Silvia Richter:} Writing – review \& editing, Investigation, Formal analysis, Data curation.

\textbf{Annika Baum:} Writing  – review \& editing, Formal analysis, Software,  Data Curation.
\textbf{Khalil Rejiba:} Writing  – review \& editing, Software,  Data Curation.
\textbf{Joshua Spille:} Writing – review \& editing, Investigation, Formal analysis, Data curation.
\textbf{Siyuan Zhang:} Writing – review \& editing, Investigation, Formal analysis, Data curation.
\textbf{Christina Scheu:} Writing – review \& editing, Funding acquisition.
\textbf{Joachim Mayer:} Writing – review \& editing, Funding acquisition.
\textbf{Benjamin Berkels:} Writing – review \& editing, Software, Supervision.
\textbf{Ben Britton:} Writing – review \& editing, Supervision, Software, Project administration, Methodology, Conceptualisation.

\textbf{Sandra Korte-Kerzel:} Writing – review \& editing, Supervision, Project administration, Funding acquisition, Conceptualisation.

 \subsection*{Declaration of Generative AI and AI-assisted technologies in the
writing process}
During the preparation of this work, the authors used Grammarly and 
RWTH-GPT in order to check for spelling and grammar
errors as well as to improve language conciseness and style. These tools were not
used for the generation of new text, but only for the modification of existing
text passages. Further, after using Grammarly and RWTH-GPT,
the respective author reviewed and edited the content as needed, and the author takes
full responsibility for the content of the published article. For final checks of latex errors during the submission process to Elsevier, Claude Code was used.

\subsection*{Data Availability}
EBSD Data used in the present study is available at
\doi{10.5281/zenodo.21703145}

Dynamical Template Simulations used in the present study are found at  \doi{10.5281/zenodo.21703051}

Other data used in the present study is available at Zenodo : \doi{10.5281/zenodo.19349148}

\subsection*{Code Availability}
The Source Code used in this publication is available at the AstroEBSD repository \url{https://github.com/ExpMicroMech/AstroEBSD}

%% Bibliography is read from references.tex (BibTeX output) because the journal system does not run BibTeX.
%% To regenerate after editing references.bib: see the instructions at the top of references.tex.
% \bibliographystyle{elsarticle-num} 
%  \bibliography{references}
%% Bibliography generated by BibTeX (elsarticle-num) from references.bib.
%% The journal build does not run BibTeX, so main.tex \input's this file instead.
%% To regenerate: in main.tex comment in \bibliographystyle/\bibliography (and comment out
%% \input{references}), run pdflatex + bibtex, then copy main.bbl over this file's body.

 % bibliography (generated from references.bib)
%% https://www.overleaf.com/learn/how-to/Cross_referencing_with_the_xr_package_in_Overleaf
% \input{Supplementary.tex}
% 
% \begin{supplementary}
% \ifshowsupplementary

% \documentclass[000main]{subfiles}
% \begin{document}

\clearpage
\setcounter{page}{1}
\renewcommand{\thepage}{S-\arabic{page}}
\renewcommand{\thefigure}{S\arabic{figure}}

\setcounter{figure}{0}

\renewcommand{\thetable}{S\arabic{table}}
\setcounter{table}{0}

\renewcommand{\thesection}{S\arabic{section}}
\setcounter{section}{0}
\renewcommand{\thesubsection}{\thesection.\arabic{subsection}}
\setcounter{subsection}{0}
\clearpage
\onecolumn
\section{Supplementary}

% \section{Supplementary}
% \onecolumn
\subsection{Additional data for WDS}
\begin{figure}[!h]
    \centering
    \includegraphics[width=\linewidth, trim = 0cm 3cm 0cm 0cm,clip]{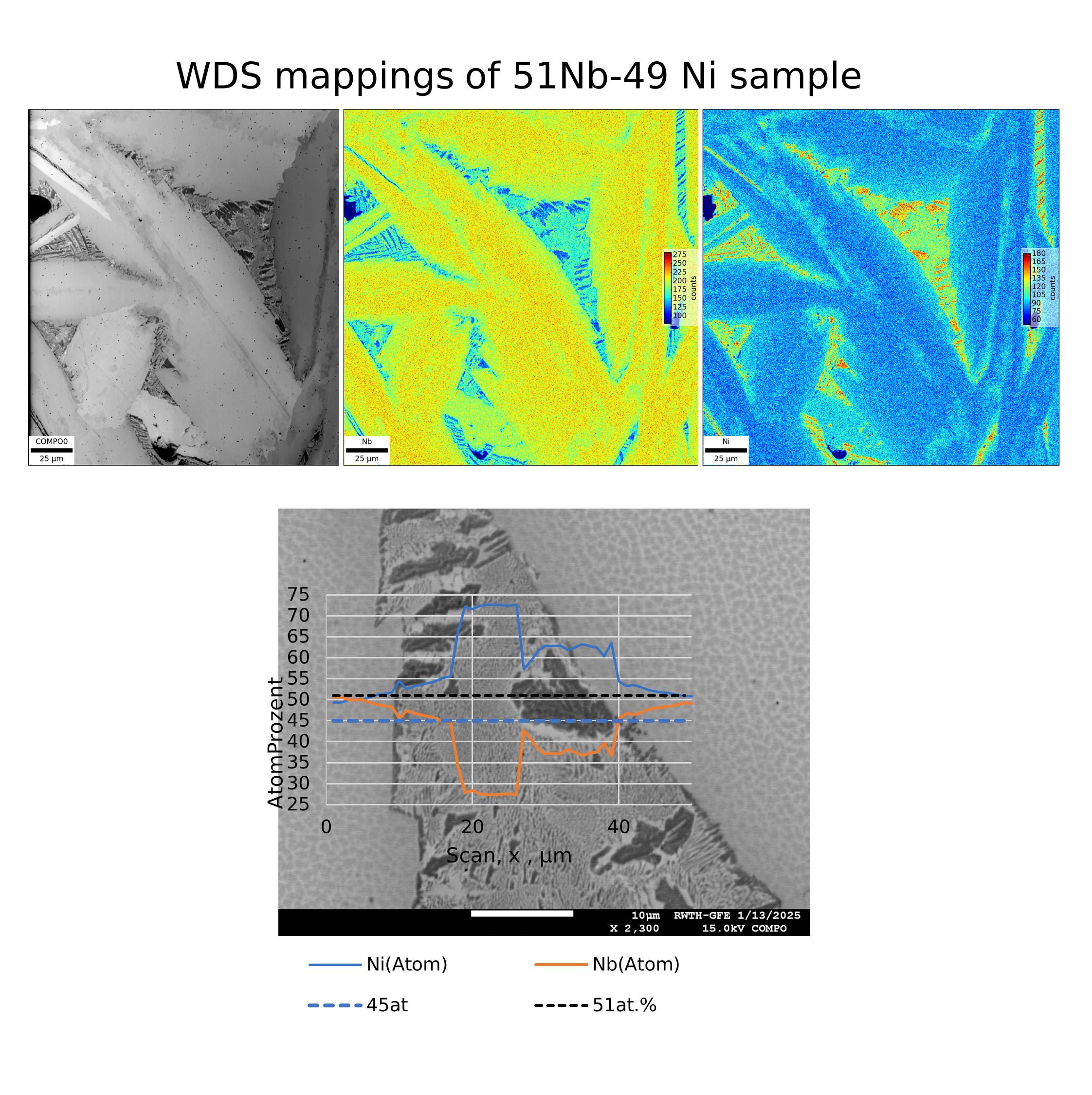}
    \caption{Additional Data from WDS scanning. The line scan shows a gradient of atomic composition towards the secondary phase region. orange = Nb, blue = Ni.}
    \label{fig:WDS_gradients}
\end{figure}

\clearpage
\subsection{Testing of best matching acceleration Voltage}
\begin{figure}[!htbp]
    \centering
    \includegraphics[width=\linewidth, trim = 0cm 5cm 0cm 0cm]{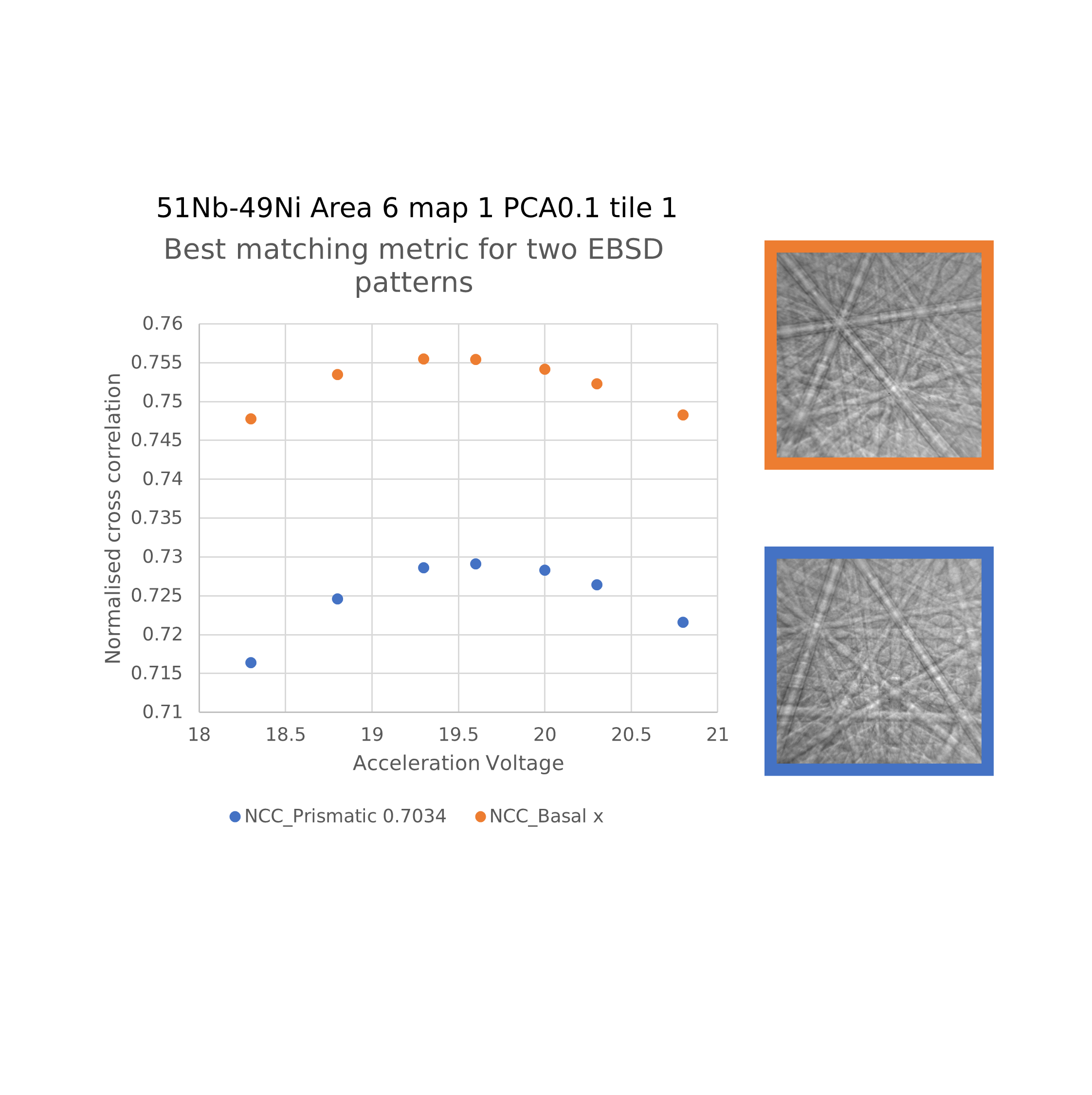}
    \caption{Initial testing for optimal Voltage of the simulated structures using experimental structure, as observed by Joubert et al. 2004.}
    \label{fig:VoltageTesting}
\end{figure}
\FloatBarrier
\clearpage

\subsection{Additional Information for noise and PC offsets, when site lattice occupancy is varied. }
\begin{figure*}[!hbp]
    \centering
    \includegraphics[width=\linewidth, trim=0cm 0cm 0cm 0cm,clip ]{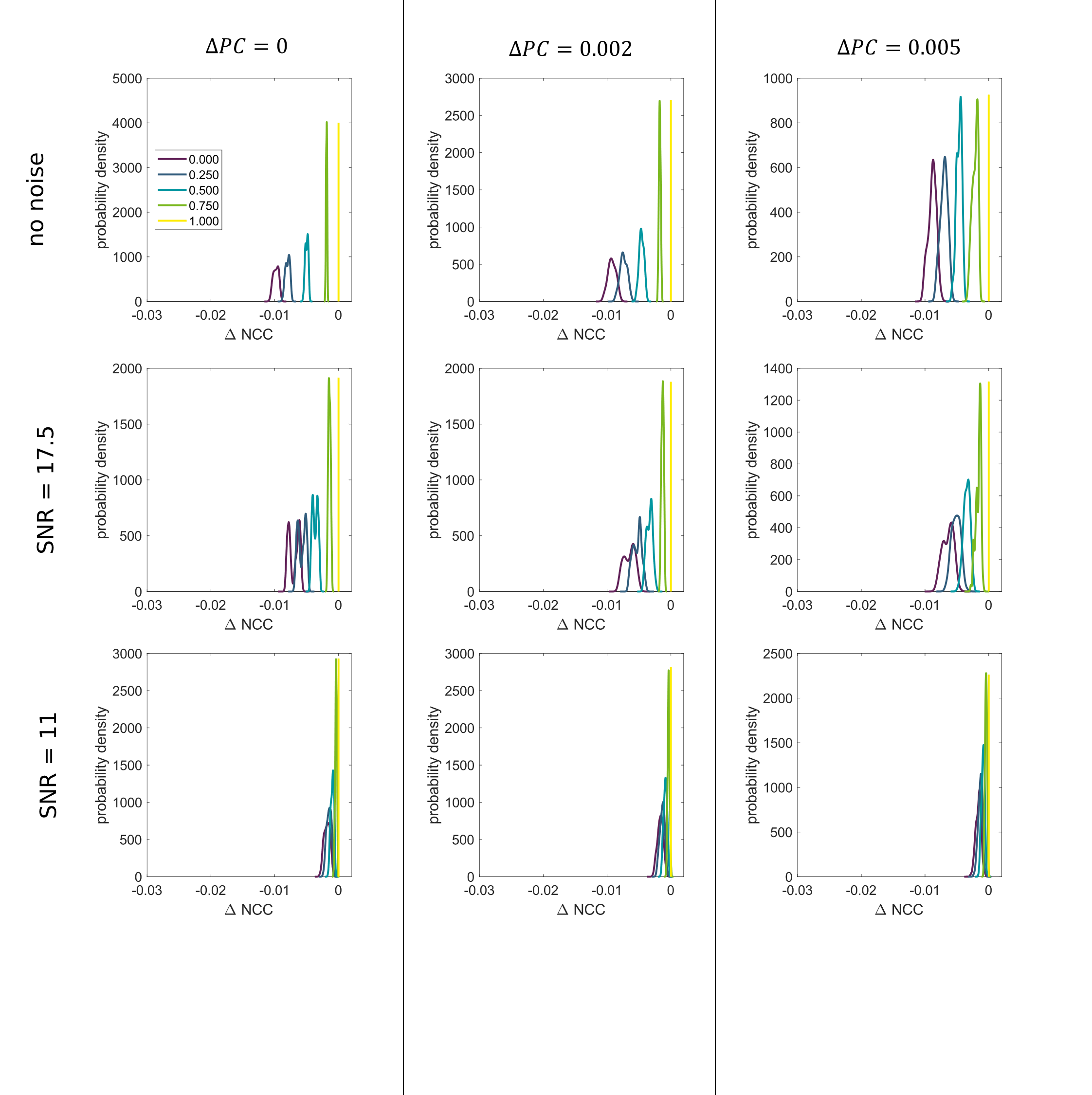}
    \caption{Addition to the assessment of the influence of noise and PC offsets on the NCC difference. In addition to the comparisons shown in the main text, we also give the 4 additional comparisons for PCX offset = [0.002, 0.005] and noise = [17.5 dB, 11 dB].}
    \label{fig:accAssesmentOCCADD}    
\end{figure*}
\clearpage
\subsection{Noise and Offset assessments for IL spacing}
\begin{figure*}[!hbp]
    \centering
    \includegraphics[width=\linewidth, trim=0cm 9.8cm 0cm 0cm,clip ]{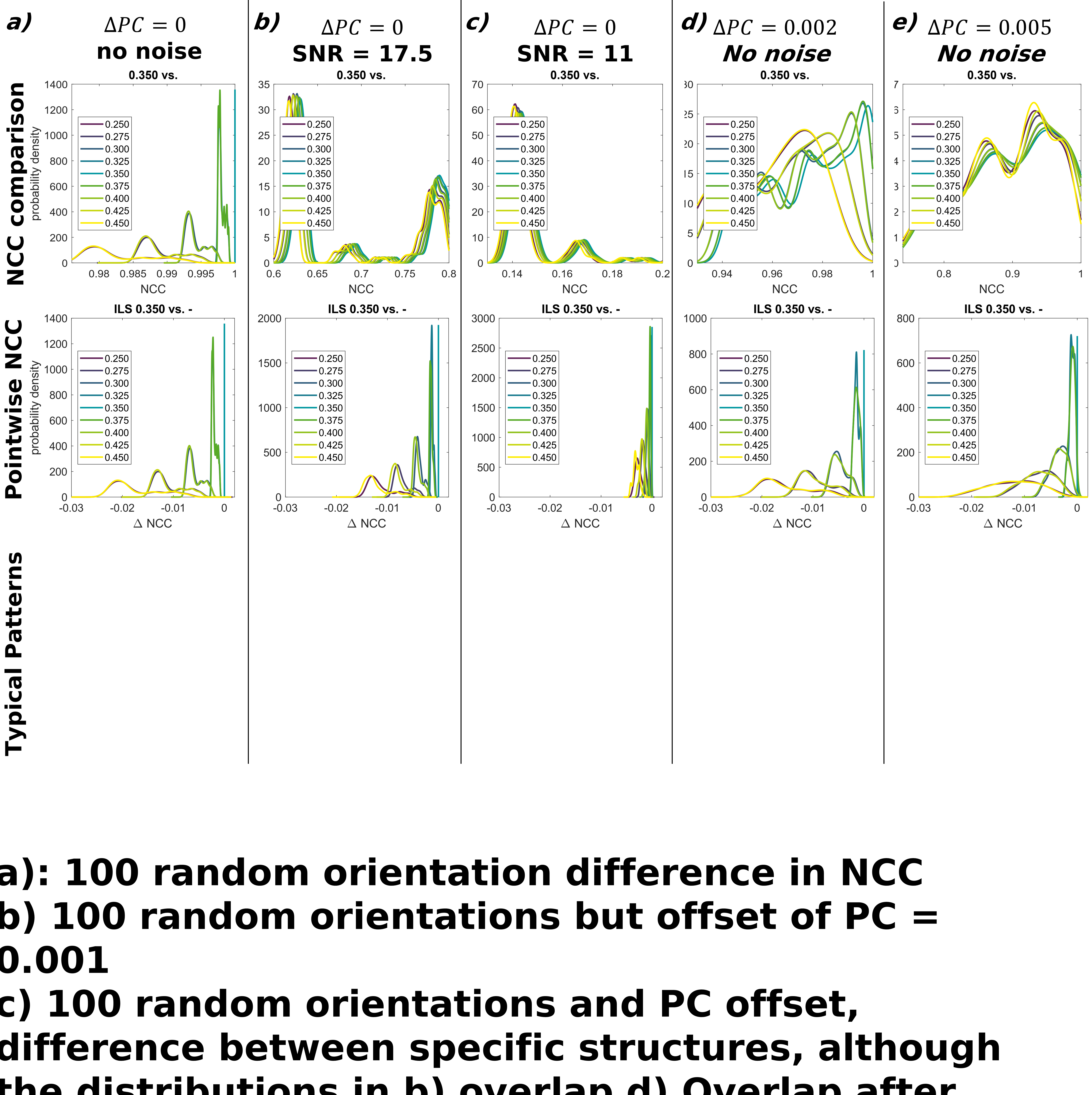}
    \caption{Comparison of distinguishability of \ac{ILS} ($d_t$) in dependence of noise and PC misalignment. Rows: Top: Comparison of NCC values of 100 random patterns against an "unaltered" reference pattern, Bottom: "Point-To-Point" comparison of the same data. The bottom row is omitted in comparison to \autoref{fig:accAssesment}, as noise ratios and offsets are the same. No noise and offset applied. \textit{b)}: Average SNR of 17.5, no offset. \textit{c)}: Average SNR of 11, no offset.
    \textit{d)}: No Noise. Offset of 0.2 \% = 1px offset. \textit{e)}: No Noise. Offset of 0.5 \% = 2.5px offset.}
    \label{fig:accAssesmentIL}    
\end{figure*}
\FloatBarrier
\clearpage
\subsection{Working Principle of PCA}
\begin{figure*}[!hbp]
    \centering
    \includegraphics[width=\linewidth,trim = 0cm 2.5cm 0cm 0cm, clip]{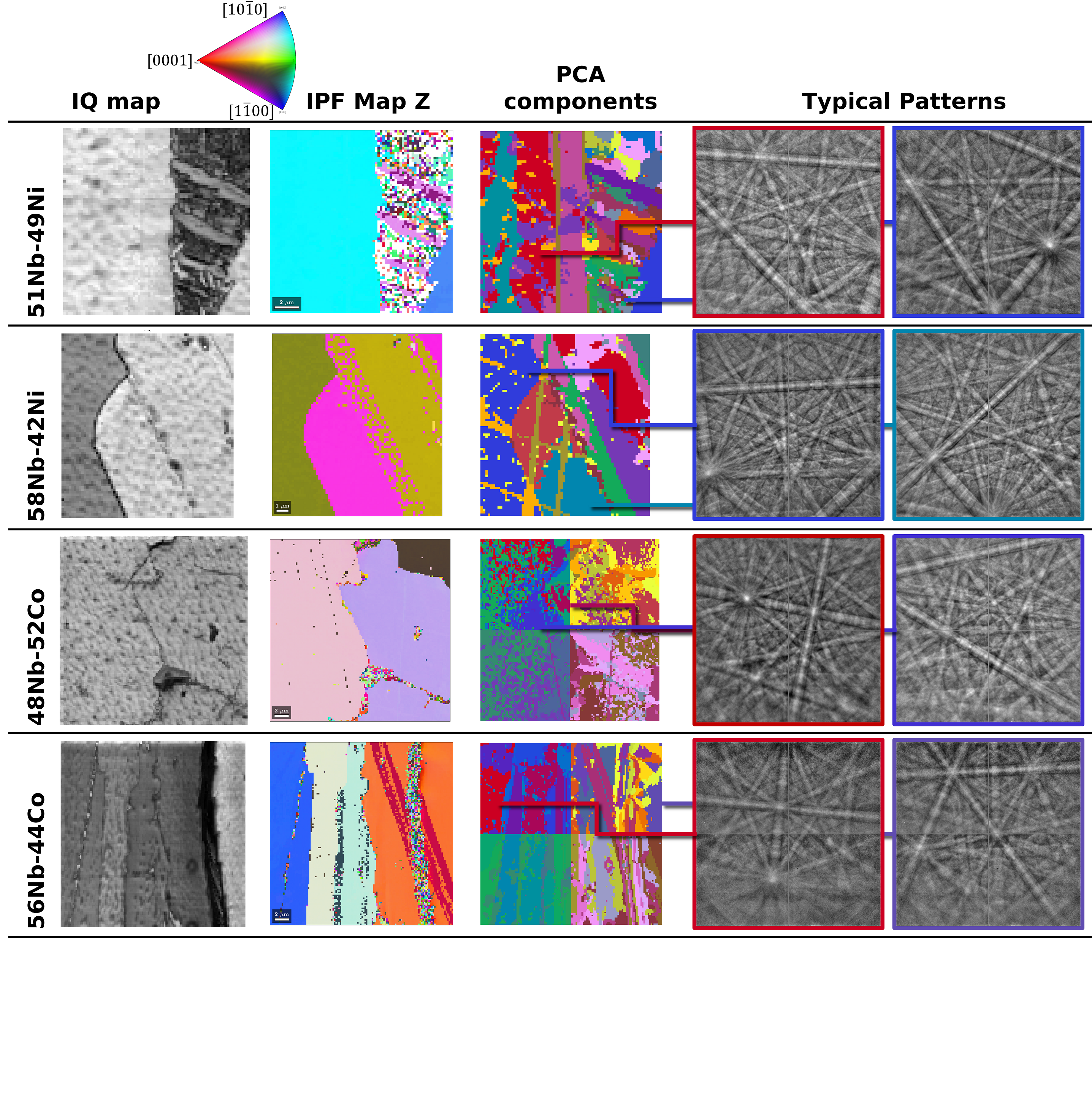}
    \caption{Working principle of PCA, each colour represents a component identified by the PCA, each component has a representative pattern associated with it, utilised for further analysis. IQ on the left side for comparison. In the 56Nb-44Co and 48Nb-52Co samples, cropping of the EBSD map to 2x2 tiles was necessary as the full area was too large to be loaded into the RAM. The PCA was therefore run on the tiles individually.}
    \label{fig:PCAexp}
\end{figure*}
\clearpage
\FloatBarrier
\subsection{Visual Influence of PCA on Pattern Noise}
\begin{figure}[htbp!]
    \centering
    \includegraphics[width=0.7\linewidth,trim= 0cm 8cm 0cm 0cm, clip]{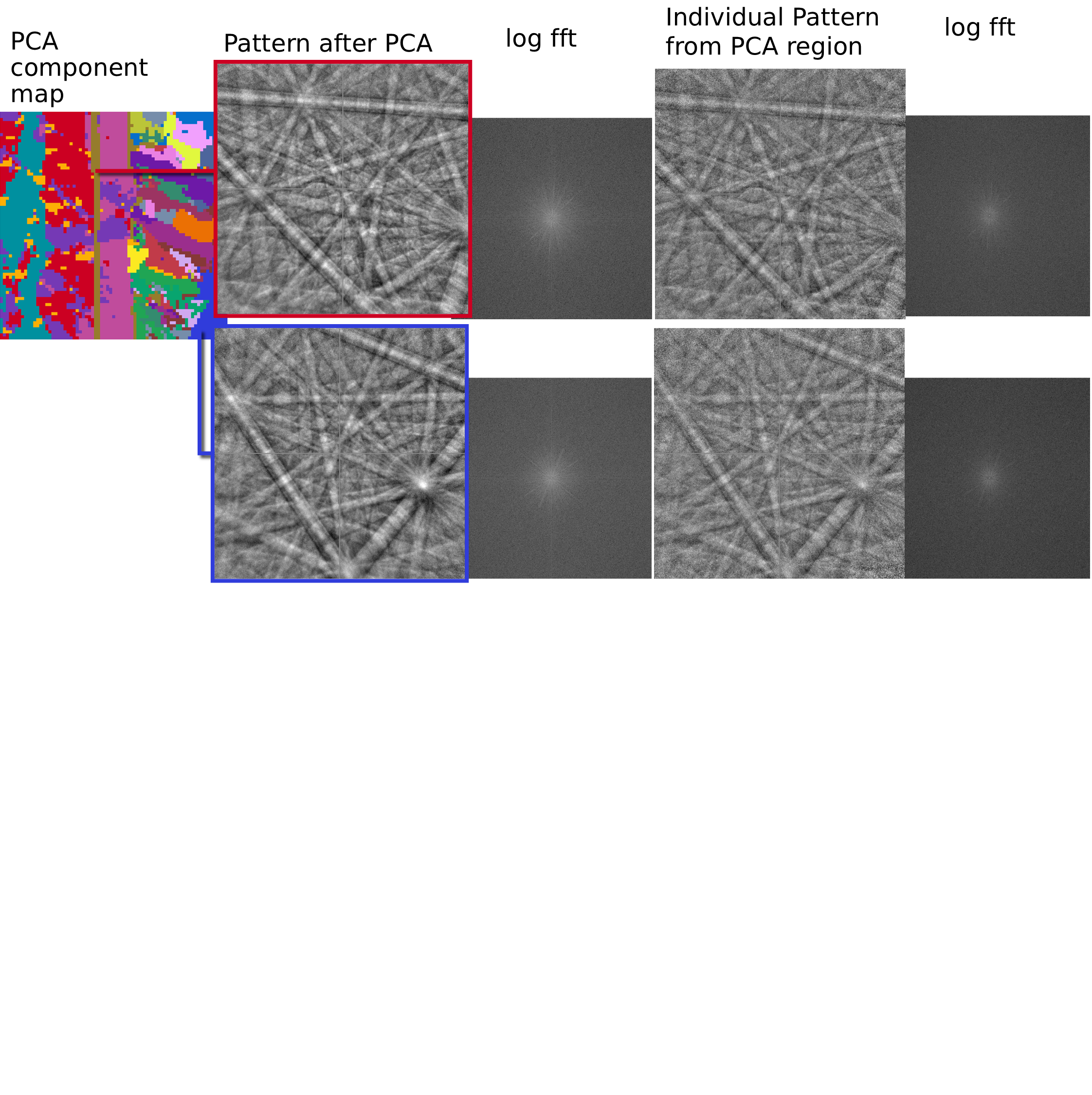}
    \caption{Visualisation of the pattern improvements from the PCA. \textit{Left Column}: PCA component map of Sample 51Nb-49Ni. \textit{Centre Column}: Two exemplary PCA patterns, with the log10 power spectrum shown next to it \textit{Right Column}: Two patterns corresponding to individual points from the EBSD map belonging to the respective PCA regions, with the log10 power spectrum shown next to it.}
    \label{fig:PCA_pat_improv}
\end{figure}

\clearpage
\FloatBarrier

\subsection{Convergence of Algorithm}
\begin{figure*}[!htbp]
    \centering
    \includegraphics[width=\linewidth, trim = 0cm 3.2cm 0cm 1cm,clip]{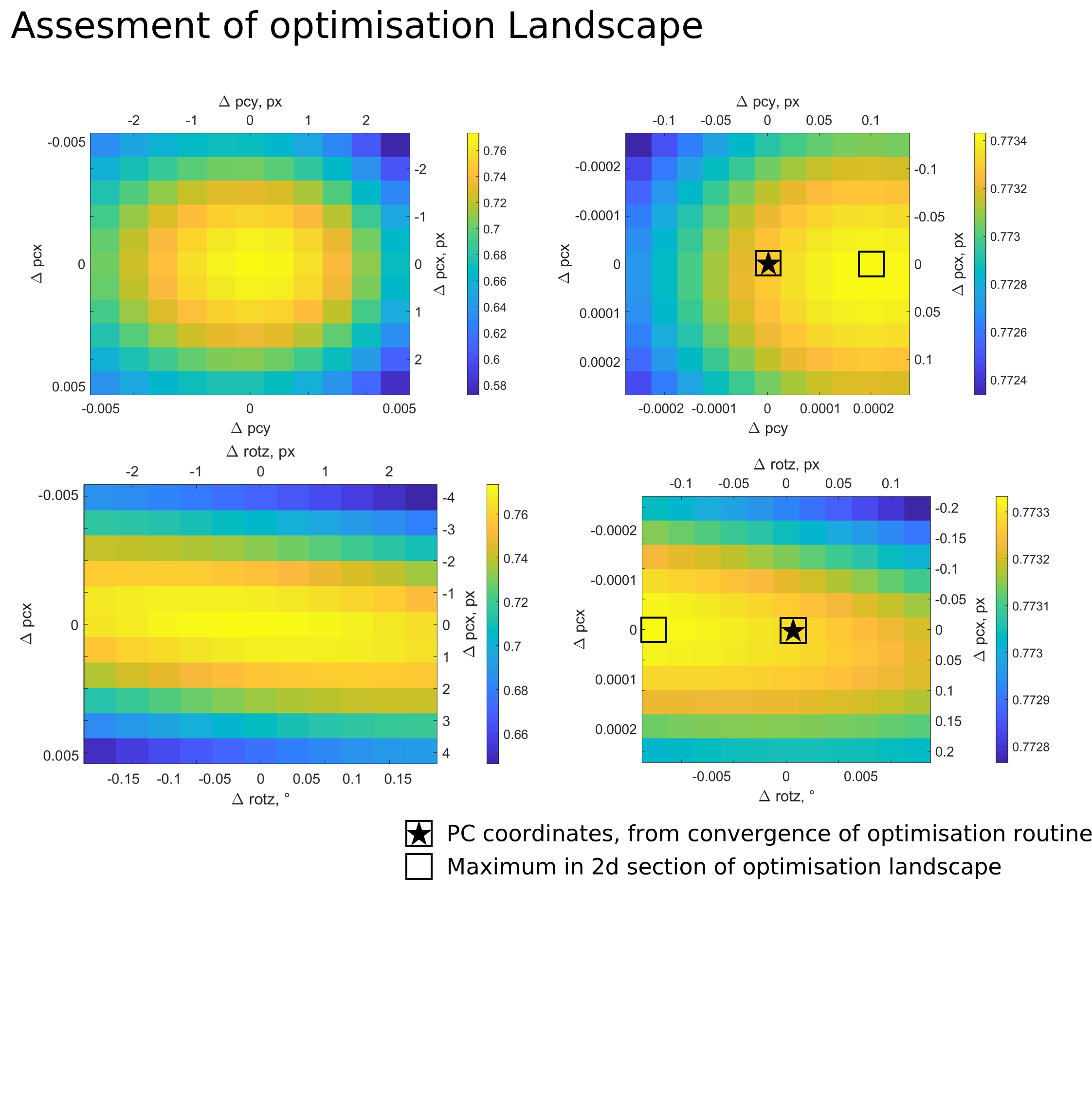}
    \caption{Comparison of experimental optimisation Landscape based on the NCC metric. A coarse and a fine region around the $PC$ and orientation values the algorithm converged to is shown for the example pairs of $\Delta PC_x$, $\Delta PC_y$ and $\Delta rot_z$, $\Delta PC_y$. The offset between the highest metric and the actual pattern centre can be deduced by finding the maximum of the distribution of the NCC values and the centre of the plot at (0,0), further indicated by a star and a square, respectively}
    \label{fig:Landscape_Exp}
\end{figure*}

% \subsection{Convergence of the optimisation routine}
We tested how close the optimisation routine can get to the maximum of the comparison metric between 
the experimental and simulated patterns. For this, we screened all possible two-dimensional sections of the 6-dimensional parameter space of projection centre coordinates and Euler angles (This is visualised for the combination of 
$PCX$ and $PCY$ as well as for $PCX$ and a rotation of the crystallographic unit cell around the sample normal in \autoref{fig:Landscape_Exp}). From this, we can see that for the 
$PCX$, $PCY$ combination, the maximum of our comparison metrics is off by \qty{0.1}{px} in the direction of $PCX$. The optimisation landscape for $PCX$ and rotation around the sample normal is "sloppy", as was observed before \cite{Pang.2020}. Here, the maximum is missed by \qty{0.5}{px}. These deviations can be due to the high interdependence of parameters in combination with "slop" and flatness of the optimisation landscape close to the optimum. The difference between the maximum $NCC$ value and the value at the convergence result of the optimisation routine is about 0.0002.
\FloatBarrier
\clearpage
\subsection{Additional Comparison of XRD data with literature values}
\begin{figure*}[!hbp]
    \centering
    \includegraphics[width=\linewidth,trim = 0cm 8cm 0cm 0cm,clip]{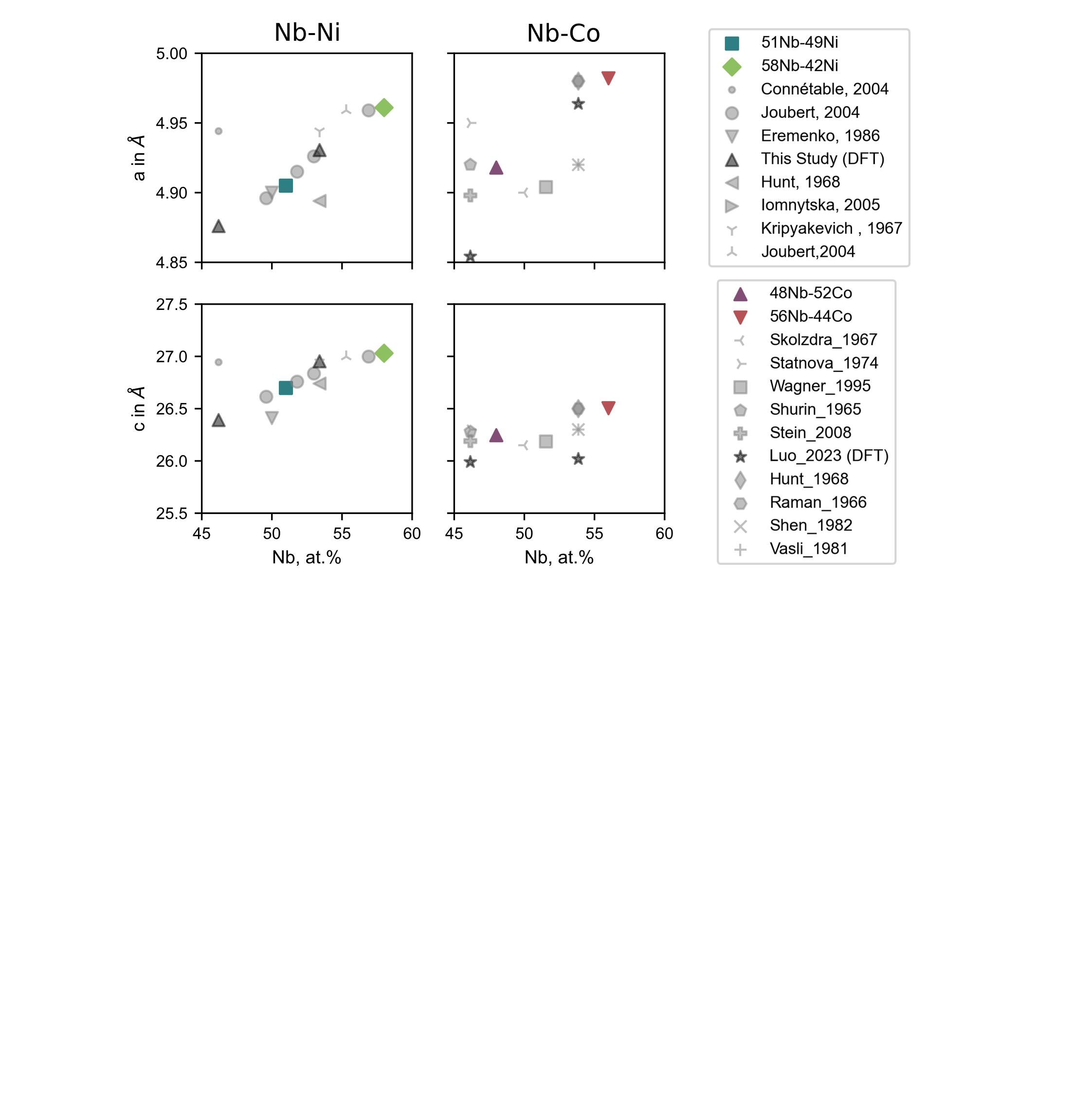}
    \caption{Comparison of lattice parameter measurements by XRD with literature values}
    \label{fig:XRDLitComparison}
\end{figure*}
\clearpage
\subsection{Investigation of broadening from XRD data}
\begin{figure*}[!hbp]
    \centering
    \includegraphics[width=\linewidth,trim = 0cm 0cm 0cm 0cm,clip]{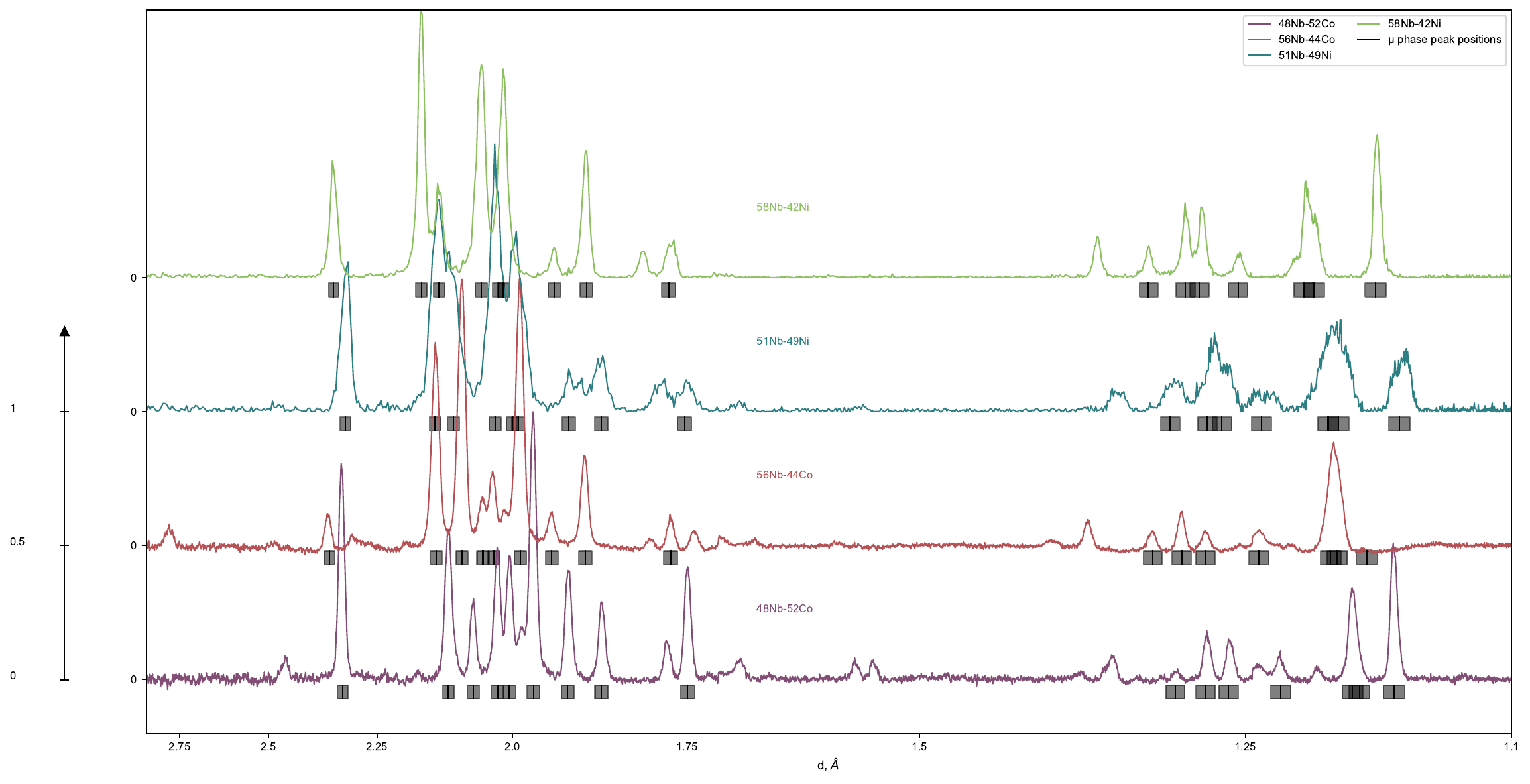}
    \caption{XRD data with refined mu phase positions overlaid. The grey box around the black lines marks the maximum expected variation of lattice parameter for a change of lattice parameters $a$ and $c$ by 0.998 or 1.002 corresponding to 0.2 \% change in lattice parameter.}
    \label{fig:XRDLitComparison_broadening}
\end{figure*}
\FloatBarrier
\clearpage

\subsection{Template matching of individual Patterns}
\begin{figure}[h]
    \centering
    \includegraphics[width=\linewidth,trim = 0cm 10cm 0cm 0cm, clip]{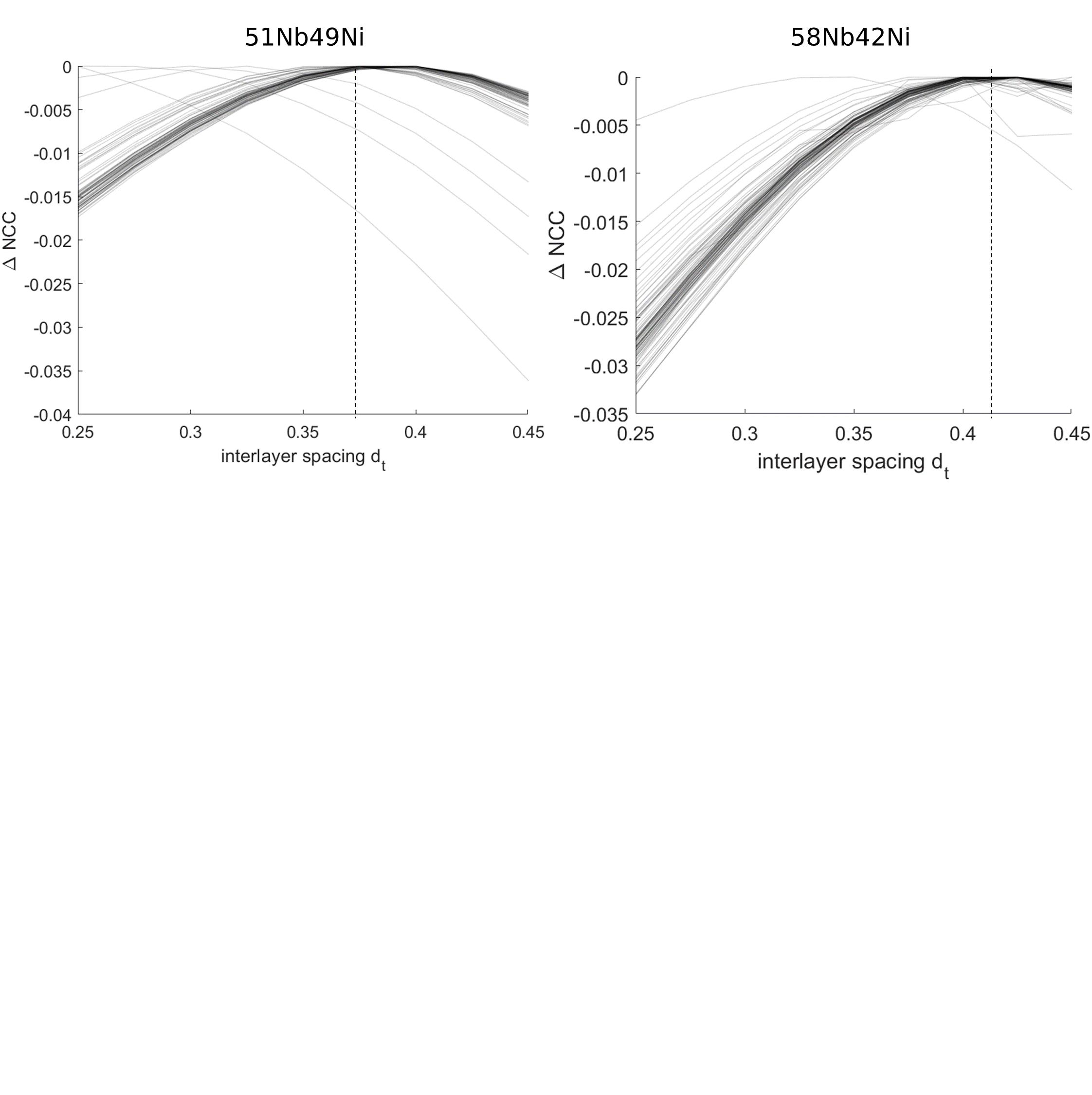}
    \caption{Pattern matching of 100 random patterns from 51Nb-49Ni and 58Nb-42Ni. Outliers in the 51Nb-49Ni are caused by overlap of patterns that contain intensities from two variants of the basal twin, leading to wrong prediction of $d_t$ spacing.}
    \label{fig:PatMatchIndividual}
\end{figure}

\FloatBarrier
\clearpage
\subsection{Influence of NCC on $d_t$ spacing}
\begin{figure}[h]
    \centering
    \includegraphics[width=\linewidth,trim = 0cm 0cm 0cm 0cm, clip]{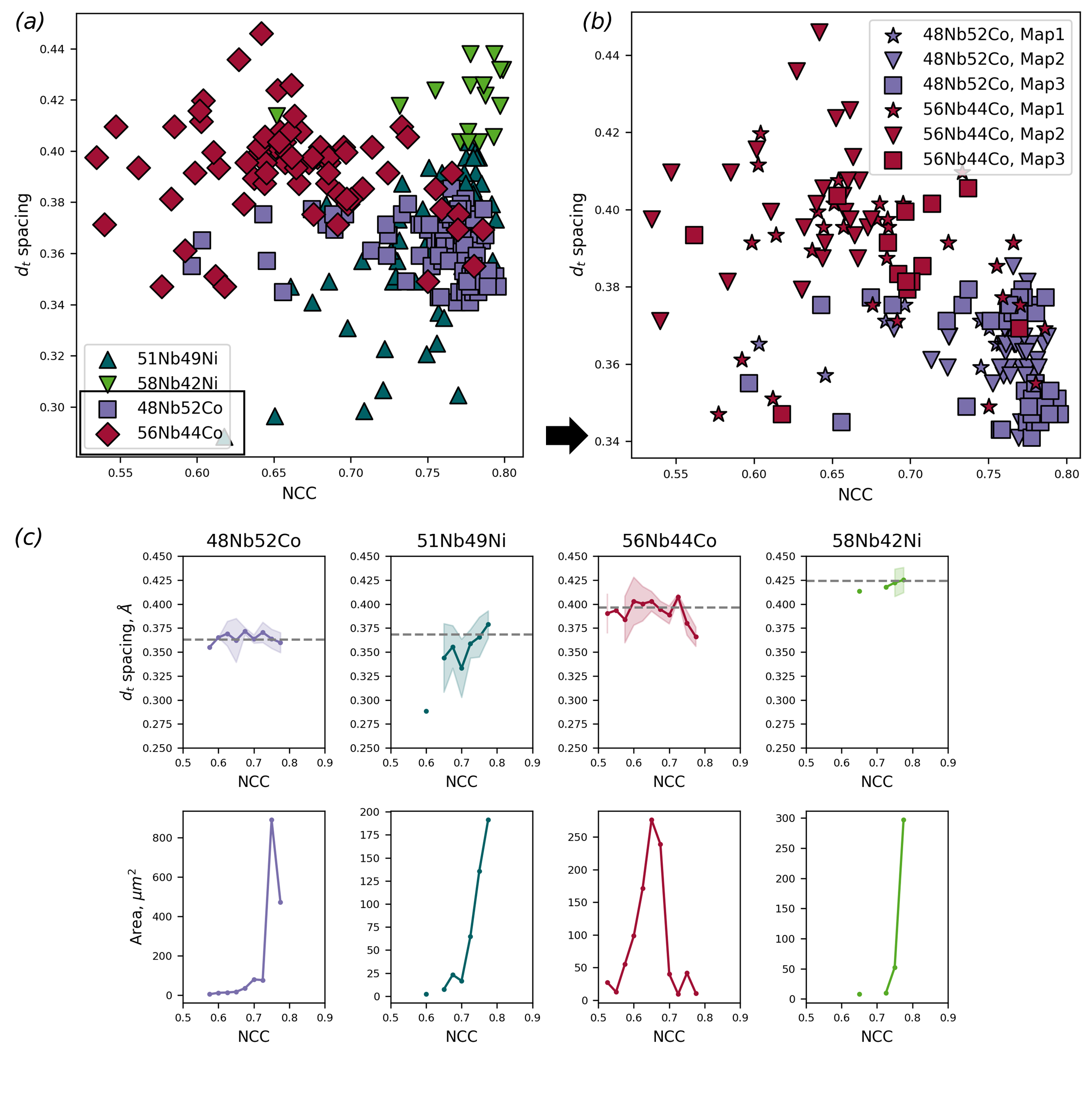}
    \caption{Influence of the normalised cross correlation on the determined $d_t$ spacing: \textit{(a):} Dependency between NCC and $d_t$ for all samples. \textit{(b):} Dependency between NCC and $d_t$, for Nb-Co samples, resolved by map. \textit{(c):} Dependency of $d_t$ spacing on NCC bins, each point marks the beginning of an NCC bin with range 0.025. In addition, the measured area is given for each bin to give context about the density of information. The grey dashed line gives the mean \ac{ILS} value as reported in the main text.}
    \label{fig:NCC_corr_dt}
\end{figure}

% \clearpage
\FloatBarrier
\clearpage
\subsection{Influence of spatial position in PCA on $d_t$ spacing}
\begin{figure}[htbp!]
    \centering
    \includegraphics[width=\linewidth]{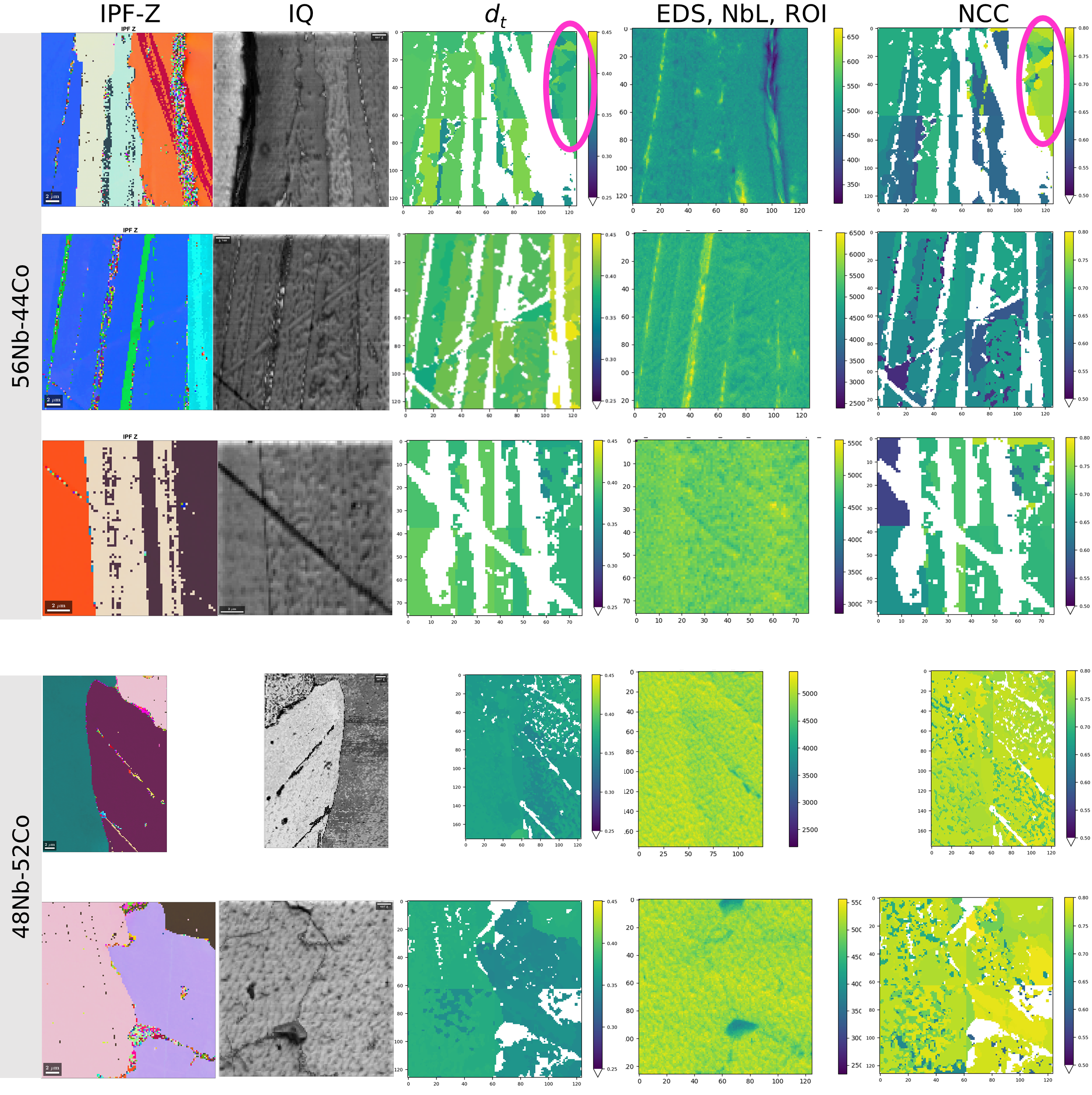}
    \caption{Mosaic panel, showing in rows, three EBSD Maps from 56Nb-44Co and two EBSD maps from 48Nb-52Co. The columns give from left to right: The IPF in the sample z direction, the image quality, the $d_t$ spacing, the EDS signal of the Nb L shell and the NCC values per PCA component. The tiling of the maps is still visible from the ILS and the NCC column. For some grains, there are discontinuities along the tile boundaries. In general, the NCC values seem to be more dependent on the local grain orientation than the tile or PCA component itself. There seems to be no observable correlation with the variation in EDS signal. The region showing high NCC and low $d_t$ values from \autoref{fig:NCC_corr_dt}, is highlighted with a pink circle. }
    \label{fig:PCA_NbCO_spatial}
\end{figure}
\FloatBarrier
\begin{figure}[htbp!]
    \centering
    \includegraphics[width=\linewidth]{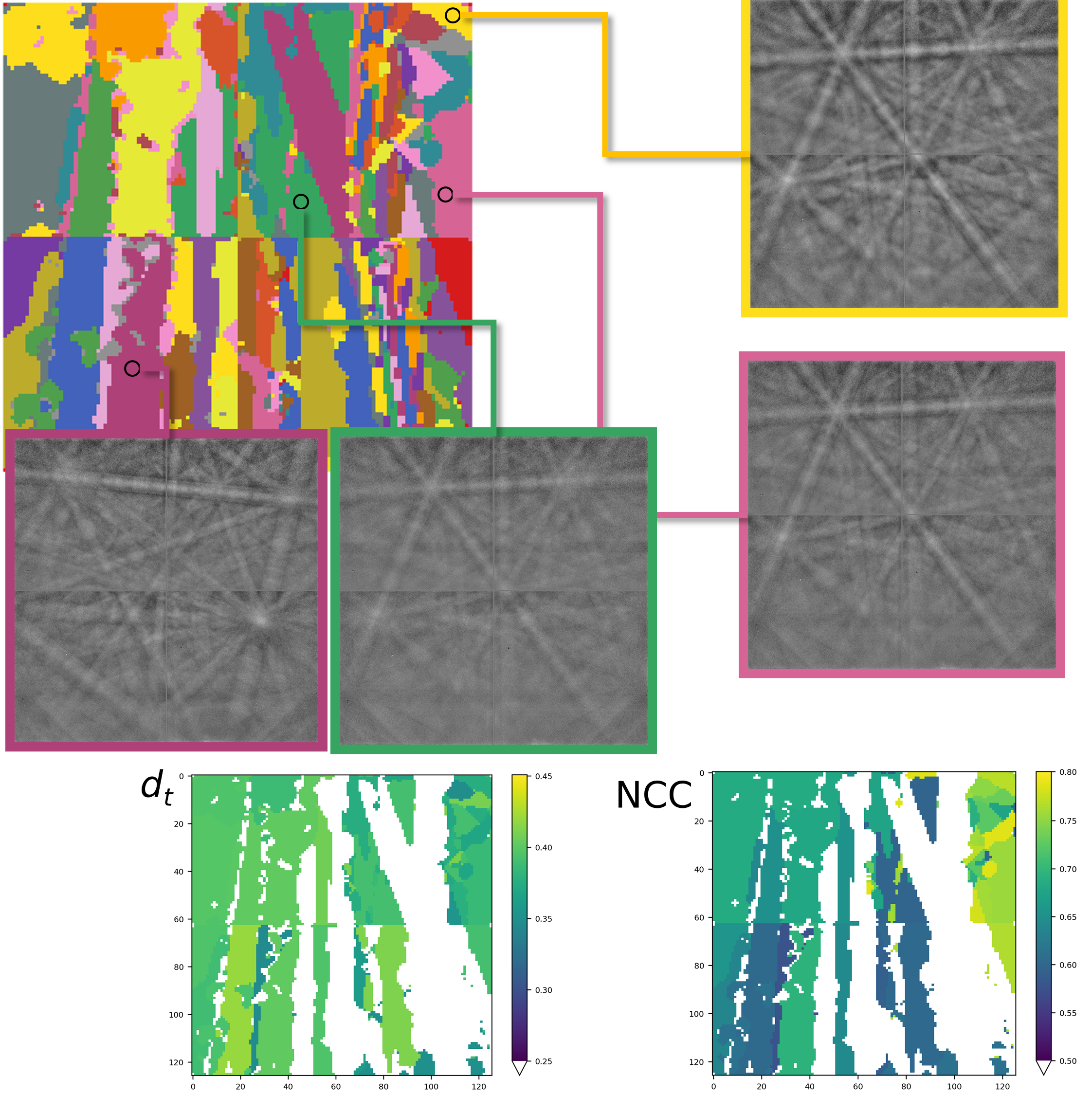}
    \caption{Detailed view of some patterns from the 56Nb44Co sample, correlated with their respective PCA components. In addition the $d_t$ and NCC plots from \autoref{fig:PCA_NbCO_spatial}, are given for comparison}
    \label{fig:56Nb44Co_details}
\end{figure}
\FloatBarrier
\clearpage
\subsection{Influence of Bethe parameters}
\begin{figure*}[!hbp]
   \centering
    \includegraphics[width=0.5\linewidth,trim = 0cm 14cm 8cm 0cm,clip]{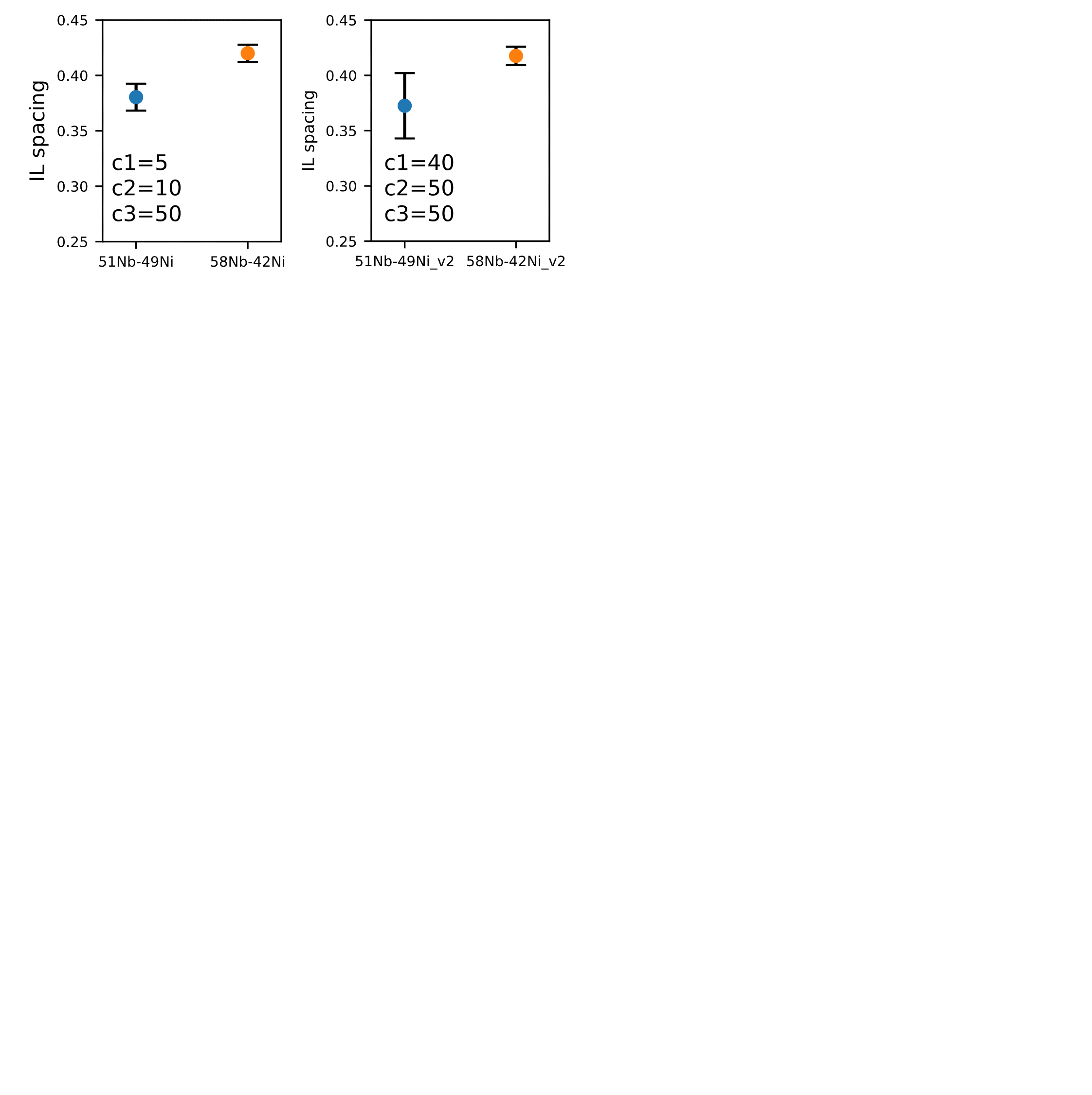}
    \caption{Comparison of ILS values measured with different Bethe parameters}
    \label{fig:BetheInfluence_IL}
\end{figure*}
\clearpage
\subsection{Influence of Energy Filtering}
\begin{figure*}[!hbp]
    \centering
    \includegraphics[width=\linewidth]{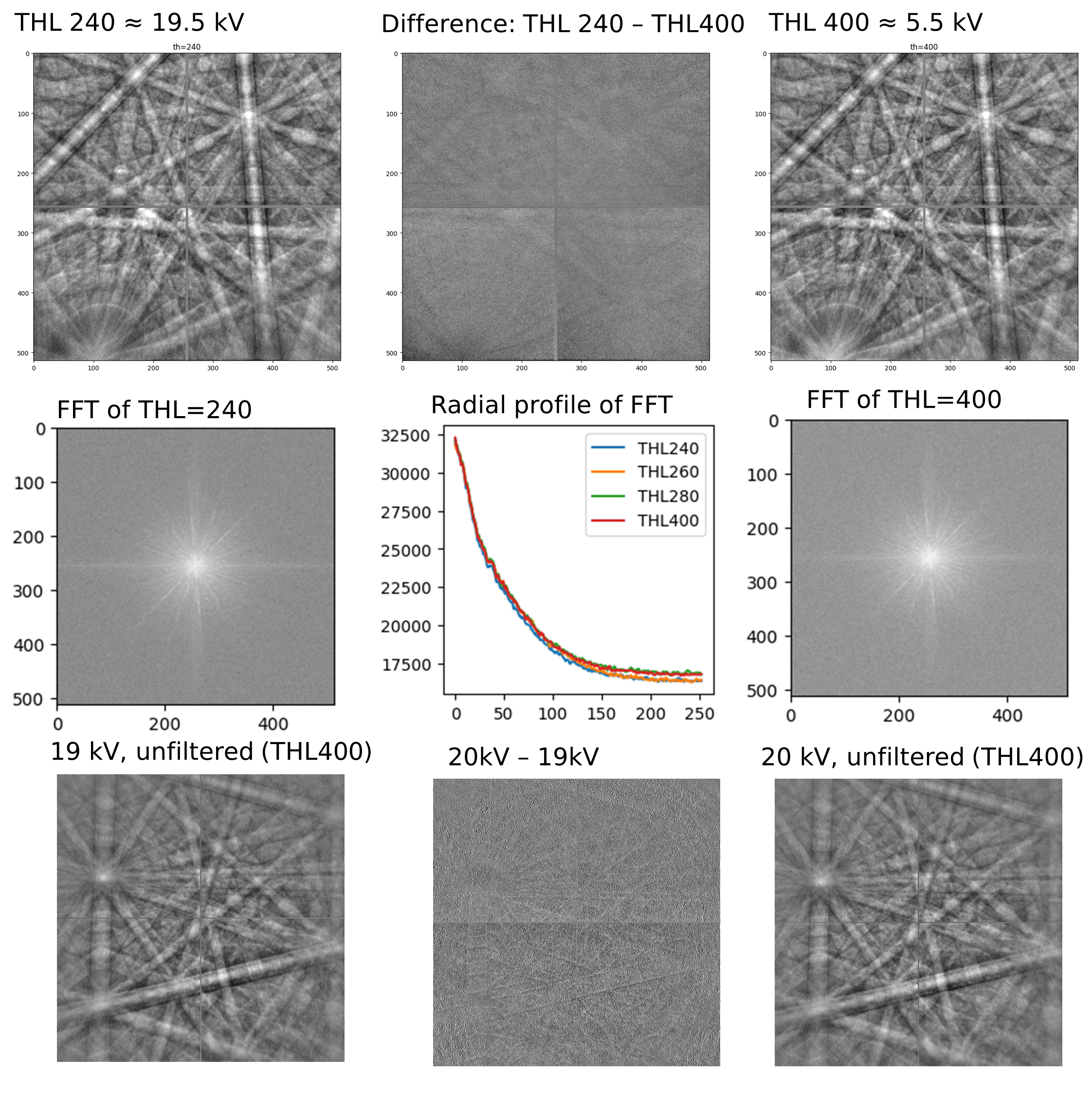}
    \caption{Energy Filtering on example of the 58Nb-42Ni sample. Left side: EBSP for a relatively high threshold of 240, corresponding to \qty{19.5}{kV}, which is close to the primary beam. Right Side. Low threshold of 400, close to the noise level of the detector. The difference Plot between  (centre top) shows the weak difference between the filtered and unfiltered EBSD pattern, mostly corresponding to diffuse (low frequency features). In contrast to this, the difference between an unfiltered EBSP, acquired at 20kV and 19kV beam voltage, shows a lot of higher-order features, which do not appear in the difference between unfiltered and filtered EBSP. }
    \label{fig:EnergyFiltering}
\end{figure*}

The patterns presented here are acquired using the ASI SoPhy software. The relative thresholds (THL) of the detector can be changed to control the filtering voltage. Patterns were acquired with varying exposure times between 300 ms for a THL of 400 and 1.8s for THL=240 and are part of a parameter sweep series over THL values of 240,260,280 and 400, where 10 patterns were acquired per step. For acquisition, a point on the sample 58Nb-42Ni was selected using the Tescan Clara spot mode. To acquire a static background, scanning in the Tescan Clara was enabled, and the same sweep was performed, with a now moving beam.  To visualise the diffraction contrast, we further give the log-power 2D FFT spectra in analogy to e.g \cite{DellaVentura.2026} and \cite{Britton.2026AngRes} and further evaluate the radial profiles of these FFT spectra, as given in the central panel of \autoref{fig:EnergyFiltering}. Similar to the visual impression of the pattern acquired at THL=400 and THL=240, the radial profile does not show significant differences between the different threshold values. As a threshold of THL = 400 allows for significantly faster acquisition rates than THL = 240, the threshold of 400 was chosen in the present study for determining \ac{ILS}.
% \clearpage
\FloatBarrier
% \section{Methods to distinguish the phases}
\clearpage
% \subsection{Visual comparison of Patterns from different site lattice occupancies}

% \begin{figure}[!hbp]
%     \centering
%     \includegraphics[width=\linewidth,trim = 0cm 4cm 0cm 0cm, clip]{S09_MissingBands_OCC.png}
%     \caption{Comparison of experimental pattern with various simulated \acp{EBSP} that show various differences compared to the experimental pattern. The most striking difference is the absence of some band edges in the structure with complete occupation of the 18h site with Ni.}
%     \label{fig:OCC_missingBands}
% \end{figure}
% \clearpage
\subsection{Band profiles of the samples}
\begin{figure}[htbp!]
    \centering
    \includegraphics[width=\linewidth,trim = 0cm 0cm 0cm 0cm, clip]{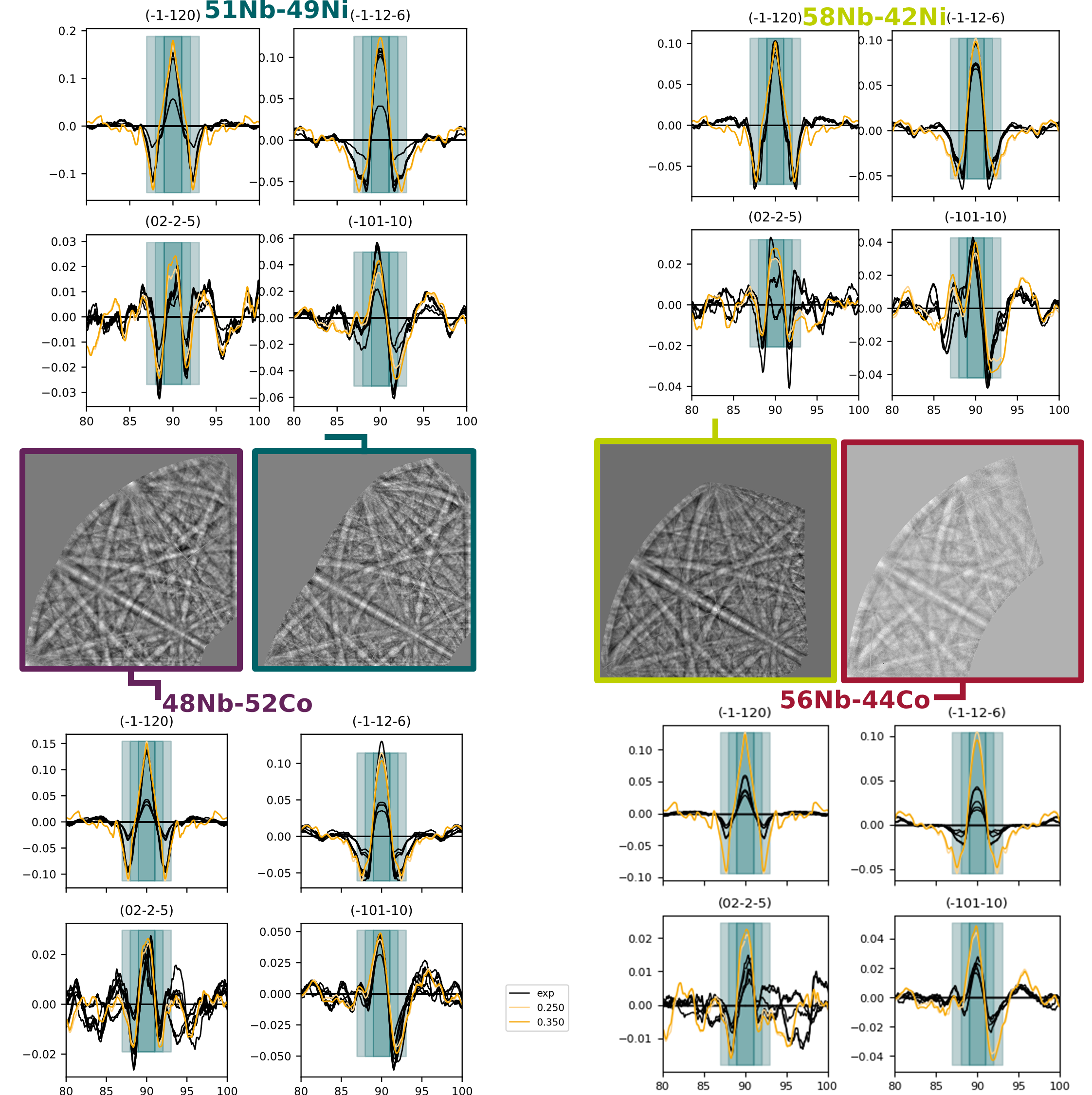}
    \caption{Band profiles from Nb-Ni and Nb-Co samples. All Profiles for the refined Patterns from the first map are shown, respectively. Differences in intensity are mostly due to different sampling (different orientations of the EBSD pattern, while the simulated band profile is only plotted for the first pattern.}
    \label{fig:BandProfiles}
\end{figure}

\clearpage
% \subsubsection{Difference Plots}
% \FloatBarrier
% \begin{figure*}[!h]
%     \centering
%     \includegraphics[width=\linewidth]{S01_OCCInfluenceSimulated.png}
%     \caption{Influence of site lattice occupancy on difference in pattern. The fundamental zone of stereographic projection is shown. The change in site lattice occupancy involves regions around certain bands stronger. Further, there are some specific changes to distinct zone axes.}\label{fig:placeholder}
% \end{figure*}

\begin{figure*}
    \centering
    \includegraphics[width=\linewidth,trim=0cm 12cm 0cm 0cm]{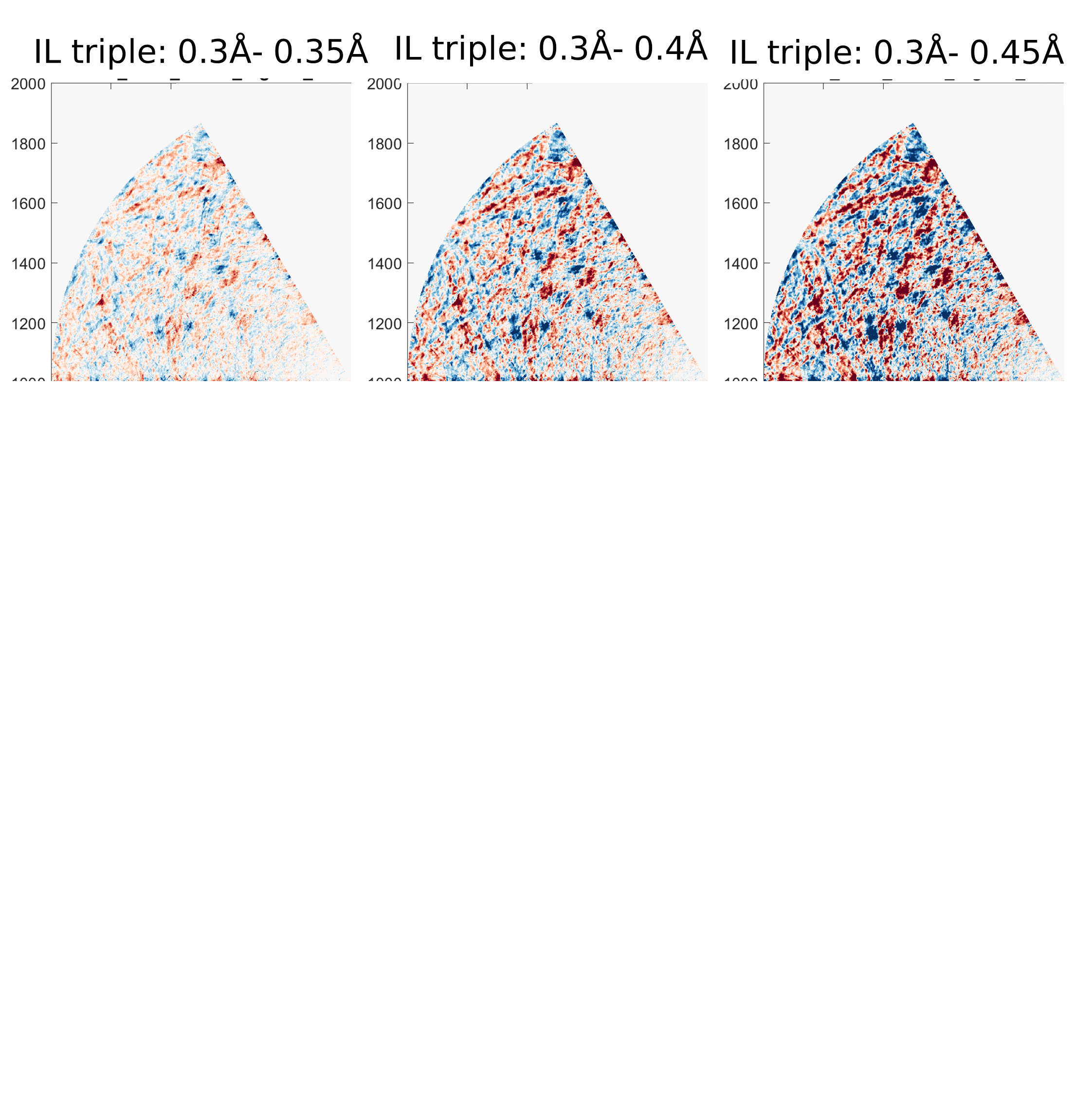}
    \caption{In comparison to the changes of site lattice occupancy, the variation of \ac{ILS} does not influence any specific zone axis as much but rather leads to shifts of band intensities. This is more pronounced at the edge of the stereographic projection.}
    \label{fig:TripleLayerInfluenceOnMasterpattern}
\end{figure*}

% \FloatBarrier

\FloatBarrier
% \clearpage
% \subsection{Band profiles}

% \begin{figure*}
%     \centering
%     \includegraphics[width = \linewidth]{S04_PerpendicularBandprofiles.png}
%     \caption{Band profiles extracted after "rectifying" each band (see top panel of \autoref{fig:parallelBandProfile} and summing values of same angle away from band center. There is a strong variation on some of the band profiles, while other intensities do not vary at all.}
%     \label{fig:placeholder}
% \end{figure*}
\FloatBarrier
\clearpage
% \subsection{Parallel band profiles}
% \begin{figure*}
%     \centering
%     \includegraphics[width=\linewidth,trim=0cm 1cm 0cm 0cm]{S03_AlongBandProfiles.png}
%     \caption{A change in interlayer spacing can be visualised by shifts in intensity maxima of bands crossing the (11-20) band. The black arrows int the two top panels further highlight two very striking differences dependent on the Interlayer spacing for the Nb-Ni mu phase.}
    
%     \label{fig:parallelBandProfile}
% \end{figure*}
% \FloatBarrier
% \clearpage
% \subsection{Metric based (NCC)}

% \clearpage
% \subsection{Influence of Bethe Parameters}

% \begin{figure}
%     \centering
%     \includegraphics[width=\linewidth,trim = 0cm 7cm 0cm 8cm,clip]{S06_BehteComp.png}
%     \caption{Influence of Bethe paramters, a set of relatively high cutoff values for the bethe pertubation (left side does not change the qualtiative outcome of best matching structures for the different site lattice occupancies a lot}
%     \label{fig:BetheInfluenceOCC}
% \end{figure}

\FloatBarrier

% \bibliographystyle{elsarticle-num} 
%  \bibliography{references}

% \end{document} % supplementary appended for arXiv
% \input{Supplementary} % Your supplementary file (figures, sections, etc.)
% \end{supplementary}
% \fi

%%%%% comment in for final supplementary compile.
% \bibliographystyle{elsarticle-num} 
%  \bibliography{references}

% \bibliographystyle{elsarticle-num} 
 % \bibliography{references}

\end{document}